\documentclass[twocolumn,aps,prl,superscriptaddress,tightenlines]{revtex4-1}
\usepackage{amsmath, amsfonts,amssymb}
\usepackage{comment}
\usepackage{dsfont}
\usepackage{soul}
\usepackage{mathtools}			 % Math capabilities
\usepackage{physics}
\usepackage{wasysym}
\usepackage{graphicx}
\usepackage{blindtext}
\usepackage{booktabs}
\usepackage{etoolbox}
\usepackage{dcolumn}
\usepackage{epsfig}
\usepackage{comment}
\usepackage{color}
\usepackage{textcomp}
\usepackage[colorlinks,citecolor=blue]{hyperref}
\usepackage{CJKutf8}

\makeatletter
\newcommand{\smalldot}{\raisebox{0.25ex}{\scalebox{0.7}{$\bullet$}}}
\newcommand*\bigcdot{\mathpalette\bigcdot@{.5}}
\newcommand*\bigcdot@[2]{\mathbin{\vcenter{\hbox{\scalebox{#2}{$\m@th#1\bullet$}}}}}
\makeatother

\begin{document}
\title{Floquet Engineering and Harnessing Giant Atoms in Frequency-Comb Emission and Bichromatic Correlations in Waveguide QED}
\author{Qing-Yang Qiu}
\affiliation{School of Physics and Institute for Quantum Science and Engineering, Huazhong University of Science and Technology, and Wuhan institute of quantum technology, Wuhan 430074, China}

\author{Li-Li Zheng}
\affiliation{School of Artificial intelligence, Jianghan University, Wuhan 430074, China}

\author{Ying Wu}
\affiliation{School of Physics and Institute for Quantum Science and Engineering, Huazhong University of Science and Technology, and Wuhan institute of quantum technology, Wuhan 430074, China}

\author{Xin-You L\"{u}}\email{xinyoulu@hust.edu.cn}
\affiliation{School of Physics and Institute for Quantum Science and Engineering, Huazhong University of Science and Technology, and Wuhan institute of quantum technology, Wuhan 430074, China}

\date{\today}% It is always \today, today,
\begin{abstract}
The capability to design spectrally controlled photon emission is not only fundamentally interesting for understanding frequency-encoded light-matter interactions, but also is essential for realizing the preparation and manipulation of quantum states. Here we consider a dynamically modulated qubit array, and realize frequency-controlled single-photon emission focusing on the generation of a frequency comb constituted solely of even-parity or  anti-Stokes sidebands. Our system also offers parity-dependent bunching and antibunching in frequency-filtered quantum correlations. In particular, the waveguide quantum electrodynamics (QED) setup is extended to include chiral and non-local coupling architectures, thereby enhancing its versatility in Floquet engineering. Our proposal also supports the predictable generation of high-dimensional entangled quantum states, where the corresponding effective Hilbert space dimension is well controlled by energy modulation. Moreover, the utilisation of sophisticated numerical tools, such as the matrix product states (MPSs) and the discretization approach, enables the efficient simulation of multi-photon dynamics, in which the non-Markovian Floquet steady states emerge. This work fundamentally broadens the fields of collective emission, and has wide applications in implementing frequency-encoded quantum information processing and many-body quantum simulation.
\end{abstract}
%corresponding excitations

\maketitle
An optical frequency comb, characterized by a series of equally spaced discrete lines, finds broad application in various fields~\cite{Fortier2019,Chang2022,WANG2023114137}. The ability to accurately control the photonic sideband emission and process frequency-encoded quantum entanglement via  one channel is at the heart of numerous quantum information applications, including universal one-way quantum computing~\cite{PhysRevLett.86.5188,PhysRevLett.101.130501}, scalable generation of entangled cluster states~\cite{PhysRevLett.107.030505,Du2023} and high-dimensional entanglement protocols~\cite{Kues2017,Lukens17,PhysRevB.110.L220301} in optical frequency comb. In many quantum architectures, the implementation of photon-mediated quantum protocols requires highly controlled spectral distribution, which often suffers undesired spectral diffusion and inhomogeneity~\cite{Sweeney2014,PhysRevLett.116.033603}. Therefore, suppressing these imperfections or equivalently designing any target emission spectra is of paramount importance for contemporary quantum technology. With respect to this motivation, spectral-shaping techniques based on electro-optic phase modulators~\cite{Specht2009}, nonlinear frequency conversion~\cite{ Lavoie2013,sciadv.1501223}, and Floquet engineering of energy modulation~\cite{Lukin2020} have been proposed.

\begin{figure}[!ht]
  \centering
  \includegraphics[width=8.5cm]{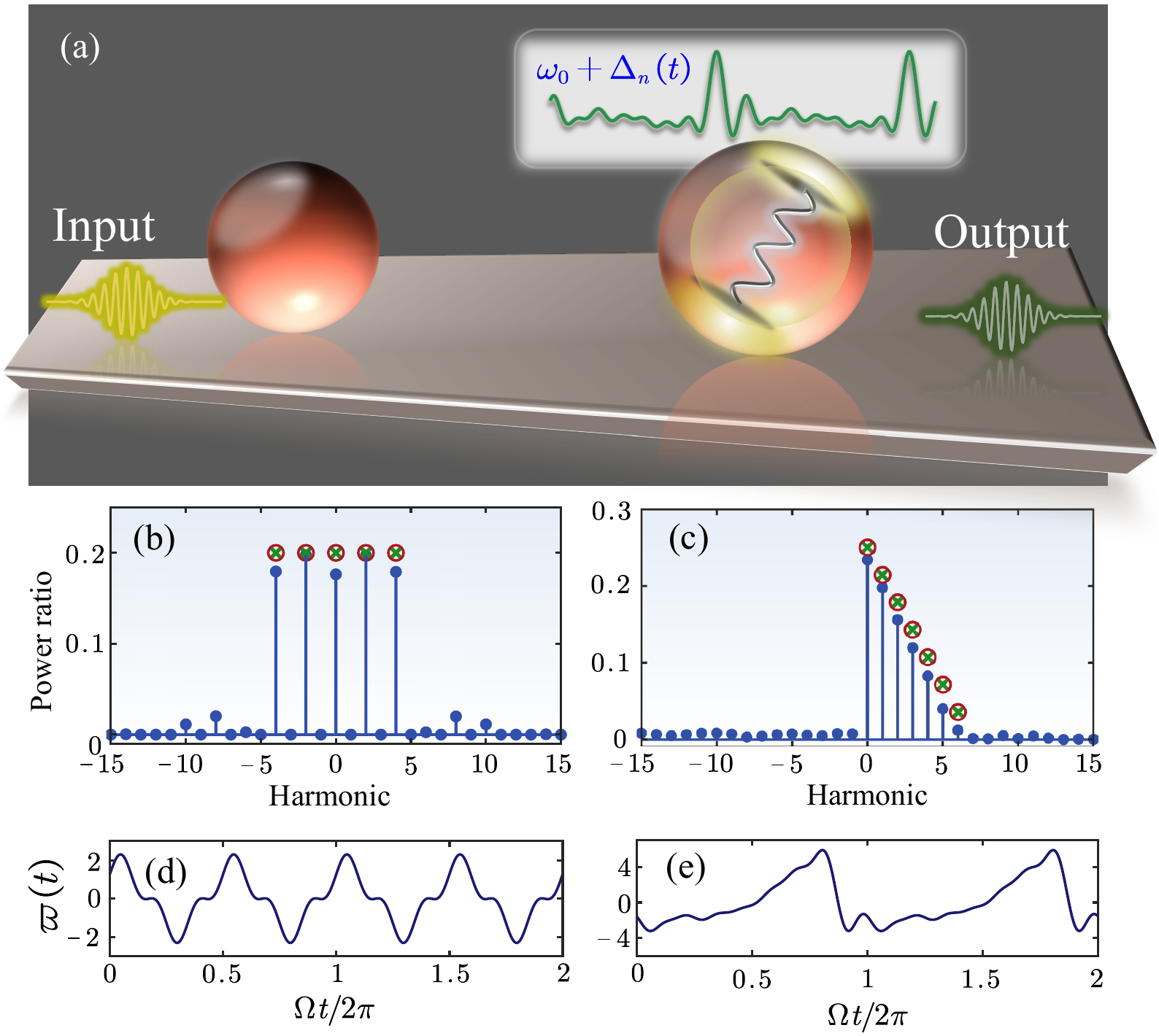}
  \caption{(a) Sketch of $N$ emitters coupling to a 1D waveguide with the dynamical modulation $\omega_{0}+\Delta_{n}(t)$ for $n^{{\rm th}}$ qubit. Panels (b) and (c) are optimized Floquet spectra corresponding to pure even harmonics and pure anti-Stokes harmonics, respectively, where the symbols marked by the red circles with green cross inside denote the target photon spectra. Blow each spectrum panel, we show the optimized modulation $\Delta(t)$ over two periods, as illustrated in panels (d) and (e).}\label{fig1}
  \vspace{-20pt}
\end{figure}

Of particular interest is deterministically entangling the sidebands in an output frequency comb, which subsequently serves as an indispensable quantum resource and has been demonstrated in various experimental platforms~\cite{PhysRevLett.114.050501,PhysRevLett.130.120601,APL0203426}. In addition to the central significance of practical applications, the frequency-resolved photon correlations also make it feasible to unveil the fundamental and hidden properties of the majority of complex quantum systems~\cite{PhysRevLett.109.183601,PhysRevA.94.033835,Schmidt_2021}.
In particular, it was recently shown that parity-protected sideband correlations can be achieved by combining temporal modulation with phase shifts~\cite{PhysRevLett.130.023601}. Due to the limited theoretical research in this emerging field, it is still of fundamental and practical interest to search for novel mechanisms to engineer highly controllable frequency-encoded quantum correlations.

Here, we investigate the frequency-resolved scattering of photons from a periodically modulated atomic chain interacting with a continuum of bosonic modes in a one-dimensional (1D) waveguide.  In the resolved-sideband regime, the single atom functions as a spectrally reconfigurable single-photon source. Particular emphasis is placed on the deterministic generation of photonic frequency comb containing only pure even-parity or pure anti-Stokes sidebands due to their wide applications in quantum information.  Building upon this, we seek to further explore approach for entangling sidebands from a frequency comb in a desired manner. By introducing non-local and chiral interaction, the parametric flexibility is greatly extended to enable the efficient generation of arbitrary photon-photon correlations. The statistical properties of photons are numerically analyzed in light of the equal-time photon-photon correlation function, with the results rigorously validated via the frequency-domain $N$ photon scattering matrix formalism. More intriguingly, the delay-induced non-Markovianity offers an enlightening perspective on light-mater interactions under dynamical modulation. We have numerically investigated the dynamics of the system plus the waveguide modes by considering both coherent and pulsed wavepacket drives, with real-time photon transport dynamics also provided to ensure a comprehensive description of the dynamics. Our work opens new possibilities for the generation and processing of high-dimensional quantum states that are encoded by distinct frequencies.

\emph{Waveguide QED setup and Floquet engineering}---The waveguide QED setup consists of $N$ two-level emitters with the ground state $\ket{g}$ and the excited state $\ket{e}$, coupled to the propagating modes in the 1D waveguide, as illustrated in Fig.\ref{fig1}(a). The radiative loss is characterized by the radiative decay rate $\gamma_{1D}$ of an individual coupled emitter. To give rise to a frequency comb equally spaced  by $\Omega$ in the scattered light spectrum, we impose a dynamical modulation on the resonance frequency of $n^{{\rm th}}$ qubit with a general form
\begin{align}\label{eq1}
\omega_{n}(t)=\omega_{0}+\Delta_{n}(t)=\omega_{0}+\sum\limits_{r=1}^{R}A^{n}_{r}\cos(r\Omega t+\alpha^{n}_{r}),
\end{align}
where $\omega_{0}$ is the equilibrium qubit resonance frequency;  $A^{n}_{r}$ and $\alpha^{n}_{r}$ are $n$-dependent amplitude and phase for $r^{{\rm th}}$ modulation tone, respectively. Here, $n$ enumerates the qubits that are spaced periodically, and the total number of the considered modulation tones is assumed to be $R$.

\begin{figure}
  \centering
  \includegraphics[width=8cm]{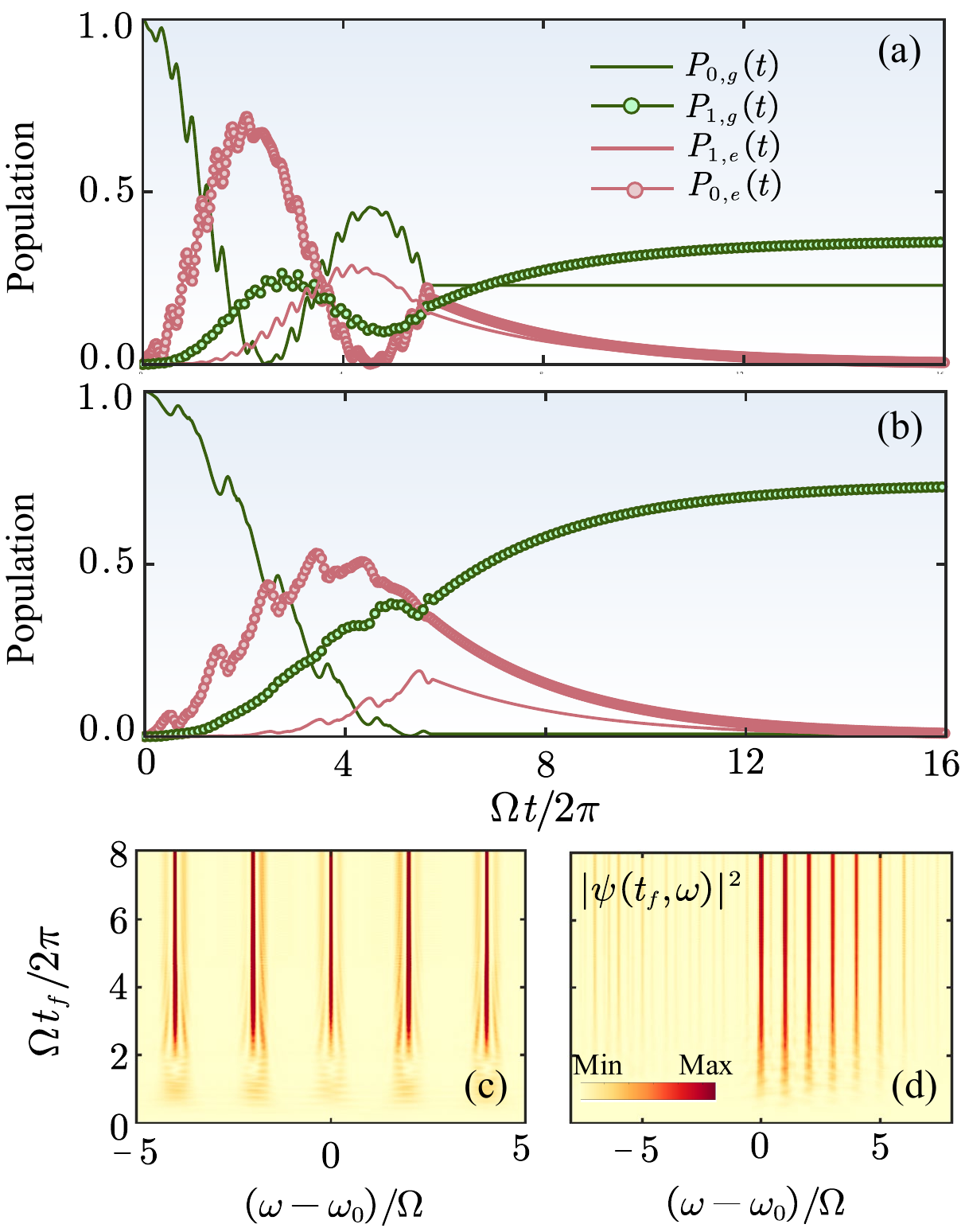}
  \caption{Panels (a) and (b) show the time evolution of the probabilities $P_{0,g}(t),P_{0,e}(t),P_{1,g}(t)$ and $P_{1,e}(t)$ for an energy-modulated atom driven by a coherent pulse. Panels (c) and (d) are single photon emission spectra determined by Eq.(\ref{eq4}). Note that the dynamical modulation parameters for panels (a) and (c) [(b) and (d)] correspond to the Floquet optimization results shown in Fig.\,\ref{fig1}(d) [e]. Additionally, the duration of the coherent pulse is set to $T_{d}=2/\gamma_{1D}$.}\label{fig2}
  \vspace{-20pt}
\end{figure}

We are interested in the fast modulation regime, i.e., $\Omega \gg \gamma_{1D}$, in which distinctive sidebands can be captured from the emission spectrum. The resulting photon spectrum typically contains three dominant types of components: elastic scattering refers to emission processes where no exchange of energy occurs with the optical mode; in contrast, during Stokes and anti-Stokes (red or blue) processes, the emitted photon either loses or absorbs energy corresponding to $l\,(l>0)$ modulation photons, leading to light emission at frequencies $\omega_{{\rm s/as}}=\omega_{0}\mp l\Omega$. We first focus on the simplest case of a single qubit ($N=1$), which is sufficient to modify any desired single-photon spectrum by optimizing periodic modulation $\Delta(t)$. Note that in the single-qubit case, the subscript in $\Delta_{n}(t)$ is omitted.  More general case of multiple qubits will be discussed in the context of engineering sideband correlations, as we will demonstrate subsequently.

We proceed by generating controllable spectrum through engineering the Floquet phase, $\varpi(t)\equiv\exp\left[-i\int_{0}^{t}\Delta(\tau)d\tau\right]$, which exhibits temporal periodicity as well. This complex $\varpi(t)$ allows the occurrence of an asymmetric Floquet spectrum, with the components $X_{k}$ determined by the discrete Fourier transform and taking the form of~\cite{SM}
\begin{align}\label{eq2}
X_{k}=&\sum\limits_{\gamma_{1}\cdots\gamma_{R};\kappa_{1}\cdots\kappa_{R}=-\infty}^{\infty}\Big\{i^{\sum\limits_{r=1}^{R}\gamma_{r}+\kappa_{r}}[\prod_{r=1}^{R}J_{\gamma_{r}}(\frac{a_{r}}{\Omega})J_{\kappa_{r}}(\frac{b_{r}}{\Omega})]\nonumber\\
&\times e^{i\sum\limits_{r=1}^{R}(\frac{1}{2}\gamma_{r}+\kappa_{r})\pi}\Big\} e^{i\sum\limits_{r=1}^{R}\frac{b_{r}}{r\Omega}},
\end{align}
where $a_{r},b_{r}$ are Fourier coefficients, corresponding to the cosine and sine components of $\Delta(t)$, respectively; $J_{n}(\smalldot)$ is the Bessel function of order $n$ and $i$ is the imaginary unit. Note that the summation over $\gamma_{r}$ and $\kappa_{r}$ in Eq.(\ref{eq2}) is constrained by $\sum\limits_{r=1}^{R}r(\gamma_{r}+\kappa_{r})=k$. With the goal of shaping modulation signal $\Delta(t)$ to align with the target's emission spectrum, we apply here the Broyden-Fletcher-Goldfarb-Shanno (BFGS) algorithm in conjunction with the particle swarm optimization ~\cite{PhysRevB.100.235452,IEEE1256,Lukin2020}.

We present two paradigmatic optimization results of Floquet spectra: pure even harmonic spectrum containing only the even-parity sidebands [see Fig.\ref{fig1}(b)];  and pure anti-Stokes harmonic spectrum containing only the blue sidebands combined with a single Rayleigh emission band [see Fig.\ref{fig1}(c)].  Figs.\ref{fig1}(d)-(e) are their corresponding optimized Floquet phases. The target spectra are marked by the red circles with green cross inside. It is evident that the spectra obtained through numerical optimization align closely with the expected ones.

The two types of Floquet spectra, i.e., pure even and pure anti-Stokes harmonics, notably, hold significant practical importance. More concretely, the former fundamentally predicts a novel form of dipole acceleration that stems from the permanent dipole moment~\cite{PhysRevLett.119.173201,Appl124}, while the latter paves the way for particle cooling by enabling the transfer of energy from particles to photons, thereby reducing the particle's thermal motion and lowering their temperature ~\cite{PhysRevLett.85.3600,science3993}.

\emph{Controllable frequency-comb emission}---We proceed by investigating the emission from a single modulated qubit, anticipating the desired photon spectra when imposing a target modulation. To this end, we consider an energy-modulated emitter driven by a resonant coherent pulse during the time interval $[0,T_{d}]$ with a driving strength $\xi$, while the qubit is initially in its ground state and the field is  in the vacuum state $\ket{{\rm vac}}$ containing empty excitations~\cite{ SM}. Leveraging the advanced features of recently developed Green’s functions techniques~\cite{PhysRevB.98.144112,PhysRevA.102.033707}, we are able to efficiently capture the entire dynamics of the time-dependent quantum system. The calculation focuses on the time evolution of the population $P_{m,\sigma}(t)$, which represents the probability of finding $m$ photons propagating along the waveguide, with the qubit in either its ground state ($\sigma=g$) or excited state ($\sigma=e$) at time $t$. The quantities of interest can be formally expressed as
\begin{align}
P_{0,g/e}(t)=&\left |\bra{{\rm vac}; g/e}U(t,0)\ket{{\rm vac}; g}\right|^{2},\nonumber\\
P_{1,g/e}(t)=&\int\left |\bra{x; g/e}U(t,0)\ket{{\rm vac}; g}\right|^{2}dx,\label{eq3}
\end{align}
where $U(\tau_{+},\tau_{-})$ is the Schr$\rm{\ddot{o}}$dinger picture propagator~\cite{PhysRevB.98.144112} giving the relationship between the state of the quantum system at time instants $\tau_{\pm}$.

\begin{figure}
  \centering
  \includegraphics[width=9cm]{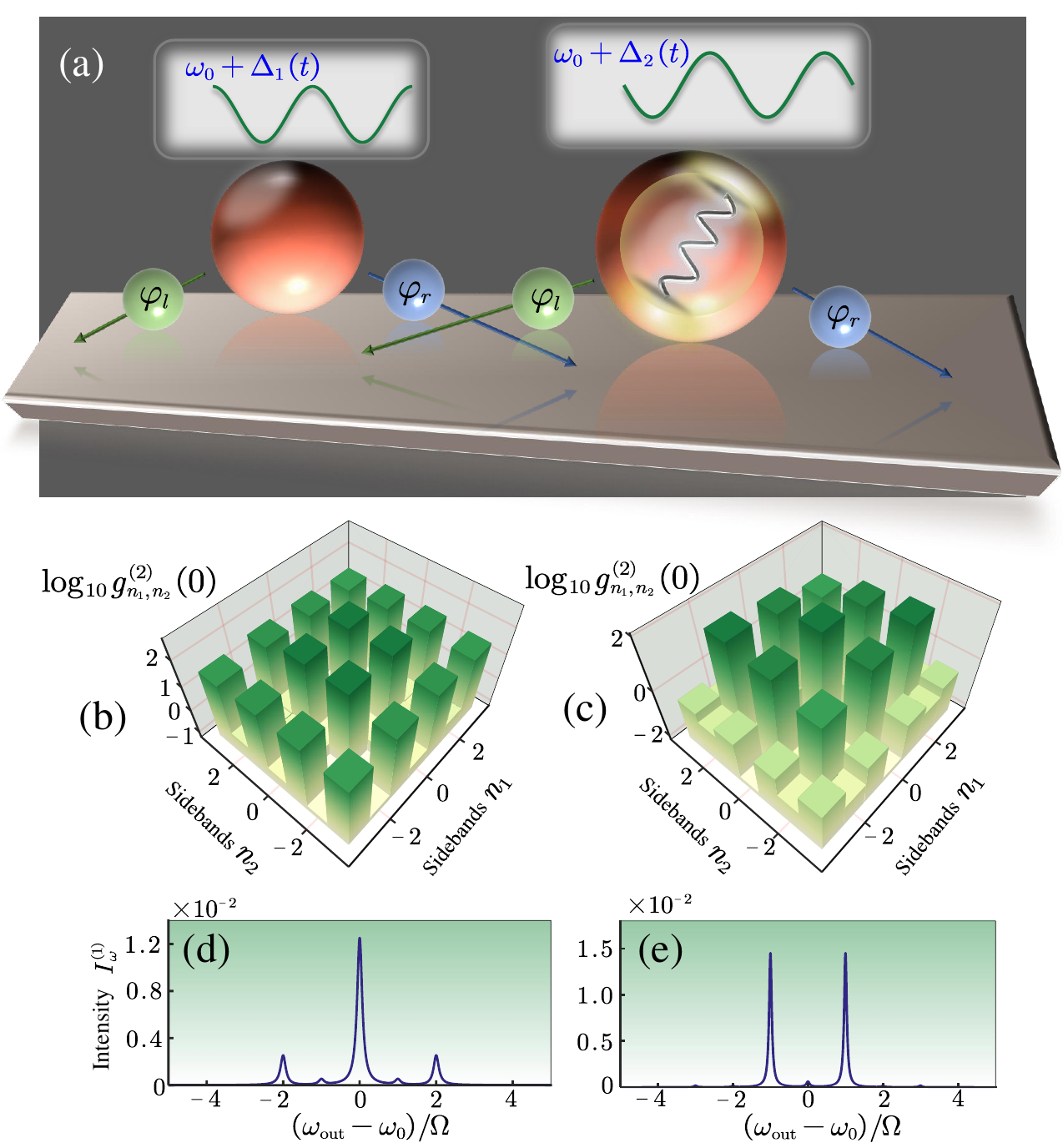}
  \caption{(a) Sketch of a linear chain of $N=2$ emitters non-locally coupling to a 1D waveguide with the dynamical modulation $\omega_{0}+A\cos(\Omega t)$ for qubit $1$ and $\omega_{0}+A\cos(\Omega t+\alpha)$ for qubit $2$. Each qubit couples the waveguide modes at two separate locations with complex coupling rate, i.e., $\varphi_{l}$ for the left coupling point marked by green ball and  $\varphi_{r}$  for the right coupling point marked by blue ball. Note that there are two additional photon detectors, not shown here, placed at the same location as the left-most coupling point of the qubit pair. Panels (b) and (c) show the bichromatic photon-photon correlations as a function of sideband number $n_{1},n_{2}$, while panels (d) and (e) are their corresponding single-photon scattering intensity. The calculations are performed using $\varphi=0,\varphi_{l}=\varphi_{r}=0$ for panels (b) and (d), and $ \varphi=0.5\pi,\varphi_{l}=0, \varphi_{r}=\pi$ for panels (c) and (e). Other parameters are $\Omega=500\gamma_{1D},A=1.5\Omega$.}\label{fig3}
\end{figure}

Figures \ref{fig2}(a)-(b) show the time dependence of the probabilities defined in Eq.(\ref{eq3}) by adopting the optimized modulations $\Delta(t)$, which generate accordingly the pure even and anti-Stokes sidebands. The curves of $P_{m,\sigma}(t)$ exhibit prominent envelope-like oscillations superimposed with smaller, sharp fluctuations due to the periodic qubit modulation. Since the observations of the population dynamics are not color-coded, we thus transform the time-dependent single-photon states into the frequency domain and simulate the emission spectra $|\psi(\omega,t_{f})|^{2}$ with
\small
\begin{align}
\!\!\psi(\omega,\!t_{f})\!=\!-i\sqrt{\frac{\gamma_{1D}}{2\pi}}\int_{0}^{t_{f}}\!\!\bra{g}U_{{\rm eff}}(t_{f},\!t)\sigma U_{{\rm eff}}(t,\!0)\ket{g}\!e^{i\omega t}dt, \label{eq4}
\end{align}
where $U_{{\rm eff}}(\tau_{+},\tau_{-})$ denotes the propagator for the effective Hamiltonian by tracing out the 1D continuum of bosonic modes~\cite{ SM}.

We show the numerical simulation of the output photon spectra in Figs.\ref{fig2}(c)-(d), which exhibit a steady spectral structure that is consistent with the Floquet engineering results shown in Figs.\ref{fig1}(b)-(c), respectively. More paradigmatic and rich spectra, along with their corresponding modulation parameters $a_{r},b_{r}$ that are obtained via Floquet optimization, are shown in ~\cite{ SM}. Up to this stage, we have demonstrated a versatile quantum strategy for producing arbitrary tunable optical frequency comb sources using just a single two-level system. In particular, this color-controlled scheme opens new possibilities for a significant revolution in the fields of information processing~\cite{PhysRevLett.124.190502, Zhang2023}, time-frequency metrology ~\cite{Pupeza2021} and sensing~\cite{Picque2019}.

\emph{Tunable frequency-filtered quantum correlations}---We are now in a position to analyze and unravel a rich landscape of frequency-resolved correlations. In order to generate strongly correlated quantum multi-color states, a modulated atomic array is essential. As sketched in Fig.\ref{fig3}(a), we consider a pair of waveguide-coupled qubits. Each qubit interacts non-locally with the waveguide modes at two separate locations, accompanied by well-defined coupling phases: $\varphi_{l}$ for the left coupling point and  $\varphi_{r}$  for the right coupling point, being reminiscent of chiral ``giant atom"~\cite{PhysRevLett.127.233601,PhysRevX.13.021039,DuL2023,PhysRevLett.133.063603} with braided configuration~\cite{PhysRevLett.120.140404,Qiu2023, PhysRevResearch.6.033243}. The atomic coupling legs are spaced equidistantly, introducing a photonic propagation phase $\varphi\equiv\omega_{0}d/v_{g}$, where $d$ is the distance between adjacent coupling points and $v_{g}$ is the group velocity of the field. And the qubits are modulated with a relative phase $\alpha$, which leads to $\Delta_{1}(t)=A\cos(\Omega t), \Delta_{2}(t)=A\cos(\Omega t+\alpha)$. This relative modulation phase is crucial in generating correlated sidebands in a frequency comb, as illustrated in~\cite{SM}.

To quantify the quantum correlations between a pair of scattered photons into the sidebands with energies $\omega_{0}+n_{1}\Omega$ and $\omega_{0}+n_{2}\Omega$, respectively, we calculate the second-order cross-correlations~\cite{PhysRevLett.109.183601}
\small
\begin{align}
g^{(2)}_{n_{1},n_{2}}(0)=I^{(2)}_{n_{1},n_{2}}/I^{(1)}_{n_{1}}I^{(1)}_{n_{2}}, \label{eq5}
\end{align}
where $I^{(1)}_{n_{1(2)}}$ denotes the scattering intensity of a single photon into sideband $n_{1(2)}$, and $I^{(2)}_{n_{1},n_{2}}$ represents the scattering intensity of a photon pair into the respective sidebands $n_{1}$ and $n_{2}$. The measurement of frequency-filtered quantum correlations is performed by positioning two photodetectors, $D_{1}$ and $D_{2}$,  on the left side of the system, where the detector frequencies are $\omega_{D_{1}}=\omega_{0}+n_{1}\Omega$ and $\omega_{D_{2}}=\omega_{0}+n_{2}\Omega$. This measurement setup provides the capability for frequency-resolved correlations $g^{(2)}_{n_{1},n_{2}}(0)=\langle\sigma^{\dagger}_{D_{1}}\sigma^{\dagger}_{D_{2}}\sigma_{D_{2}}\sigma_{D_{1}}\rangle/\langle\sigma^{\dagger}_{D_{1}}\sigma_{D_{1}}\rangle\langle \smash{\sigma^{\dagger}_{D_{2}}} \sigma_{D_{2}} \rangle$. We consider the detectors, modeled as point-like two-level atoms, to be co-located with the left coupling point of the first qubit, i.e., $\omega_{0}|x_{D_{1(2)}}-x_{11}|/v_{g}=0$ (or $2\pi$). Here, $x_{D_{1(2)}}$ and $x_{np}$ denotes the positional coordinates of the detector $D_{1(2)}$ and the $p^{{\rm th}}$ leg from $n^{{\rm th}}$ system qubit, respectively. Our quantum optics model features a spin network comprising both locally and non-locally coupled atoms with a complex coupling architecture. Thus, a more generalized master equation is necessary to carry out the associated numerical simulations as we present in~\cite{ SM}.

\begin{figure}
  \centering
  \includegraphics[width=7.0cm]{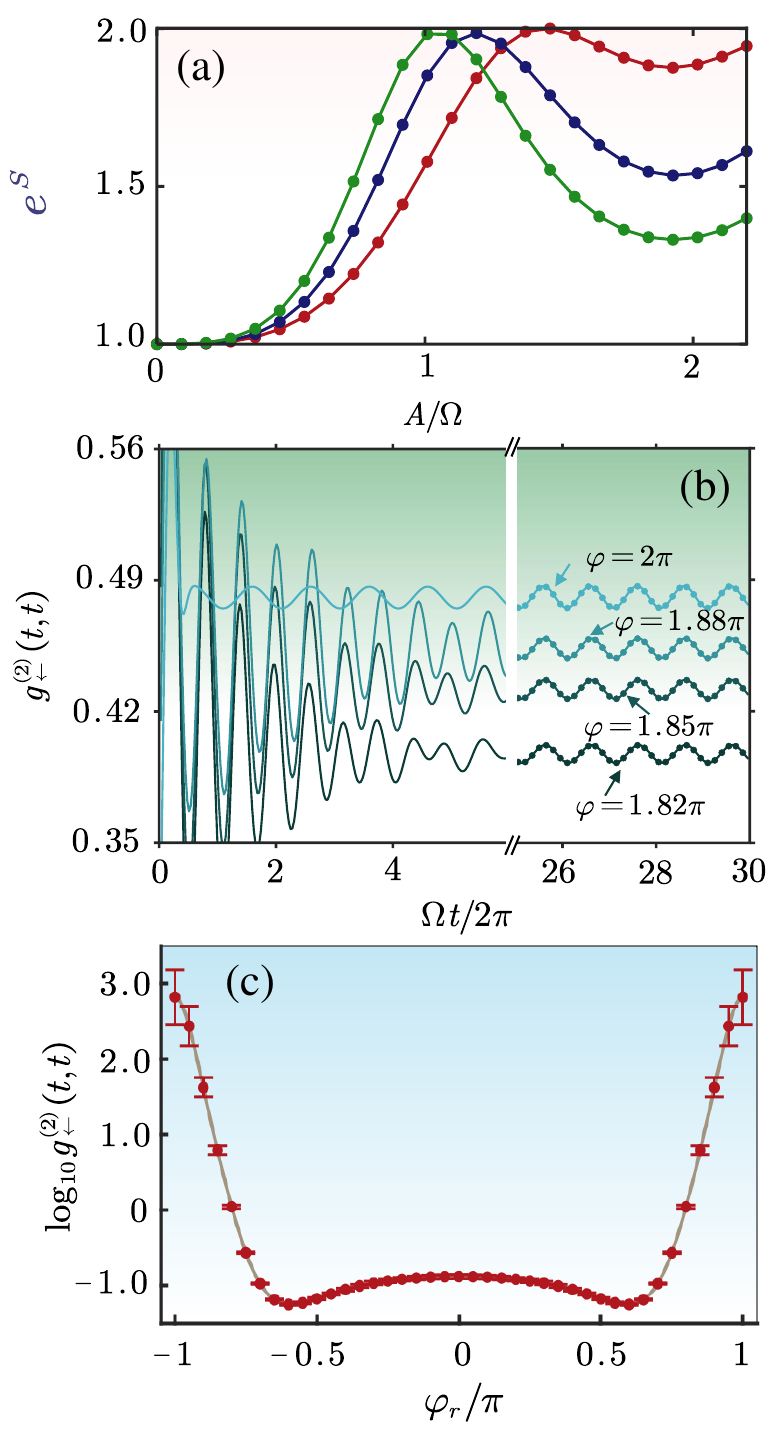}
  \caption{(a) Exponential of the entanglement entropy $e^{S}$ as a function of scaled modulation amplitude $A/\Omega$. Values of $\varphi,\varphi_{l/r}$ are distinguished by coloring: $\varphi=0,\varphi_{l/r}=0$ (red),  $\varphi=0.75\pi,\varphi_{l/r}=0$ (blue),  $\varphi=\pi/2,\varphi_{l}=0,\varphi_{r}=\pi$ (green). (b) Comparison of numerical (solid lines) and analytical (dotted lines) solutions for equal-time photon-photon correlation functions $g_{\leftarrow}^{(2)}(t,t)$. The excellent agreement between the two demonstrates the accuracy of the analytical method. The calculations are performed by gradually  changing the distance $d$ between adjacent coupling points. And the other parameters applied in this panel are $\varphi_{l/r}=0,A=0.035\gamma_{1D},\Omega=4\gamma_{1D},\varepsilon=\omega_{0}-1.5\Omega$. (c) The dependence of photon-photon correlation functions ${\rm{log}}_{10}g_{\leftarrow}^{(2)}(t,t)$ on the right coupling phase $\varphi_{r}$. The calculations are performed with parameters $\Omega=500\gamma_{1D},A=1.5\Omega,\varphi_{l}=0$.} \label{fig4}
\end{figure}

In the considered resolved-sideband regime, we plot the bichromatic photon-photon correlation function $g^{(2)}_{n_{1},n_{2}}$ in Fig.\ref{fig3}(b) without atomic chirality, i.e., $\varphi_{l}=\varphi_{r}=0$. The qubits are excited by a driving field incident from the left. The propagation phase is set to $\varphi=0$ (or $2\pi$), and the relative modulation phase is $\alpha=\pi$. Note also that the numerical simulations are performed by extracting the steady-state photon-photon correlations in the long-time limit, accounting for a weak monochromatic coherent drive at the frequency $\varepsilon=\omega_{0}$. In this scenario, we demonstrate parity-protected bunching as well as antibunching in cross-correlations. The former occurs when the photon pair scatters dominantly into even-parity sidebands, which guaranteed approximately by parity-dependent single photon scattering intensity $I^{(1)}_{n}\propto 1+\cos(n\alpha)$ [see also Fig.\ref{fig3}(d)] and photon pair scattering intensity $I^{(2)}_{n_{1},n_{2}}\propto 1+\cos(n_{1}-n_{2})\alpha$. More intriguingly, the high degree of tunability in chirality and the distance between coupling points facilitates the occurrence of a richer variety of correlation patterns. For example, a completely opposite pattern of bichromatic correlation is captured [see Fig.\ref{fig3}(c)] when the modulated atoms carry moderate chirality, with $\varphi_{l}=0,\varphi_{r}=\pi$, while the propagation phase is set to $\varphi=\pi/2$. This parameters setting renders the drastic suppression of the even-parity sidebands and a corresponding enhancement of the odd-parity sidebands as shown in Fig.\ref{fig3}(e). In light of this, the parity properties of the scattering spectrum, and consequently the corresponding high-dimensional entangled states, can be precisely controlled by only engineering the chirality of the multi-point coupling, offering a significant advantage over other schemes~\cite{PhysRevLett.130.023601}. This proposal holds significant potential for applications in efficient optical filtering and quantum precision measurement.

The emitted photons can be viewed as qudits, which are multi-level quantum systems, since each photon can occupy one of many sidebands or exist in a superposition of them. The effective Hilbert space dimension for these frequency qudits can be approximated by $2A/\Omega$. This is because sidebands with indices $|l|$ greater than $A/\Omega$ experience minimal excitation, a result of the quenching effect on the Bessel function $J_{l}(A/\Omega)$. The detection of bipartite entanglement can be realized by calculating the entanglement entropy~\cite{RevModPhys.81.865,RevModPhys.90.035007,PhysRevLett.127.173601,Poshakinskiy2021} $S=-\sum_{\lambda}|\lambda|^{2}\ln |\lambda|^{2}$, where $\lambda$ represents the singular values of the two-color wave function $\Psi_{n_{1},n_{2}}$, which is determined numerically from the correlations of detectors $D_{1}$ and $D_{2}$. The dependence of entanglement entropy, scaled by $e^{S}$, on various system parameters is presented in Fig.\ref{fig4}(a). From this, we find that the generation of frequency-encoded quantum entangled states can be flexibly controlled, with the speed of achieving maximal entanglement being tunable by adjusting atomic spacing and chirality.

\emph{Analytic time-dependent correlations}---It is also instructive to explore the analytical signatures of the correlation dynamics from the perspective of time-dependent, second-order photon-photon correlation functions~\cite{PhysRevA.108.053703}
\begin{align}
g_{\leftarrow}^{(2)}(t+\tau,t)=\frac{\langle a^{\dagger}_{\scriptstyle\leftarrow}(t+\tau)a^{\dagger}_{\gets}(t)a_{\leftarrow}(t)a_{\leftarrow}(t+\tau)\rangle}{(\langle a^{\dagger}_{\leftarrow} a_{\leftarrow}\rangle_{0})^{2}}, \label{eq6}
\end{align}

where $a_{\leftarrow}$ is the bosonic annihilation operator for the reflecting mode;  $\langle\cdots\rangle$ and $\langle\cdots\rangle_{0}$ refer to the averages calculated over the system's state with and without temporal modulation, respectively.  Notice that the denominator in Eq.(\ref{eq6}) is time-independent, while the numerator is periodic in time domain, leading to the relation $g_{\leftarrow}^{(2)}(t+\tau,t)=g_{\leftarrow}^{(2)}(t+2m\pi/\Omega+\tau,t+2m\pi/\Omega)$ for any integer $m$. This property encourages expressing the correlation function $g_{\leftarrow}^{(2)}(t+\tau,t)$ as a Fourier series: $\sum_{l=-\infty}^{\infty}e^{-il\Omega t}g^{(2)}_{l}(\tau)$.

Leveraging the advanced features of the powerful theoretical toolbox underpinned by the Feynman diagram techniques~\cite{PhysRevLett.130.023601,PhysRevA.93.033856,PhysRevLett.123.253601,PhysRevA.108.023715,RevModPhys.95.015002}, the components $g^{(2)}_{l}(\tau)$  can be described by the two-photon scattering amplitudes of Stokes or anti-Stokes processes of various orders. For instance, in the limit of small modulation amplitude, the $l=1$ harmonic is approximately given by $g^{(2)}_{1}(\tau)\propto S_{1}(\tau)S^{*}_{0}(\tau)+S_{0}(\tau)S^{*}_{-1}(\tau)$. Here, $S_{l}$ describes the two-photon scattering amplitudes where the total energy of the scattered photon pair is given by $\omega_{{\rm out}}=2\varepsilon+l\Omega$. In particular, the amplitude for the first order two-photon anti-Stokes scattering process reads~\cite{ SM}

\small
\begin{align}
S_{1}\!=\!&\sum_{ijk}A_{k}\Big\{\!2\mathcal{M}_{ij}(\varepsilon\!+\!\frac{\Omega}{2})\Sigma_{i}(0,2\varepsilon+\Omega)s_{k}^{+}(\varepsilon)s_{j}^{+}(\varepsilon)G_{kj}(\varepsilon+\Omega)\nonumber\\
&+\mathcal{M}_{ij}(\varepsilon)[\text{\ensuremath{\mathcal{A}_{ki,ki}}}(0,2\varepsilon\!+\!\Omega;\Omega)\!+\!\text{\ensuremath{\mathcal{B}_{ik,ik}}}(0,2\varepsilon\!+\!\Omega;2\varepsilon)]s_{j}^{+2}(\varepsilon) \Big\}\nonumber\\
&\times (-\!2i\gamma_{1D}^{2})-4i\gamma_{1D}^{2}\sum_{ijkl}\mathcal{M}_{ik}(\varepsilon+\frac{\Omega}{2})\chi_{kl}^{(+)}(\varepsilon)\mathcal{M}_{lj}(\varepsilon)\nonumber\\
&\times \Sigma_{i}(0,2\varepsilon+\Omega)s_{j}^{+2}(\varepsilon)+4r(\varepsilon)r_{1}(\varepsilon),\label{eq7}
\end{align}
where the quantities $\mathcal{M},\mathcal{A},\mathcal{B},\Sigma,s^{(+)},\chi^{(+)}$ are the fundamental ingredients that constitute the diagrammatic representation of two-photon scattering, while $r(\varepsilon)$$[r_{1}(\varepsilon)]$ denotes the elastic (inelastic) single-photon scattering amplitude, which can be found in~\cite{ SM}.

As shown in Fig.\ref{fig4}(b), for a pair of reflected photons, we plot the equal-time photon-photon correlation functions $g_{\leftarrow}^{(2)}(t,t)$ by gradually varying qubit spacing $d$. The analytical solutions derived from the diagrammatic technique (dotted lines) agree perfectly with the numerical results based on the Floquet quantum master equation (solid lines) at sufficiently long times, e.g., for $\Omega t/2\pi\ge 25$. Each predicted numerical curve exhibits oscillations with a decaying amplitude over time, eventually stabilizing into periodic oscillations of constant amplitude around a steady value, which depends on the system parameters.

Figure\,\ref{fig4}(c) illustrates the dependence of quantum correlation function ${\rm{log}}_{10}g_{\leftarrow}^{(2)}(t,t)$ on the chiral coupling phases. The double-well-like curve represents the average steady-state quantum correlation, while the error bars denote the fluctuations. It demonstrates that the statistical properties of the scattered photons can be flexibly switched between bunching and antibunching simply by adjusting the coupling phase at the right connecting point. The steady correlation function exhibits pronounced oscillations with increasing amplitude as $|\varphi_{r}|$ approaches $\pi$.

\emph{Non-Markovian features and photonic dynamics}--- It is intriguing to explore the dynamics of multi-photon interacting with dynamically modulated multiple emitters in the non-Markovian regime. The interference properties of the field would be modified drastically due to retardation~\cite{PhysRevLett.124.043603, PhysRevLett.131.193603}. In this scenario, the analytic structure of the dynamical description in the multi-excitation sector is extremely complex, even for radiation problem, where the complexity can be mitigated by hidden symmetries ~\cite{PhysRevResearch.6.033196}.

\begin{figure}
  \centering
  \includegraphics[width=8.9cm]{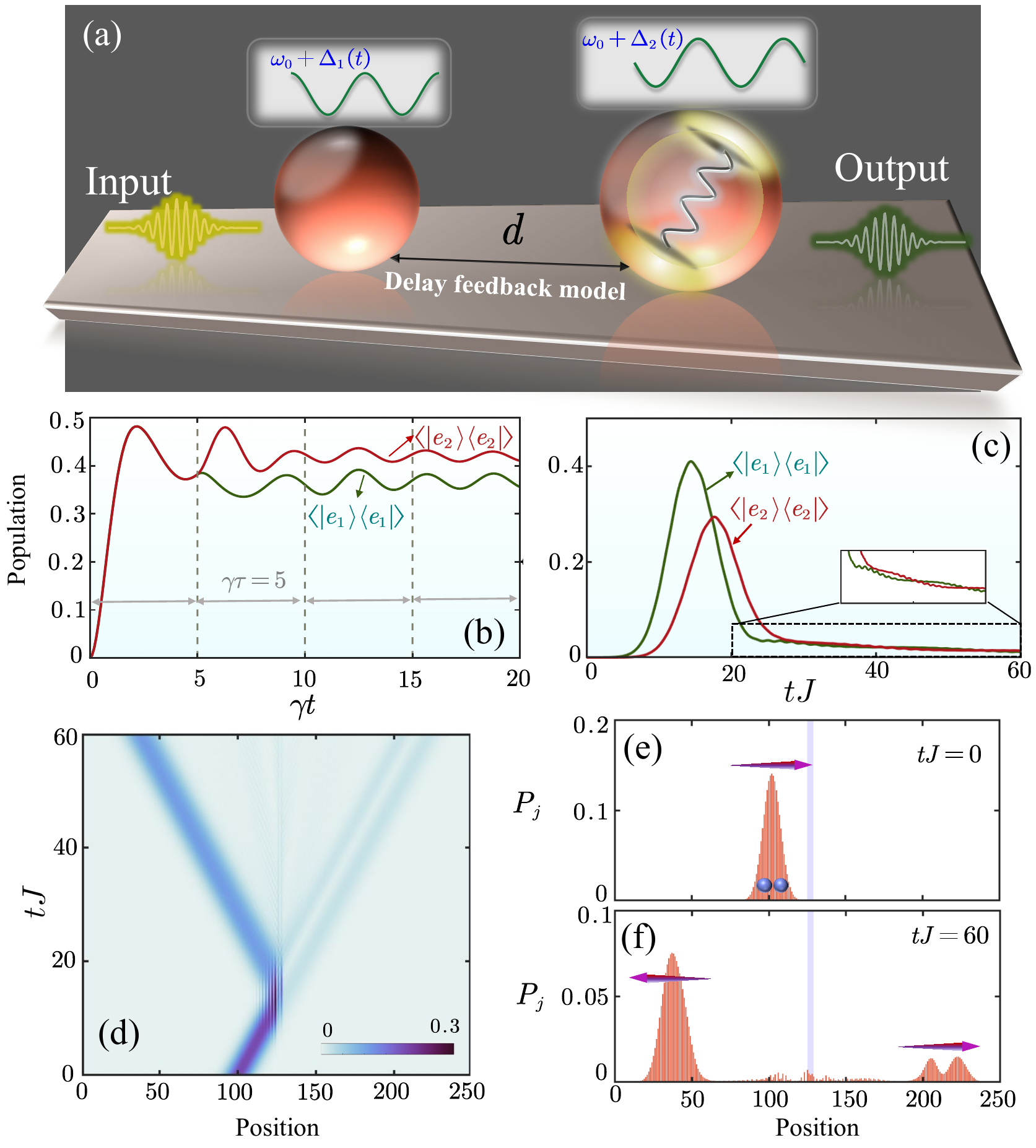}
  \caption{(a) Sketch of a pair of dynamically modulated qubits coupled to a $1$D waveguide with time-delay feedback. (b) Time evolution of the excitation probabilities of qubits driven by a coherent field with Rabi frequencies $\Omega_{1}=1.5$ and $ \Omega_{2}=1.5e^{-i\varphi}$. The photon emitted by the atom into the waveguide requires a finite time $\tau=d/v_{g}$ to reach the other atom. By considering the dynamics of two-photon wave packet incident from the left, the atomic excitation excitation are presented in (c) along with the lattice dynamics (d), which displays the lattice occupation $P_{j}(t)$ as a function of position $j$ and scaled time $tJ$. Snapshots of the expectation values  $P_{j}(t=0)$ and $P_{j}(t=60/J)$ are also shown in panels (e) and (f), respectively. The calculations are performed by choosing $\gamma_{1D}\tau=5, \varphi=0.5\pi, A=0.5\gamma_{1D}, \Omega =\gamma_{1D}$ for panel (b), and $\sigma=8, g=0.6J, A=0.4 J, k_{0}=q_{0}=\pi/2, \Omega =5g^{2}/J, N_{c}=249$ for panels (c)-(f). The definitions of $g, J, N_{c}$ can be found in~\cite{ SM}.}\label{fig5}
\vspace{-20pt}
\end{figure}

In the following, we capture the primary non-Markovian features of modulated systems by employing the versatile numerical technique of MPS~\cite{PhysRevLett.116.093601, PhysRevResearch.3.023030}. For simplicity, the considered  quantum optics model is composed of a pair of qubits locally coupled to a continuum of bosonic modes within a $1$D waveguide, where the qubits are excited by a coherent field with Rabi frequencies $\Omega_{n}$ at the resonant frequency $\omega_{0}$, and subjected to additional energy modulations $\omega_{1,2}=\omega_{0}+A\cos(\Omega t)$. The qubits are separated by a finite distance $d$ as shown in Fig.\,\ref{fig5}(a), such that they cannot immediately respond to the photons emitted by the other atom.  Notably, although the light-matter interaction is local in the real space, the presence of time-delay feedback renders the interaction fundamentally non-local. This can be revealed intuitively by the dynamical map $\ket{\Psi_{I}(t_{k+1})}\!=\!\exp[-iH_{{\rm sys},I}(t)\Delta t +V_{k,1}(t_{k})+V_{k,2}(t_{k})]\ket{\Psi_{I}(t_{k})}$ obtained through the quantum stochastic Schrödinger equation with the rotated system Hamiltonian $H_{{\rm sys},I}(t)\!=\!\sum_{n=1}^{2}[\Delta_{n}(t)\sigma_{n}^{\dagger}\sigma_{n}-\frac{\Omega_{n}}{2}(\sigma_{n}+\sigma_{n}^{\dagger})]$ and the non-local interaction~\cite{ SM}
\begin{align}
V_{k,1}(t_{k})&=\sqrt{\gamma_{1D}}[\Delta B_{R}^{\dagger}(t_{k})+\Delta B_{L}^{\dagger}(t_{k-l})e^{i\varphi}]\sigma_{1}-{\rm H.c.}\nonumber\\
V_{k,2}(t_{k})&=\sqrt{\gamma_{1D}}[\Delta B_{R}^{\dagger}(t_{k-l})e^{i\varphi}+\Delta B_{L}^{\dagger}(t_{k})]\sigma_{2}-{\rm H.c.}, \label{eq8}
\end{align}
Here, $B_{\lambda}(t_{k})$ denotes the time-bin noise operators, $\Delta t=t_{k+1}-t_{k}$ represents the time increment, and $l=\tau/\Delta t=d/v_{g}\Delta t$ indicates the index of the delay time bin, respectively.The temporal evolution of excitation probabilities for a pair of qubits is depicted in Fig.\,\ref{fig5}(b), when the drive phase difference is set to $\pi/2$. Initially, the atom follows single-atom Rabi dynamics before receiving the delayed signals emitted by the other atom. Subsequently, the interference between the driving field and the delayed field induces periodic oscillations in the dynamics of the first (second) atom, centered around lower (higher) population. Such an interference effect will emerge instantaneously at the onset of coherent driving in the Markovian limit.

The MPS simulations in the time-bin representation absorb the positional information of photons propagating along the waveguide. In order to visualize the detailed multi-photon transport, a spatial discretization approach for the waveguide is applied to simulate the dynamics of two-photon scattering~\cite{PhysRevLett.122.073601}. The discretized bosonic modes are denoted by $a_{j}$. We consider a two-photon Gaussian wave packet incident on a pair of dynamically modulated qubits with the initial quantum state $\sum_{j,j'}[\psi_{1}^{(k_{0},\sigma)}(j,n_{c})\psi_{2}^{(q_{0},\sigma')}(j',m_{c})a_{j}^{\dagger}a_{j'}^{\dagger}\!+\!1\leftrightarrow 2]\ket{G}$. Here, $\psi_{s}^{(k_{0},\sigma)}(j,n_{c})$ denotes the single-photon wave function following a Gaussian distribution~\cite{PhysRevLett.104.023602,Longo_2009}, with $k_{0},n_{c}$, and $\sigma$ denoting the carrier wave's wavenumber, the wavepacket's center, and its width, respectively. The dynamical excitation probabilities of qubits and the ``fictitious" cavities introduced from the discretization process are presented in Figs.\,\ref{fig5} (c) and (d), followed by corresponding snapshots [see Figs.\,\ref{fig5}(e) and (f)] for the latter. During the scattering process, the atoms are excited sequentially due to finite retardation, with the maximal excitation probabilities depending on the delayed quantum feedback and the envelope shape of the incident wave packet~\cite{ SM}. Meanwhile, most of the energy in the incident wave packet is reflected, and the waveform of the scattered wave packets undergoes significant changes. The scattering dynamics under other two-photon excitation schemes, along with the effect of the relative modulation phase, are presented in~\cite{ SM}.

\emph{Experimental implementation and conclusions}--- Regarding experimental implementations, artificial atoms modeled by transmon qubits coupled to a superconducting transmission line provide an ideal candidate for our proposal~\cite{Andersson2019, Kannan2020}. Such a system provides a sufficiently large resonant frequency compared to the radiative decay rate (e.g., $\omega_{0}/2\pi=5.23{\rm GHz}, \gamma\sim1{\rm MHz}$ in \cite{Kannan2020}), thereby justifying the applicability of the rotating wave approximation. Non-local coupling architectures are achieved by capacitively coupling the qubits to a meandering coplanar waveguide at multiple points. Furthermore, the complex coupling phases are engineered by connecting each coupling point to an independent dissipation port and employing time-modulated parametric couplings ~\cite{PhysRevX.13.021039}. Finally, the temporal modulation can be realized by coupling individual superconducting qubits to a local flux bias line, which is connected to an arbitrary waveform generator channel~\cite{Redchenko2023}. The modulation frequency and amplitude are highly flexible, allowing for a wide range of adjustments. For example, the experimental parameters of Ref.~\cite{LiJian2020}correspond to $ A/2\pi=250{\rm MHz}, \Omega/2\pi\sim 500{\rm MHz}$.

In conclusion, we have shown how to exquisitely manipulate sidebands emission and  efficiently generate frequency-encoded quantum entanglement via a dynamic modulation proposal. Parity-dependent multicolor quantum correlations can be achieved by carefully designing the relative modulation phase, and they are highly tunable when the nonlocal coupling waveguide QED setup is employed. Moreover, we have also analyzed the non-Markovian quantum dynamics of two distant, energy-modulated qubits, either by employing the tensor product states or by modeling the continuous waveguide as a coupled-cavity array.

This work is supported by the National Science Fund for Distinguished Young Scholars of China (Grant No. 12425502) and the National Key Research and Development Program of China (Grant No. 2021YFA1400700). The computation was completed in the HPC Platform of Huazhong University of Science and Technology.
\bibliographystyle{myapsrev4-1}
%\bibliography{refermain}

%\let\addcontentsline\oldaddcontentsline% Restore \addcontentsline

%%%%%%%%%% Merge with supplemental materials %%%%%%%%%%
\onecolumngrid

%%%%%%%%%% Prefix a "S" to all equations, figures, tables and reset the counter %%%%%%%%%%
\newcommand\specialsectioning{\setcounter{secnumdepth}{-2}}
\setcounter{equation}{0} \setcounter{figure}{0}

\setcounter{table}{0}
\renewcommand{\theequation}{S\arabic{equation}}
\renewcommand{\thefigure}{S\arabic{figure}}
\renewcommand{\bibnumfmt}[1]{[S#1]}
\renewcommand{\citenumfont}[1]{S#1}
\renewcommand\thesection{S\arabic{section}}
%%%%%%%%%% Prefix a "S" to all equations, figures, tables and reset the counter %%%%%%%%%%
\renewcommand{\baselinestretch}{1.2}

%\renewcommand{\theequation}{S\arabic{equation}}

%%%%%%%%%%%%%%%%%%%%%%%%%%%%%%%%%%%%%%%%%%%%%%%%%%%%%%%%%%%%%%%%%
\newpage

\setcounter{page}{1}\setcounter{secnumdepth}{3} \makeatletter
\begin{center}
    {\Large \textbf{ Supplemental Material for\\
            ``Floquet Engineering and Harnessing Giant Atoms in Frequency-Comb Emission and Bichromatic Correlations in Waveguide QED"}}
\end{center}

\begin{center}
Qing-Yang Qiu$^{1}$, Li-Li Zheng$^{2}$,Ying Wu$^{1}$, Xin-You L\"{u}$^{1}$
\end{center}

\begin{minipage}[]{16cm}
\small{\it
	\centering $^{1}$School of Physics and Institute for Quantum Science and Engineering, Huazhong University of Science and Technology, and Wuhan Institute of Quantum Technology, Wuhan 430074, China \\
$^{2}$School of Artificial intelligence, Jianghan University, Wuhan 430074, China\\}
\end{minipage}

\vspace{8mm}


\tableofcontents

%%%%%%%%%%%%%%%%%%%%%%%%%%%%%%%%%%%%%%%%%%%%%%
\section{FLOQUET ENGINEERING OF PHOTONIC SPECTRUM}
\setcounter{equation}{0}
\renewcommand\theequation{S\arabic{equation}}
\makeatletter
\renewcommand{\thefigure}{S\@arabic\c@figure}
\makeatother
In this section, we explore the controllable sideband emission from a modulated two-level emitter. Below, by applying the standard numerical optimization techniques, we present the detailed optimization process of Floquet eigenstates for our considered quantum optics model. Note that the numerical optimization results will cover those presented in Fig. 1 of the main text.

The motion of a single excitation  is governed by the effective Hamiltonian $H_{\rm{eff}}(t)=[\omega_{e}(t)-\frac{\gamma_{1D}}{2}i]\sigma^{\dagger}\sigma$, where $\gamma_{1D}$ is the decay rate of spontaneous emission into the 1D waveguide.

We now assume that the atomic modulation is periodic in time with the fundamental frequency $\Omega$ such that $\omega_{e}(t)=\omega_{e}(t+2\pi/\Omega)$, and the average value of such modulation is set to be zero. Without loss of generality, the qubit resonance frequency is typically written as a finite number of Fourier components
\begin{align}
\omega_{e}(t)=\omega_{0}+\Delta(t)=\omega_{0}+\sum_{r=1}^{R}[a_{r}\cos(r\Omega t)-b_{r}\sin(r\Omega t)],\label{S1}
\end{align}
where $R$ denotes the total number of the considered modulation tones. In order to obtain the scattering properties of the time-dependent Hamiltonian $H_{\rm{eff}}(t)$, it is necessary to determine the Floquet spectrum of the modulated quantum system by solving the Floquet equation $[H_{\rm{eff}}(t)-i\partial_{t}]\ket{\phi_{n}(t)}=\varepsilon_{n}\ket{\phi_{n}(t)}$, which leads to
\begin{align}
\ket{\phi_{1}(t)}=\ket{g},\varepsilon_{1}=0;\ket{\phi_{2}(t)}=\exp\left[-i\int_{0}^{t}\Delta(\tau)d\tau\right]\ket{e},\varepsilon_{2}=\omega_{0}-\frac{\gamma_{1D}}{2}i.\label{S2}
\end{align}
The Floquet excited state phase $\varpi(t)\equiv\exp\left[-i\int_{0}^{t}\Delta(\tau)d\tau\right]$ is also periodic in time and thus can be expanded into a Fourier series $\varpi(t)=\sum\limits _{k=-\infty}^{\infty}X_{k}e^{ik\Omega t}$. Note that the Fourier coefficients $X_{k}$ are closely responsible for the scattering off the modulated emitter and are completely governed by the temporal shape of the modulation $\Delta(t)$. In order to numerically optimize the output photon spectrum, it is convenient to express the Floquet phase $\varpi(t)$ as
\begin{align}
\varpi(t)=&\exp\left (-i\sum_{r=1}^{R}[\frac{a_{r}}{r\Omega}\sin(r\Omega t)+\frac{b_{r}}{r\Omega}\cos(r\Omega t)-\frac{b_{r}}{r\Omega}]\right)\nonumber\\
=&\prod_{r=1}^{R}(\sum_{\gamma_{r}=-\infty}^{\infty}i^{\gamma_{r}}J_{\gamma_{r}}(\frac{a_{r}}{r\Omega})e^{i\gamma_{r}(r\Omega t+\frac{\pi}{2})})(\sum_{\kappa_{r}=-\infty}^{\infty}i^{\kappa_{r}}J_{\kappa_{r}}(\frac{b_{r}}{r\Omega})e^{i\kappa_{r}(r\Omega t+\pi)})\exp(i\sum_{r}\frac{b_{r}}{r\Omega})\nonumber\\
=&\sum_{\gamma_{r}=-\infty}^{\infty}\sum_{\kappa_{r}=-\infty}^{\infty}J_{\gamma_{1}}(\frac{a_{1}}{\Omega})J_{\beta_{1}}(\frac{b_{1}}{\Omega})J_{\gamma_{2}}(\frac{a_{2}}{2\Omega})
J_{\beta_{1}}(\frac{b_{2}}{2\Omega})...J_{\gamma_{R}}(\frac{a_{R}}{\Omega})J_{\beta_{R}}(\frac{b_{R}}{\Omega})]\exp[i\sum_{r=1}^{R}(\frac{\pi}{2}\gamma_{r}+\kappa_{r}\pi)]\exp(i\sum_{r}\frac{b_{r}}{r\Omega})\nonumber\\
&\times\exp\bigg(i[(\gamma_{1}+\kappa_{1})+2(\gamma_{2}+\kappa_{2})+...+R(\gamma_{R}+\kappa_{R})]\Omega t\bigg),\label{S3}
\end{align}
where $J_{l}(\bullet)$ is the Bessel function of order $l$. We immediately find that the Fourier coefficients $X_{k}$ have the form of
\begin{figure}
  \centering
  \includegraphics[width=17cm]{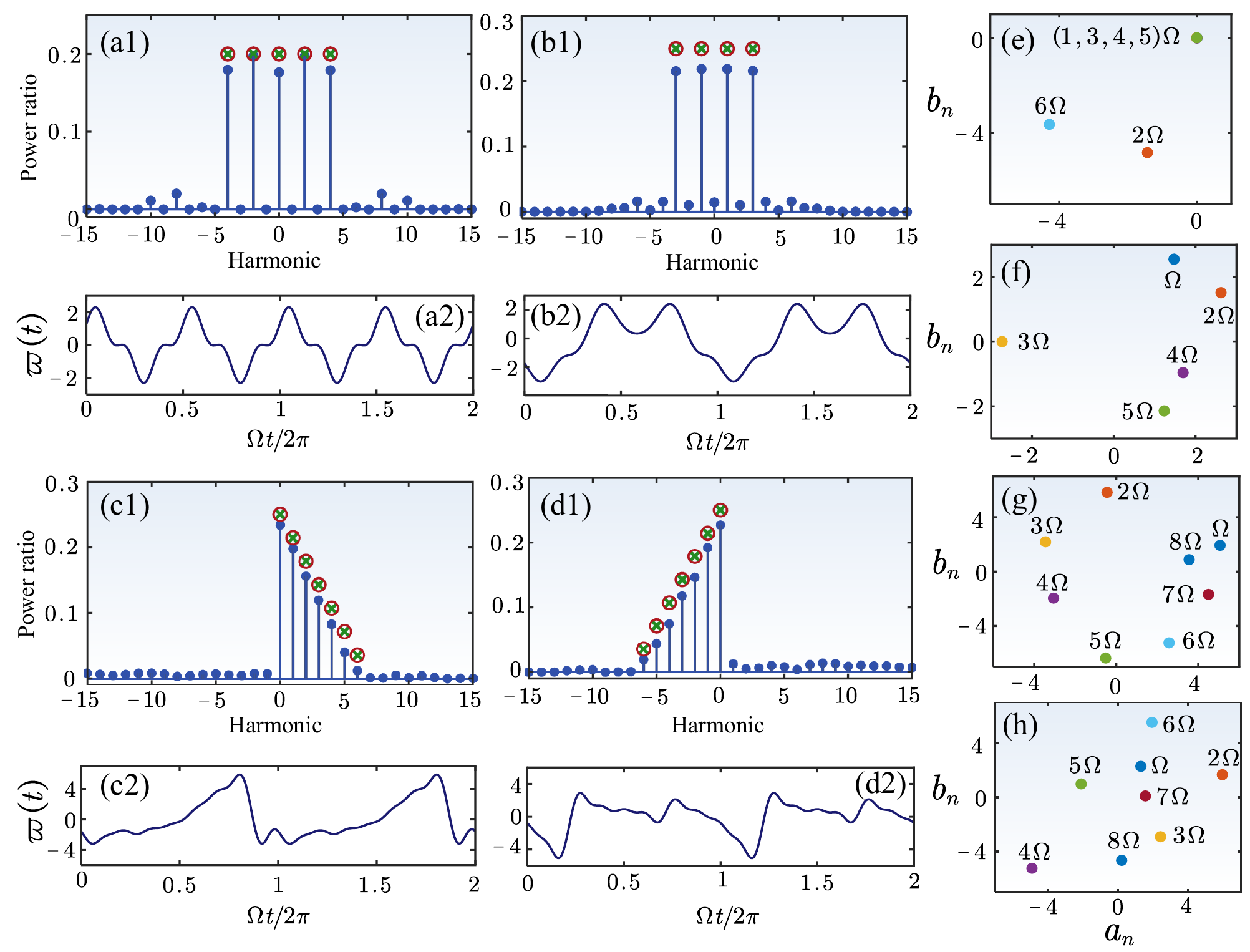}
  \caption{Four paradigmatic spectral optimizations: (a1) the optimised spectra containing only even-parity harmonics together with the corresponding Floquet phase (a2) and  modulation parameters (e);  (b1) the optimised spectra containing only odd-parity harmonics together with the corresponding Floquet phase (b2) and  modulation parameters (f);  (c1) the optimised spectra containing only anti-Stokes harmonics together with the corresponding Floquet phase (c2) and  modulation parameters (g); (d1) the optimised spectra containing only Stokes harmonics together with the corresponding Floquet phase (d2) and modulation parameters (h). Note that the target spectrum is marked by the red circles with green crosses inside. }\label{figS1}
\end{figure}
\begin{align}
X_{k}=&\underset{(\gamma_{1}+\kappa_{1})+2(\gamma_{2}+\kappa_{2})+...+R(\gamma_{R}+\kappa_{R})=k}{\sum_{\gamma_{r}=-\infty}^{\infty}\sum_{\kappa_{r}=-\infty}^{\infty}}\Big\{i^{\sum\limits _{r=1}^{R}\gamma_{r}+\kappa_{r}}[J_{\gamma_{1}}(\frac{a_{1}}{\Omega})J_{\beta_{1}}(\frac{b_{1}}{\Omega})J_{\gamma_{2}}(\frac{a_{2}}{2\Omega})J_{\beta_{2}}(\frac{b_{2}}{2\Omega})...J_{\gamma_{R}}(\frac{a_{R}}{\Omega})J_{\beta_{R}}(\frac{b_{R}}{\Omega})]\nonumber\\
&\times e^{i\sum_{r=1}^{R}(\frac{1}{2}\gamma_{r}+\kappa_{r})\pi}\Big\}\exp(i\sum_{r}\frac{b_{r}}{r\Omega}).\label{S4}
\end{align}
Since the Floquet phase $\varpi(t)$ is a complex function, it follows that the obtained Floquet components of $\varpi(t)$ are not necessarily symmetric, that is to say, $|X_{k}|\neq |X_{-k}|$. This property enables us to design arbitrary desired Floquet spectrum. For this purpose, we optimize the parameters $a_{r}$ and $b_{r}$ to match the target Fourier components $X^{(0)}_{k}$ by utilizing the powerful Broyden-Fletcher-Goldfarb-Shanno algorithm. As shown in Fig.\,\ref{figS1}, we present four paradigmatic Floquet spectra by optimizing the atomic modulation frequency $\Delta(t)$. The target spectra are marked by red circles with green crosses inside. Apparently, the numerical optimization results are almost as expected.

We are now in a position to analyze the Floquet dynamics of the emitter and bath quantum system using the optimized qubit modulation. The theoretical formalism we apply here is facilitated by the versatile techniques of interaction picture propagator\,\cite{smPhysRevB.98.144112, smPhysRevA.102.033707}, a computational tool for capturing the exact dynamics of the time-dependent, low-dimensional system.  We consider the atom to be initially in its ground state $\ket{g}$ and the waveguide in its vacuum state $\ket{{\rm vac}}$. To induce the intrinsic sideband structure from the emission of the modulated atom, we incorporate a pulsed coherent drive into the effective Hamiltonian:
\begin{align}
H_{\rm{eff}}(t)=\left(\sum_{r=1}^{R}[a_{r}\cos(r\Omega t)-b_{r}\sin(r\Omega t)]-\frac{\gamma_{1D}}{2}i \right)\sigma^{\dagger}\sigma+\xi\Theta(t)\Theta(T_{d}-t)(\sigma^{\dagger}+\sigma),\label{S5}
\end{align}
where $\xi$ is the the amplitude of the coherent drive; $T_{d}$ is the duration of this driving pulse; $\Theta(\bullet)$ is the Heaviside step function satisfying $\Theta(x)=1$ for $x\ge 0$ and $\Theta(x)=0$ for $x< 0$.

\begin{figure}
  \centering
  \includegraphics[width=15cm]{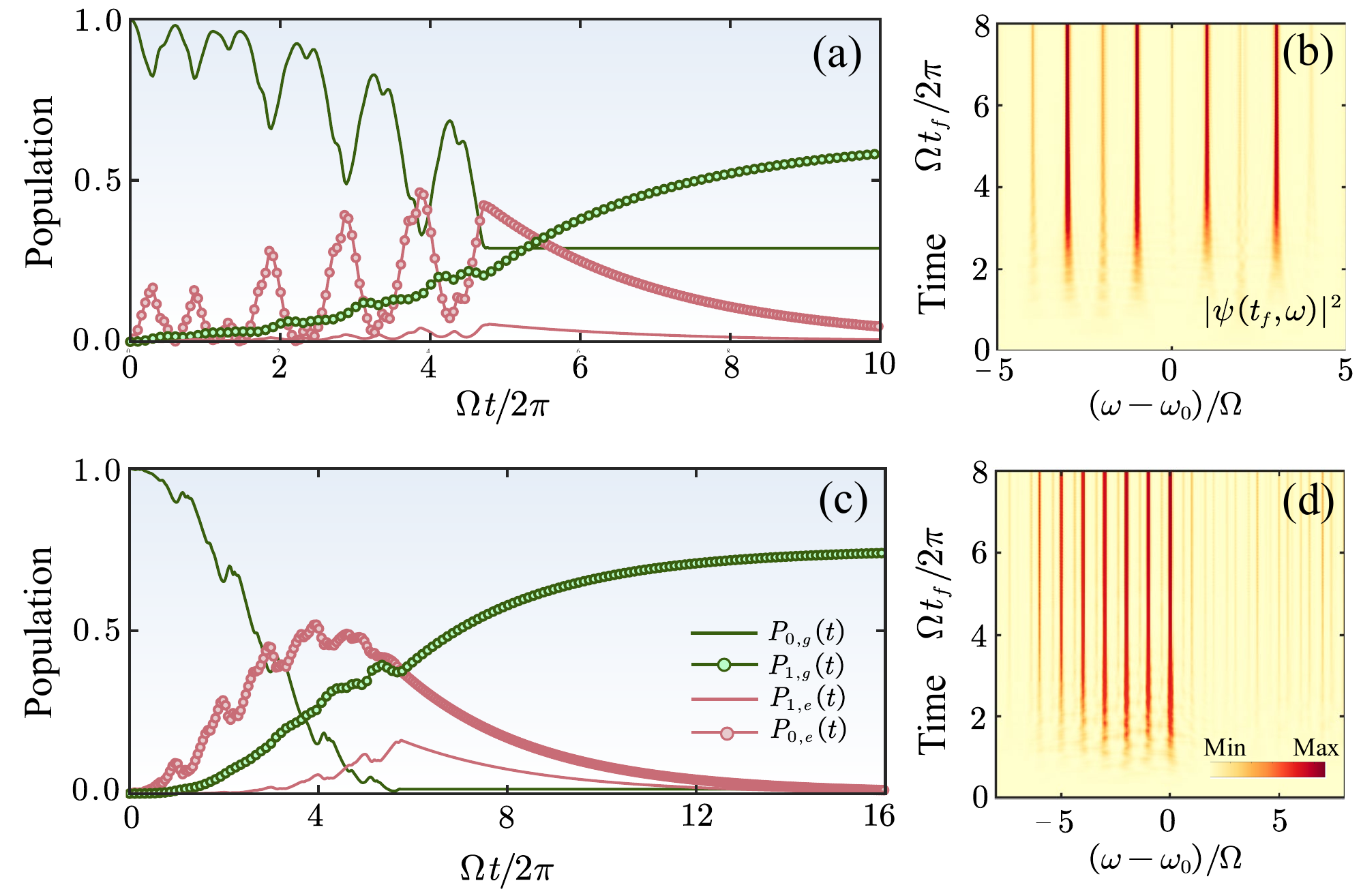}
  \caption{Panels (a) and (c) are time evolution of the probabilities $P_{0,g}(t),P_{0,e}(t),P_{1,g}(t)$ and $P_{1,e}(t)$ for a modulated atom driven by a coherent pulse. (b) and (d) are spectra of the emitted single photon from the modulated atom. The modulation parameters for panels (a)-(b) are same as that from Fig.\,\ref{figS1}(b1) while the ones for panels (c)-(d) are same as that from Fig.\,\ref{figS1}(d1). The duration of the coherent pulse is assumed to be $T_{d}=2/\gamma_{1D}$.} \label{figS2}
\end{figure}

We then simulate the population dynamics of finding 0 or 1 excitation in the waveguide when the modulated emitter is in a lower or upper energy state at time $t$. The general expressions of these probabilities are given by:
\begin{align}
P_{0,g/e}(t)=&\left |\bra{{\rm vac}; g/e}U(t,0)\ket{{\rm vac}; g}\right|^{2},\nonumber\\
P_{1,g/e}(t)=&\int\left |\bra{x; g/e}U(t,0)\ket{{\rm vac}; g}\right|^{2}dx,\label{S6}
\end{align}
where $U(t,0)$ is the propagator for the system in the Schr$\rm{\ddot{o}}$dinger picture. As shown in Figs.\,\ref{figS2} (a) and (c), we plot the time evolution of the population quantities defined in Eq.\,(\ref{S6}) using the optimized qubit modulation parameters obtained in Figs.\,\ref{figS1}(f) and (h), respectively. These time-dependent population curves exhibit Rabi oscillations as a whole, and locally show fluctuations in the amplitude induced by the atomic dynamical modulation. As time evolves, $P_{0,e}(t)$ and $P_{1,e}(t)$ gradually decrease to 0 while $P_{0,g}(t)$ and $P_{1,g}(t)$ stabilize at certain finite values. To verify that the output Floquet spectra are consistent with the numerical optimization results shown in  Figs.\,\ref{figS1}(b1) and (d1), it is necessary to transform the time-dependent single-photon states into the frequency domain and calculate the emission spectrum $|\psi(\omega,t_{f})|^{2}$ with
\begin{align}
\psi(\omega,t_{f})=-i\sqrt{\frac{\gamma_{1D}}{2\pi}}\int_{0}^{t_{f}}\bra{g}U_{{\rm eff}}(t_{f},t)\sigma U_{{\rm eff}}(t,0)\ket{g}e^{i\omega t}dt, \label{S7}
\end{align}
where $U_{{\rm eff}}(t_{2},t_{1})\equiv\mathcal{T}\exp\left [-i\int_{t_{1}}^{t_{2}}H_{{\rm eff}}(t)dt\right ]$ is the propagator of the effective Hamiltonian with $\mathcal{T}$ the chronological time-ordering operator. We simulate the spectra of the emitted single photon, i.e., $\left|\psi(\omega,t_{f})\right|^{2}$, as shown in Figs.\,\ref{figS2} (b) and (d) for above two considered cases. The resulting spectra contain indeed only odd-parity harmonics for the former and contain only Stokes harmonics for the letter.
%%%%%%%%%%%%%%%%%%%%%%%%%%%%%%%%%%%%%%%%%%%%%
\section{ DERIVATION OF THE GENERAL MASTER EQUATION FOR CHIRAL SPIN NETWORKS}
\setcounter{equation}{7}
\renewcommand\theequation{S\arabic{equation}}
\makeatletter
\renewcommand{\thefigure}{S\@arabic\c@figure}
\makeatother
In this section, we present the detailed derivation of master equation for a collection of $N$ two-level systems or spins coupled to a common bath, which in our case is the 1D waveguide, with tunable chirality. The two-level systems are modeled by point-like atoms, giant atoms or their arbitrary combination. Note that this derivation is similar to the strategy proposed in \,\cite{smPhysRevA.91.042116, smPhysRevResearch.2.013369}. For the $n^{{\rm th}}$ atom, we denote the upper and lower energy states as $\ket{e}_{n}$ and $\ket{g}_{n}$, respectively. Additionally, we assume that the nodes or spins in the network share a common resonant frequency $\omega_{0}$ for simplicity. The total Hamiltonian for the emitters-plus-bath quantum system is given by $H=H_{0}+H_{{\rm int}}$, where
\begin{align}
H_{0}&=\sum_{n=1}^{N}\omega_{0}\sigma_{n}^{\dagger}\sigma_{n}+\sum_{\lambda=L,R}\int d\omega\,\omega b_{\lambda}^{\dagger}(\omega)b_{\lambda}(\omega),\nonumber\\
H_{{\rm int}}&=i\sum_{\lambda=L,R}\sum_{n=1}^{N}\sum_{p=1}^{M_{n}}\int d\omega\sqrt{\frac{\gamma_{\lambda}}{2\pi}}e^{i\phi_{p}^{(n)}}b_{\lambda}^{\dagger}(\omega)\sigma_{n}e^{-i\omega x_{np}/v_{\lambda}}+{\rm H.c}.\label{S8}
\end{align}
Here $\sigma_{n}\equiv \ket{g}_{n}\bra{e} $ is the atomic coherence operator and $b_{\lambda}(\omega)$ are bosonic annihilation operators for the right-moving ($\lambda=R$) and left-moving ($\lambda=L$) modes of frequency $\omega$, satisfying the commutation relation $[b_{\lambda}(\omega),b^{\dagger}_{\lambda'}(\omega')]=\delta_{\lambda,\lambda'}\delta(\omega-\omega')$. Furthermore, we extend the flexibility of the model Hamiltonian by considering tunable atomic configuration with a collection of controllable parameters: $M_{n}$ is the number of coupling points for $n^{{\rm th}}$ atom ($M_{n}=1$ for small atom); $\phi_{p}^{(n)}$ the phase of the complex coupling for the $p^{{\rm th}}$ leg from $n^{{\rm th}}$ atom that is located at $x_{np}$.  Note that $v_{\lambda}$ is the group velocity for a single photon propagating to the left ($v_{L}<0$) or right ($v_{R}>0$) along the 1D waveguide.

We proceed by applying rotating-frame transformation for Eq.\,(\ref{S8}), i.e., $H=UHU^{\dagger}-i\frac{dU}{dt}U^{\dagger}$ with $U=e^{iH_{0}t}$. After that, we obtain
\begin{align}
H(t)=i\sum_{\lambda=L,R}\sum_{n=1}^{N}\sum_{p=1}^{M_{n}}\int d\omega\sqrt{\frac{\gamma_{\lambda}}{2\pi}}e^{i\phi_{p}^{(n)}}b_{\lambda}^{\dagger}(\omega,t)\sigma_{n}(t)e^{i(\omega-\omega_{0})t}e^{-i\omega x_{np}/v_{\lambda}}+{\rm H.c}.\label{S9}
\end{align}
The Heisenberg equations of motion for bosonic modes $b_{\lambda}(\omega,t)$ and arbitrary atomic operator $X(t)$ are accessible according to Eq.\,(\ref{S9}):
\begin{align}
\dot{b}_{\lambda}(\omega,t)=&i[H(t),b_{\lambda}(\omega,t)]=\sum_{\lambda=L,R}\sum_{n=1}^{N}\sum_{p=1}^{M_{n}}\sqrt{\frac{\gamma_{\lambda}}{2\pi}}e^{i\phi_{p}^{(n)}}\sigma_{n}(t)e^{i(\omega-\omega_{0})t}e^{-i\omega x_{np}/v_{\lambda}},\label{S10}\\
\dot{X}(t)=&i[H(t),X(t)]=\sum_{\lambda=L,R}\sum_{n=1}^{N}\sum_{p=1}^{M_{j}}\int d\omega\sqrt{\frac{\gamma_{\lambda}}{2\pi}}\Big\{e^{i\phi_{p}^{(n)}}b_{\lambda}^{\dagger}(\omega,t)e^{i(\omega-\omega_{0})t-i\omega x_{np}/v_{\lambda}}[X(t),\sigma_{n}(t)]\nonumber\\
&-e^{-i\phi_{p}^{(n)}}[X(t),\sigma_{n}^{\dagger}(t)]e^{-i(\omega-\omega_{0})t+i\omega x_{np}/v_{\lambda}}b_{\lambda}(\omega,t)\Big\}.\label{S11}
\end{align}
The dynamical solutions of $b_{\lambda}(\omega,t)$ can be obtained by formally solving Eq.\,(\ref{S10}), which leads to
\begin{align}
b_{\lambda}(\omega,t)=b_{\lambda}(\omega,0)+\int_{0}^{t}ds\sum_{n=1}^{N}\sum_{p=1}^{M_{n}}\sqrt{\frac{\gamma_{\lambda}}{2\pi}}e^{i\phi_{p}^{(n)}}\sigma_{j}(s)e^{i(\omega-\omega_{0})s}e^{-i\omega x_{np}/v_{\lambda}}.\label{S12}
\end{align}
Inserting Eq.\,(\ref{S12}) into Eq.\,(\ref{S11}) and defining quantum noise operators $b_{\lambda}(t)	\equiv	\frac{1}{\sqrt{2\pi}}\int d\omega b_{\lambda}(\omega,0)e^{-i(\omega-\omega_{0})t}$, we have
\begin{align}
\dot{X}(t)=&\sum_{\lambda=L,R}\sum_{n=1}^{N}\sum_{p=1}^{M_{n}}\sqrt{\gamma_{\lambda}}\left\{e^{i\phi_{p}^{(n)}}b_{\lambda}^{\dagger}(t-x_{np}/v_{\lambda})e^{-i\omega_{0}x_{np}/v_{\lambda}}[X(t),\sigma_{n}(t)]-e^{-i\phi_{p}^{(n)}}e^{i\omega_{0}x_{np}/v_{\lambda}}[X(t),\sigma_{n}^{\dagger}(t)]b_{\lambda}(t-x_{np}/v_{\lambda})\right\}\nonumber\\
&+\sum_{\lambda=L,R}\sum_{m,n=1}^{N}\sum_{p=1}^{M_{n}}\sum_{q=1}^{M_{m}}\int d\omega\int_{0}^{t}ds\frac{\gamma_{\lambda}}{2\pi}\Big\{e^{i(\phi_{p}^{(n)}-\phi_{q}^{(m)})}e^{i(\omega-\omega_{0})(t-s)-i\omega(x_{np}-x_{mq})/v_{\lambda}}\sigma_{m}^{\dagger}(s)[X(t),\sigma_{n}(t)]\nonumber\\
&-e^{-i(\phi_{p}^{(n)}-\phi_{q}^{(m)})}e^{-i(\omega-\omega_{0})(t-s)+i\omega(x_{np}-x_{mq})/v_{\lambda}}[X(t),\sigma_{n}^{\dagger}(t)]\sigma_{m}(s)\Big\}.\label{S13}
\end{align}
We then perform the well-known Born-Markov approximation\,\cite{smPhysRevA.23.3118, smPhysRevA.10.1096} by neglecting the propagation time $|x_{np}-x_{mq}|/v_{\lambda}$ between the atomic coupling points, i.e., $\sigma_{n}(t-|x_{np}-x_{mq}|/v_{\lambda})\approx \sigma_{n}(t) $. Such an approximation is justified in the regime of $\gamma_{\lambda}\ll |v_{\lambda}|/|x_{np}-x_{mq}|$, in which the time scale for a single photon traveling through the waveguide is negligible compared to the characteristic evolution time for system operators. By defining $k_{\lambda}\equiv\omega_{0}/v_{\lambda}$ and $x_{np,mq}\equiv x_{np}-x_{mq}$, the equation of motion in Eq.\,(\ref{S13}) can be rewritten by taking averages as follows
\begin{align}
\langle\dot{X}(t)\rangle &=\sum_{\lambda=L,R}\sum_{n=1}^{N}\sum_{p=1}^{M_{n}}\frac{\gamma_{\lambda}}{2}\left\{\langle\sigma_{n}^{\dagger}(t)[X(t),\sigma_{n}(t)]\rangle-\langle[X(t),\sigma_{n}^{\dagger}(t)]\sigma_{n}(t)\rangle\right\}\nonumber\\
&+\mathop{\sum_{\lambda=L,R}\sum_{m,n=1}^{N}\sum_{p=1}^{M_{n}}\sum_{q=1}^{M_{m}}}\limits_{(k_{\lambda}x_{np}>k_{\lambda}x_{mq})}\gamma_{\lambda}\left\{e^{i(\phi_{p}^{(n)}-\phi_{q}^{(m)})}e^{-ik_{\lambda}x_{np,mq}}\langle\sigma_{m}^{\dagger}(t)[X(t),\sigma_{n}(t)]\rangle-e^{-i(\phi_{p}^{(n)}-\phi_{q}^{(m)})}e^{ik_{\lambda}x_{np,mq}}\langle[X(t),\sigma_{n}^{\dagger}(t)]\sigma_{m}(t)\rangle\right\}.\label{S14}
\end{align}
Note that the first line in Eq.\,(\ref{S13}) has been omitted since the bath is initially in a vacuum state, i.e., $\langle b_{\lambda}(t)\rangle=0$. To obtain the master equation, we rotate Eq.\,(\ref{S14}) to the Schr$\rm{\ddot{o}}$dinger picture by moving the time dependence of the average values of system operators to the atomic density operator, for instance, $\langle\dot{X}(t)\rangle={\rm Tr}[X\dot{\rho}(t)]$ and $\langle[X(t),\sigma_{n}^{\dagger}(t)]\sigma_{m}(t)\rangle={\rm Tr}(X[\sigma_{n}^{\dagger},\sigma_{m}\rho(t)])$. After some algebra, the chiral master equation reads
\begin{align}
\dot{\rho}(t)=&-i[H_{{\rm sys}},\rho(t)]+\sum_{\lambda=L,R}\sum_{n=1}^{N}\sum_{p=1}^{M_{n}}\frac{\gamma_{\lambda}}{2}\left\{[\sigma_{n},\rho(t)\sigma_{n}^{\dagger}]-[\sigma_{n}^{\dagger},\sigma_{n}\rho(t)]\right\}\nonumber\\
&+\mathop{\sum_{\lambda=L,R}\sum_{m,n=1}^{N}\sum_{p=1}^{M_{n}}\sum_{q=1}^{M_{m}}}\limits_{(k_{\lambda}x_{np}>k_{\lambda}x_{mq})}\gamma_{\lambda}\left\{e^{i(\phi_{p}^{(n)}-\phi_{q}^{(m)})}e^{-ik_{\lambda}x_{np,mq}}[\sigma_{n},\rho(t)\sigma_{m}^{\dagger}]-e^{-i(\phi_{p}^{(n)}-\phi_{q}^{(m)})}e^{ik_{\lambda}x_{np,mq}}[\sigma_{n}^{\dagger},\sigma_{m}\rho(t)]\right\},\label{S15}
\end{align}
where $H_{{\rm sys}}=\sum_{n=1}^{N}\omega_{0}\sigma_{n}^{\dagger}\sigma_{n}$. It is instructive to decompose the above master equation into different contributions with respect to left and right moving modes:
\begin{align}
\dot{\rho}(t)=-i[H_{{\rm sys}}+H_{L}+H_{R},\rho(t)]+\gamma_{L}\mathcal{D}[c_{L}]\rho+\gamma_{R}\mathcal{D}[c_{R}]\rho,\label{S16}
\end{align}
where $\mathcal{D}[O]\rho\equiv O\rho O^{\dagger}-(\rho O^{\dagger}O+O^{\dagger}O\rho )/2$, and $H_{L/R}$ denotes the long-range spin coherent interaction mediated by the left/right-going reservoir modes and has the form of
\begin{align}
H_{L}&=-\frac{\gamma_{L}}{2}i\mathop{\sum_{m,n=1}^{N}\sum_{p=1}^{M_{n}}\sum_{q=1}^{M_{m}}}\limits_{x_{mq}>x_{np}}(e^{-i(\phi_{p}^{(n)}-\phi_{q}^{(m)})}e^{ik_{0}|x_{np,mq}|}\sigma_{n}^{\dagger}\sigma_{m}-{\rm H.c.}),\label{S17}\\
H_{R}&=-\frac{\gamma_{R}}{2}i\mathop{\sum_{m,n=1}^{N}\sum_{p=1}^{M_{n}}\sum_{q=1}^{M_{m}}}\limits_{x_{mq}<x_{np}}(e^{-i(\phi_{p}^{(n)}-\phi_{q}^{(m)})}e^{ik_{0}|x_{np,mq}|}\sigma_{n}^{\dagger}\sigma_{m}-{\rm H.c.}).\label{S18}
\end{align}
Note that we have defined $k_{R}=-k_{L}\equiv k_{0}$ for simplicity. In addition, the jump operators $c_{L}\equiv \sum\limits_{n=1}^{N}\sum\limits_{p=1}^{M_{n}}e^{ik_{0}x_{np}}\sigma_{n}e^{i\phi_{p}^{(n)}}$ and $c_{R}\equiv \sum\limits_{n=1}^{N}\sum\limits_{p=1}^{M_{n}}e^{-ik_{0}x_{np}}\sigma_{n}e^{i\phi_{p}^{(n)}}$ in Eq.\,(\ref{S16}) describe the collective decay of atomic excitation into the left and right output channels, respectively. When the symmetry between left and right propagating excitations is conservable, i.e., in the limit of $\gamma_{L}=\gamma_{R}\equiv\gamma$, we obtain the bidirectional master equation
\begin{align}
\dot{\rho}(t)=&-i[H_{{\rm sys}}+\gamma\sum_{m,n=1}^{N}\sum_{p=1}^{M_{n}}\sum_{q=1}^{M_{m}}e^{-i(\phi_{p}^{(n)}-\phi_{q}^{(m)})}\sin k_{0}|x_{np}-x_{mq}|\sigma_{n}^{\dagger}\sigma_{m},\rho(t)]\nonumber\\
&+2\gamma\sum_{m,n=1}^{N}\sum_{p=1}^{M_{n}}\sum_{q=1}^{M_{m}}\cos k_{0}|x_{np}-x_{mq}|e^{-i(\phi_{p}^{(n)}-\phi_{q}^{(m)})}(\sigma_{m}\rho(t)\sigma_{n}^{\dagger}-\frac{1}{2}\{\sigma_{n}^{\dagger}\sigma_{m},\rho(t)\}).\label{S19}
\end{align}
The resulting master equation in Eq.\,(\ref{S19}) is readily utilized to simulate the frequency-filtered quantum correlations [see Figs.3 (b)-(e)], the entanglement entropy [see Fig.4 (a)], as well as the numerical equal-time photon-photon correlation functions [see Figs.4 (b)] in the main text.
%%%%%%%%%%%%%%%%%%%%%%%%%%%%%%%%%%%%%%%%%%%%%%
\section{GREEN FUNCTION METHOD FOR FEW-PHOTON SCATTERING IN GIANT EMITTERS ARRAY}
\setcounter{equation}{19}
\renewcommand\theequation{S\arabic{equation}}
\makeatletter
\renewcommand{\thefigure}{S\@arabic\c@figure}
\makeatother
In this section, we present the detailed derivation of few-photon Stokes/anti-Stokes scattering amplitudes for a linear chain of $N$ giant emitters. The theoretical toolbox we applied here mainly encapsulates the diagrammatic Green function approach that is recently developed in\,\cite{smPhysRevLett.130.023601, smPhysRevA.108.023715}.

Before providing a detailed illustration of general scattering matrix theory, here we give a clear Hamiltonian description for an array of $N$ giant atoms coupled to a 1D waveguide, and several notations are introduced for convenience.  The corresponding Hamiltonian of the considered waveguide-QED setup reads
\begin{align}
H=\sum_{n}[\omega_{0}+\Delta_{n}(t)]\sigma_{n}^{\dagger}\sigma_{n}+\sum_{k}\omega_{k}b_{k}^{\dagger}b_{k}+\sum_{k}g[\sigma_{n}^{\dagger}b_{k}(\sum_{p=1}^{M_{n}}e^{ikx_{np}})+\sigma_{n}b_{k}^{\dagger}(\sum_{p=1}^{M_{n}}e^{-ikx_{np}})],\label{S20}
\end{align}
where $\Delta_{n}(t)\equiv A_{n}e^{-i\Omega t}+A_{n}^{*}e^{i\Omega t}$ is the dynamically modulated frequency for the $n^{{\rm th}}$ giant atom around the equilibrium spin resonance frequency $\omega_{0}$; $A_{n}$ is the modulation amplitude; $\Omega$ is the modulation frequency; $\omega_{k}=|k|v_{g}$ is the photonic dispersion. Note that, for simplicity, the chiral response arising from the coupling phase $\phi_{p}^{(n)}$ or the symmetry breaking between right- and left-moving excitations is not considered here. Considering the bosonized excitations of the giant qubits, we rewrite the Hamiltonian \,(\ref{S20}) by replacing $\sigma_{n}$ with $c_{n}$:
\begin{align}
H=\sum_{n}[\omega_{0}+A_{n}(t)]c_{n}^{\dagger}c_{n}+\sum_{k}\omega_{k}b_{k}^{\dagger}b_{k}+\sum_{k}g[c_{n}^{\dagger}b_{k}(\sum_{p=1}^{M_{n}}e^{ikx_{np}})+c_{n}b_{k}^{\dagger}(\sum_{p=1}^{M_{n}}e^{-ikx_{np}})]+\frac{U}{2}\sum_{n}c_{n}^{\dagger}c_{n}^{\dagger}c_{n}c_{n},\label{S21}
\end{align}
where $U$ stems from the effective on-site photon-photon repulsion. The Hamiltonian\,(\ref{S21}) is completely equivalent to (\ref{S20}) in the limit of $U \rightarrow\infty$, i.e., the system lies on the so-called hard-core limit\,\cite{smPhysRevLett.124.093604, smPhysRevLett.126.203602}.

\subsection{Single-photon elastic and inelastic scattering}
In this subsection, we give the analytical single-photon scattering amplitudes that are governed by the Green function for single qubit excitation. The dressed Green's function $\bf{G}(\omega)$ of the qubit excitations satisfies the following Dyson-like equation
\begin{align}
G_{mn}(\omega)=G_{mn}^{(0)}(\omega)+\sum_{k}\sum_{i,j}G_{mi}^{(0)}(\omega)g^{2}D_{k}(\omega)(\sum_{p=1}^{M_{i}}e^{ikx_{ip}})(\sum_{q=1}^{M_{j}}e^{-ikx_{jq}})G_{jn}(\omega),\label{S22}
\end{align}
where $G_{mn}^{(0)}(\omega)\equiv\frac{\delta_{mn}}{\omega-\omega_{0}+i0^{+}}$ and $D_{k}(\omega)\equiv\frac{1}{\omega-\omega_{k}+i0^{+}}$ are bare exciton Green function and bare photon Green function, respectively. The second term in the right hand of Eq.\,(\ref{S22}) describes the effective propagation of the excitation from qubit $n$ to qubit $m$ which can be expressed as a compact form
\begin{align}
V_{ij}\equiv\sum\limits_{k}\frac{g^{2}(\sum\limits_{p=1}^{M_{i}}e^{ikx_{ip}})(\sum\limits_{q=1}^{M_{j}}e^{-ikx_{jq}})}{\omega-\omega_{k}+i0^{+}}=-i(g^{2}/v_{g})\sum_{p,q=1}^{M}e^{i(\omega/c)|x_{ip}-x_{jq}|},\label{S23}
\end{align}
where we have assumed that the coupled giant atoms are identical sharing the same number of coupling points, i.e., $M_{m}=M_{n}=M$ for $m\neq n$. Inserting Eq.\,(\ref{S23}) into Eq.\,(\ref{S22}), we obtain a standard Dyson formulation
\begin{align}
(\omega-\omega_{0})G_{mn}(\omega)+i\sum_{j}^{N}(g^{2}/v_{g})\sum_{p,q=1}^{M}e^{i(\omega/v_{g})|x_{jp}-x_{nq}|}G_{jn}(\omega)=\delta_{mn}.\label{S24}
\end{align}
We can immediately obtain the dressed matrix Green function $\bf{G}(\omega)=(\omega-{\boldsymbol{H}}_{{\rm eff}})^{-1}$ according to Eq.\,(\ref{S24}),  where the entries of the effective Hamiltonian  ${\boldsymbol{H}}_{{\rm eff}}$ read $({\boldsymbol{H}}_{{\rm eff}})_{mn}=\omega_{0}\delta_{mn}-i\gamma_{1D}\sum\limits_{p,q=1}^{M}e^{ik_{0}|x_{mp}-x_{nq}|}$. Here, $\gamma_{1D}\equiv g^{2}/v_{g}$ is the decay rate of spontaneous emission into the waveguide. Further more, the Born-Markovian approximation is applied by replacing $\omega/v_{g}$ with $k_{0}$ that is conditioned on $\gamma_{1D}\ll \omega_{0}$. Based on this dressed Green function, the single-photon elastic reflection and transmission amplitudes can be easily obtained as follows
\begin{align}
r(\omega_{k})&=-i\gamma_{1D}\sum\limits_{m,n=1}^{N}G_{mn}(\omega_{k})\sum_{p,q=1}^{M}e^{ik_{0}(x_{mp}+x_{nq})},\label{S25}\\
t(\omega_{k})&=1-i\gamma_{1D}\sum_{m,n=1}^{N}G_{mn}(\omega_{k})\sum_{p,q=1}^{M}e^{-ik_{0}(x_{mp}+x_{nq})},\label{S26}
\end{align}
and they satisfy $|r(\omega_{k})|^{2}+|t(\omega_{k})|^{2}=1$. Similarly, the single-photon in-elastic reflection coefficients are given by
\begin{align}
r_{1}(\omega_{k})&=-i\gamma_{1D}\sum_{i,j,k=1}^{N}G_{ik}(\omega_{k}+\Omega)A_{k}G_{kj}(\omega_{k})\sum_{p,q=1}^{M}e^{ik_{0}(x_{ip}+x_{jq})}=-i\gamma_{1D}\sum_{k=1}^{N}A_{k}s_{k}^{+}(\omega_{k}+\Omega)s_{k}^{+}(\omega_{k}),\label{S27}\\
r_{-1}(\omega_{k})&=-i\gamma_{1D}\sum_{i,j,k=1}^{N}G_{ik}(\omega_{k}-\Omega)A^{*}_{k}G_{kj}(\omega_{k})\sum_{p,q=1}^{M}e^{ik_{0}(x_{ip}+x_{jq})}=-i\gamma_{1D}\sum_{k=1}^{N}A^{*}_{k}s_{k}^{+}(\omega_{k}-\Omega)s_{k}^{+}(\omega_{k}),\label{S28}
\end{align}
where $s_{n}^{+}(\omega)\equiv\sum\limits_{m=1}^{N}G_{nm}(\omega)\sum\limits_{p=1}^{M}e^{ ik_{0}x_{mp}}$ is defined, which in fact corresponds to the external line factors of the relevant Feynman diagrams, as we will show below. The quantities $|r_{1/-1}(\omega_{k})|^{2}$ describe the probability to find a reflecting photon of frequency $\omega_{k}\pm\Omega$ when a single photon of frequency $\omega_{k}$ is incident. The factor $s_{n}^{+}(\omega)$ can be simplified when the specific configuration of the qubits array is given. For example, considering a linear chain of $N$ separate giant atoms, where all coupling points of each atom are outside those of another atom, the factor $s_{n}^{+}(\omega)$ reads
\begin{align}
s_{n}^{+}(\omega)=\frac{e^{ik_{0}x_{m_{0}1}}}{i\gamma_{1D}(1+e^{-i\varphi})}[\frac{\omega_{0}+2\gamma_{1D}\sin(\varphi)-\boldsymbol{H}_{{\rm eff}}}{\omega-\boldsymbol{H}_{{\rm eff}}}]_{m_{0},n}\label{S29}
\end{align}
for $\varphi\neq\pi+2p\pi$ with $p$ an arbitrary integer, where $m_{0}$ is the index of the atom with the minimal $x$-coordinate value $x_{m_{0}p}$. Here,  $\varphi\equiv k_{0}\Delta x$ is the field phase difference upon propagation, with $\Delta x$ denoting the distance between neighboring coupling points. The external line factors are determined similarly when we consider a linear chain of $N$ braided giant atoms, where part of the coupling points from each atom is inside those of another atom, with a form of
\begin{align}
s_{n}^{+}(\omega)=\frac{e^{ik_{0}x_{m_{0}1}}}{i\gamma_{1D}(1+e^{-i2\varphi})}\left\{[\frac{\omega_{0}+2\gamma_{1D}\sin(2\varphi)- \boldsymbol{H}_{{\rm eff}}}{\omega-\boldsymbol{H}_{{\rm eff}}}]_{m_{0},n}+2\gamma_{1D}\sin(\varphi)G_{m_{0}+1,n}(\omega)\right\}\label{S30}
\end{align}
for $\varphi\neq\pi/2+p\pi$ with $p$ an arbitrary integer, where $m_{0}$ and $\varphi$ follow the same definitions as in Eq.\,(\ref{S29}).

The single-photon in-elastic scattering amplitudes in Eqs.\,(\ref{S27}-\ref{S28}) are valid in the regime of weak modulation. More general expressions of scattering amplitudes of arbitrary order are available by transferring the time-domain scattering matrix (S-matrix) $S(t',t)=S_{0}(t',t)e^{-i\int_{t}^{t'}2A\cos\Omega\tau d\tau}$ into the frequency-domain. Here, we have assumed that the giant atoms in the array are modulated homogeneously, i.e., $A_{n}=A$, and $S_{0}(t',t)$ is the time-domain S-matrix without modulation. The obtained frequency-domain S-matrix is given by
\begin{align}
S(\omega',\omega)=\sum_{kk'}S_{0}(\omega'+k'\Omega,\omega-k\Omega)(-1)^{k'}J_{k'}(\frac{2A}{\Omega})J_{k}(\frac{2A}{\Omega}).\label{S31}
\end{align}
One can extract scattering amplitudes of any order by simply setting the frequency $\omega'$ of the output field to $\omega+n\Omega$. This result reads
\begin{align}
S(\omega+n\Omega,\omega)=\sum_{k}^{\infty}S_{0}(\omega-k\Omega,\omega-k\Omega)J_{k+n}(\frac{2A}{\Omega})J_{k}(\frac{2A}{\Omega}).\label{S32}
\end{align}
The analytical coefficients in Eqs.\,(\ref{S27}-\ref{S28}) and Eq.\,(\ref{S32}) are consistent in the limit of weak modulation, i.e., $A/\Omega\ll 1$.

\subsection{Two-photon elastic scattering}
In this subsection, we give the analytical two-photon scattering amplitudes that are governed by the Green function for qubit excitation-excitation interaction. The utilized approach is similar to the proposed strategy in \cite{smPhysRevA.93.033856, smPhysRevLett.123.253601}.

\begin{figure}
  \centering
  \includegraphics[width=17cm]{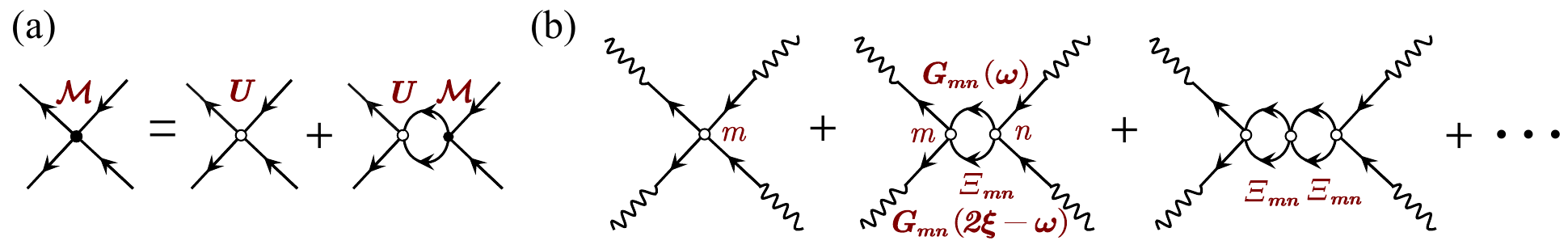}
  \caption{ (a) Diagrammatic representation of the Dyson equation for a pair of qubit excitations. Open dots and solid dots denote respectively the bare vertex and the dressed interaction vertex of excitation interaction; solid lines represent the Green functions of emitter excitations, and wavy lines represent the incoming or outgoing photons. (b) The series corresponding to the two-photon scattering.}\label{figS3}
\end{figure}

We first consider the two-photon elastic scattering process, where the two incident photons with the energies $\omega_{1}$ and $\omega_{2}$ are scattered elastically and converted into a pair of photons with the energies $\omega_{1}^{\prime}$ and $\omega_{2}^{\prime}$ satisfying $\omega_{1}+\omega_{2}=\omega_{1}^{\prime}+\omega_{2}^{\prime}\equiv 2\varepsilon$.  In order to obtain the two-photon scattering amplitude, we need to calculate the dressed vertex characterizing the interaction between qubit excitations. The dressed vertex $\mathcal{M}$ can be read directly from the Feynman diagram  as shown in Fig.\,\ref{figS3}(a), and satisfies the following Dyson equation
\begin{align}
-i\mathcal{M}=-2iU+2iU\int\frac{d\omega}{2\pi}G(\omega)G(2\varepsilon-\omega)(-i\mathcal{M}).\label{S33}
\end{align}
In the regime of two-level approximation ($U\rightarrow\infty$), the dressed vertex $\mathcal{M}$ can further be simplified as $\mathcal{M}=(-i\varXi)^{-1}$ with $\varXi_{ij}\equiv\int G_{ij}(\omega)G_{ij}(2\varepsilon-\omega)\frac{d\omega}{2\pi}$. It is instructive to make a matricization for $\varXi$ as follows
\begin{align}
\int G_{ij}(\omega)G_{kl}(2\varepsilon-\omega)\frac{d\omega}{2\pi}&=&\int\frac{d\omega}{2\pi}[\frac{1}{\boldsymbol{H}_{{\rm eff}}-\omega-i\epsilon}]_{mn}[\frac{1}{\boldsymbol{H}_{{\rm eff}}+\omega-2\varepsilon-i\epsilon}]_{kl}=&[\frac{i}{\boldsymbol{H}_{{\rm eff}}\otimes I+I\otimes \boldsymbol{H}_{{\rm eff}}-2\varepsilon}]_{mk,nl}.\label{S34}
\end{align}
Therefore, the matrix elements of  $\varXi_{mn}$ are given by
\begin{align}
\varXi_{mn}=[\frac{i}{\boldsymbol{H}_{{\rm eff}}\otimes I+I\otimes \boldsymbol{H}_{{\rm eff}}-2\varepsilon}]_{mm,nn}.\label{S35}
\end{align}
Based on these fundamental ingredients that constitute the diagram of two-photon scattering, we find the two-photon elastic scattering amplitude
\begin{align}
S_{0}^{{\rm }}(\omega'_{1},\omega'_{2};\omega_{1},\omega_{2})&=2\pi\delta(\omega'_{1}+\omega'_{2}-\omega_{1}-\omega_{2})\left[S_{0}^{{\rm coherent}}(\omega'_{1},\omega'_{2};\omega_{1},\omega_{2})+S_{0}^{{\rm incoherent}}(\omega'_{1},\omega'_{2};\omega_{1},\omega_{2})\right],\label{S36}
\end{align}
where $S_{0}^{{\rm coherent}}(\omega'_{1},\omega'_{2};\omega_{1},\omega_{2})$ and $S_{0}^{{\rm incoherent}}(\omega'_{1},\omega'_{2};\omega_{1},\omega_{2})$ represent the contributions from two-photon coherent and incoherent scattering processes, respectively. They have the following compact forms
\begin{align}
S_{0}^{{\rm coherent}}(\omega'_{1},\omega'_{2};\omega_{1},\omega_{2})&=\text{2\ensuremath{\pi}[\ensuremath{\delta(\omega'_{1}-\omega_{1})}+\ensuremath{\delta(\omega'_{1}-\omega_{2})}]}r(\omega_{1})r(\omega_{2}),\label{S37}\\
S_{0}^{{\rm incoherent}}(\omega'_{1},\omega'_{2};\omega_{1},\omega_{2})&=2\gamma_{1D}^{2}\sum_{m,n=1}^{N}s_{m}^{+}(\omega'_{1})s_{m}^{+}(\omega'_{2})[-iU\delta_{mn}+(-iU)\varXi_{mn}(-iU)+...]s_{n}^{+}(\omega_{1})s_{n}^{+}(\omega_{2})\nonumber\\
&=-2i\gamma_{1D}^{2}\sum_{m,n=1}^{N}s_{m}^{+}(\omega'_{1})s_{m}^{+}(\omega'_{2})\mathcal{M}_{mn}(\frac{\omega_{1}+\omega_{2}}{2})s_{n}^{+}(\omega_{1})s_{n}^{+}(\omega_{2}).\label{S38}
\end{align}
The amplitude in Eq.\,(\ref{S38}) accounts for the correlated excitation-excitation interaction, which is described by the summation series over indices $m$ and $n$ as shown in Fig.\,\ref{figS3}(b).

\subsection{Two-photon inelastic scattering}
In this subsection, we give the analytical two-photon scattering amplitudes in the presence of dynamical modulation. In this context, the total energy of the incident photons $\omega_{1}+\omega_{2}$ would be modulated as $\omega_{1}^{\prime}+\omega_{2}^{\prime}\pm n\Omega$ with $n$ the positive integers.

\begin{figure}
  \centering
  \includegraphics[width=17cm]{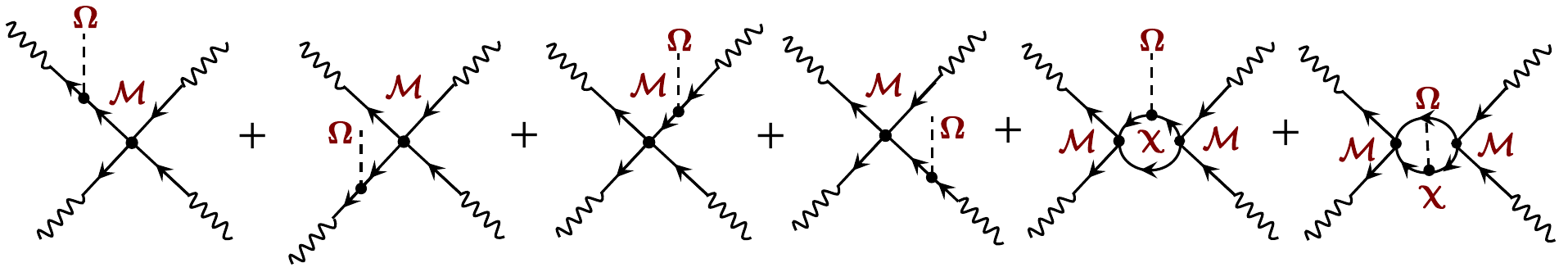}
  \caption{ Diagrammatic representation of the two-photon anti-Stokes scattering for a pair of qubit excitations. The vertically incident dashed line characterizes the absorbtion of a modulated photon. Open dots and solid dots denote respectively the bare vertex and the dressed interaction vertex of excitation interaction; solid lines indicate the Green functions of emitter excitations, and wavy lines indicate the incoming or outgoing photons.}\label{figS4}
\end{figure}

The two-photon scattering process can be truncated to the first order Stokes or anti-Stokes scattering under weak modulation. In order to obtain the two-photon inelastic scattering amplitudes, we need to calculate the dressed vertex characterizing the interaction between modulated qubit excitations. The dressed vertices for anti-Stokes and Stokes two-photon scattering are denoted by $\chi^{(+)}(\varepsilon)$ and $\chi^{(-)}(\varepsilon)$. Here, $\varepsilon$ is the average energy of a single incident photon. The vertices can be read directly from the Feynman diagram as shown in Fig.\,\ref{figS4}. The corresponding entries for $\chi^{(\pm)}(\varepsilon)$ are
\begin{align}
\chi^{(+)}_{mn}(\varepsilon)=&i\int\sum_{k}A_{k}G_{mn}(\omega)G_{mk}(2\varepsilon-\omega+\Omega)G_{kn}(2\varepsilon-\omega)\frac{d\omega}{2\pi}\nonumber\\
=&i\int\sum_{k}A_{k}[\frac{1}{\boldsymbol{H}_{{\rm eff}}-\omega-i\epsilon}]_{mn}[\frac{1}{\boldsymbol{H}_{{\rm eff}}+\omega-2\varepsilon-\Omega-i\epsilon}]_{mk}[\frac{1}{\boldsymbol{H}_{{\rm eff}}+\omega-2\varepsilon-i\epsilon}]_{kn}\frac{d\omega}{2\pi}\nonumber\\
=&[\frac{1}{2\varepsilon+\Omega-\boldsymbol{H}_{{\rm eff}}\otimes I-I\otimes \boldsymbol{H}_{{\rm eff}}}\otimes\frac{{\rm diag}(\boldsymbol{A})\otimes I}{2\varepsilon-\boldsymbol{H}_{{\rm eff}}\otimes I-I\otimes H_{{\rm eff}}}]_{mm,nn},\label{S39}\\
\chi^{(-)}_{mn}(\varepsilon)=&[\frac{1}{2\varepsilon-\Omega-\boldsymbol{H}_{{\rm eff}}\otimes I-I\otimes \boldsymbol{H}_{{\rm eff}}}\otimes\frac{{\rm diag}(\boldsymbol{A}^{*})\otimes I}{2\varepsilon-\boldsymbol{H}_{{\rm eff}}\otimes I-I\otimes \boldsymbol{H}_{{\rm eff}}}]_{mm,nn}.\label{S40}
\end{align}
After that, the two-photon anti-Stokes scattering amplitude could be determined straightforwardly from the relevant Feynman's diagram as shown in Fig.\,\ref{figS4}, and yields
\begin{align}
S_{1}^{{\rm }}(\omega'_{1},\omega'_{2};\omega_{1},\omega_{2})&=2\pi\delta(\omega'_{1}+\omega'_{2}-\omega_{1}-\omega_{2}-\Omega)\left[S_{1}^{{\rm coherent}}(\omega'_{1},\omega'_{2};\omega_{1},\omega_{2})+S_{1}^{{\rm incoherent}}(\omega'_{1},\omega'_{2};\omega_{1},\omega_{2})\right],\label{S41}
\end{align}
where $S_{1}^{{\rm coherent}}(\omega'_{1},\omega'_{2};\omega_{1},\omega_{2})$ and $S_{1}^{{\rm incoherent}}(\omega'_{1},\omega'_{2};\omega_{1},\omega_{2})$ represent the contributions from two-photon coherent and incoherent anti-Stokes scattering processes, respectively. They have the following compact forms
\begin{align}
S_{1}^{{\rm coherent}}(\omega_{1}^{\prime},\omega_{2}^{\prime};\omega_{1},\omega_{2})=&\text{2\ensuremath{\pi}[\ensuremath{\delta(\omega'_{1}-\omega_{1})}+\ensuremath{\delta(\omega'_{2}-\omega_{1})}]}r(\omega_{1})r_{1}(\omega_{2})+\text{2\ensuremath{\pi}[\ensuremath{\delta(\omega'_{1}-\omega_{2})}+\ensuremath{\delta(\omega'_{2}-\omega_{2})}]}r_{1}(\omega_{1})r(\omega_{2}),\label{S42}
\end{align}
\begin{align}
\!S_{1}^{{\rm incoherent}}(\omega_{1}^{\prime},\omega_{2}^{\prime};\!\omega_{1},\omega_{2})\!=\!&\!-2i\text{\ensuremath{\gamma}}_{1D}^{2}\sum_{m,n,k}A_{k}\{\mathcal{M}_{mn}(\varepsilon)[s_{k}^{+}(\omega'_{1})G_{km}(\omega'_{1}\!-\!\Omega)s_{m}^{+}(\omega'_{2})\!+\!s_{m}^{+}(\omega'_{1})G_{km}(\omega'_{2}\!-\!\Omega)s_{k}^{+}(\omega'_{2})]s_{n}^{+}(\omega_{1})s_{n}^{+}(\omega_{2})\nonumber\\
&+\mathcal{M}_{mn}(\varepsilon+\frac{\Omega}{2})s_{m}^{+}(\omega'_{1})s_{m}^{+}(\omega'_{2})[s_{k}^{+}(\omega{}_{1})G_{kn}(\omega{}_{1}+\Omega)s_{n}^{+}(\omega{}_{2})+s_{n}^{+}(\omega{}_{1})G_{kn}(\omega{}_{2}+\Omega)s_{k}^{+}(\omega{}_{2})]\}\nonumber\\
&-4i\gamma_{1D}^{2}\sum_{m,n,k,l}\mathcal{M}_{mk}(\varepsilon+\frac{\Omega}{2})\chi_{kl}(\varepsilon)\mathcal{M}_{ln}(\varepsilon)s_{n}^{+}(\omega'_{1})s_{n}^{+}(\omega'_{2})s_{m}^{+}(\omega_{1})s_{m}^{+}(\omega_{2}).\label{S43}
\end{align}

The amplitude in Eq.\,(\ref{S43}) accounts for the correlated interaction of modulated excitation-excitation. More concretely, the first line in the right-hand side of Eq.\,(\ref{S43}) describes the modulation after the interaction between two emitter excitations; the second line describes the modulation before the interaction; and the last line describes the modulation for the excitations scattered from the first interaction after which this photon pair interacts again. These contributions of  $S_{1}^{{\rm incoherent}}$ are sketched vividly in Fig.\,\ref{figS4}. Following a similar way, the two-photon Stokes scattering amplitudes could also be obtained.

General expressions of inelastic scattering amplitudes for arbitrary magnitude of $A/\Omega$ are available by transforming the two-photon S-matrix in time-domain  $S(t_{1}^{\prime},t_{2}^{\prime};t_{1},t_{2})=S_{0}(t_{1}^{\prime},t_{2}^{\prime};t_{1},t_{2})\displaystyle e^{-i\int_{t_{1}}^{t'_{1}} \int_{t_{2}}^{t'_{2}}2A(\cos\Omega\tau_{1}+\cos\Omega\tau_{2}) d\tau_{1}d\tau_{2}}$ into the frequency-domain. Note that we have assumed that the giant atoms in the array are modulated homogeneously, i.e., $A_{n}=A$, and $S_{0}(t_{1}^{\prime},t_{2}^{\prime};t_{1},t_{2})$ is the time-domain S-matrix without modulation. The result yields
\begin{align}
S(\omega'_{1},\omega'_{2};\omega_{1},\omega_{2})=\sum_{k'_{1}k'_{2}k_{1}k_{2}=-\infty}^{\infty}J_{k'_{1}}(\frac{2A}{\Omega})J_{k'_{2}}(\frac{2A}{\Omega})J_{k_{1}}(\frac{2A}{\Omega})J_{k_{2}}(\frac{2A}{\Omega})S_{0}(\omega'_{1}-k'_{1}\Omega,\omega'_{2}-k'_{2}\Omega;\omega{}_{1}-k{}_{1}\Omega,\omega_{2}-k{}_{2}\Omega).\label{S44}
\end{align}
One can immediately extract the scattering amplitudes of arbitrary inelastic scattering process from  Eq.\,(\ref{S44}). As an example, by setting $\omega_{1}^{\prime}+\omega_{2}^{\prime}=\omega_{1}+\omega_{2}+\Omega$, the first-order anti-Stokes scattering amplitude reads
\begin{align}
S_{1}(\omega'_{1},\omega'_{2};\omega_{1},\omega_{2})&=\frac{A}{\Omega}\left [S_{0}(\omega'_{1}-\Omega,\omega'_{2};\omega_{1},\omega_{2})+S_{0}(\omega'_{1},\omega'_{2}-\Omega;\omega_{1},\omega_{2})-S_{0}(\omega'_{1},\omega'_{2};\omega_{1}+\Omega,\omega_{2})-S_{0}(\omega'_{1},\omega'_{2};\omega_{1},\omega_{2}+\Omega)\right ].\label{S45}
\end{align}
The analytical results in Eq.\,(\ref{S41}) and Eq.\,(\ref{S45}) are consistent in the limit of weak modulation, i.e., $A/\Omega\ll 1$. Building on above theoretical preparations, we can now turn our attention to calculating the physical quantities of interest, as we present in Sec.S4

%%%%%%%%%%%%%%%%%%%%%%%%%%%%%%%%%%%%%%%%%%%%%%
\section{DYNAMICS OF THE PHOTON-PHOTON CORRELATIONS}
\setcounter{equation}{45}
\renewcommand\theequation{S\arabic{equation}}
\makeatletter
\renewcommand{\thefigure}{S\@arabic\c@figure}
\makeatother
In this section, based on the scattering amplitudes obtained in Sec.S3, we present the detailed derivation of the analytical time evolution of the correlations for a pair of reflected photons. The resulting analytical descriptions will cover the first-order two-photon scattering amplitude $S_{1}$ as presented in Eq.(7) of the main text.

We consider two modulated giant atoms that are excited by a pair of incident photons.  Since the array of giant atoms is modulated dynamically, the steady photon-photon correlation of the scattered photons is also periodic in time. The time-dependent second-order photon-photon correlation function of interest has a definition of
\begin{align}
g^{(2)}_{\leftarrow}(t+\tau,t)=\frac{\langle a_{\leftarrow}^{\dagger}(t+\tau)a_{\leftarrow}^{\dagger}(t)a_{\leftarrow}(t)a_{\leftarrow}(t+\tau)\rangle}{(\langle a_{\leftarrow}^{\dagger}a_{\leftarrow}\rangle_{0})^{2}},\label{S46}
\end{align}
where $a_{\leftarrow}$ is the annihilation operator for reflection modes, which can be expressed using system operators as $a_{\leftarrow}=i[\sigma_{1}(t)\sum\limits_{p=1}^{M}e^{ik_{0}x_{1p}}+\sigma_{2}(t)\sum\limits_{p=1}^{M}e^{ik_{0}x_{2p}}]$; $\langle{\cdots\rangle}$ and $\langle{\cdots\rangle}_{0}$ represent the average over the system state with and without modulation, respectively.  As the giant emitters are modulated periodically, we expect that the photon-photon correlation could be expanded as following Fourier series
\begin{align}
g^{(2)}_{\leftarrow}(t+\tau,t)=|\sum_{n}S_{n}(\tau)e^{-in\Omega t}|^{2}/(\langle a_{\leftarrow}^{\dagger}a_{\leftarrow}\rangle_{0})^{2}=\sum_{n=-\infty}^{\infty}e^{-in\Omega t}g_{n}^{(2)}(\tau),\label{S47}
\end{align}
where the contributions $g_{n}^{(2)}(\tau)$ are defined as $\sum\limits_{k=-\infty}^{\infty}S_{n+k}(\tau)S_{k}^{*}(\tau)/(\langle a_{\leftarrow}^{\dagger}a_{\leftarrow}\rangle_{0})^{2}$. In the limit of small enough modulation amplitude, i.e., $A_{n}/\gamma_{1D}\ll 1$ and weak driving strength, we truncate the time-dependent equal-time correlation function to the first-order sidebands as follows
\begin{align}
g^{(2)}_{\leftarrow}(t,t)\approx g_{0}^{(2)}(0)+g_{1}^{(2)}(0)e^{-i\Omega t}+g_{-1}^{(2)}(0)e^{i\Omega t}.\label{S48}
\end{align}

We are now in a position to determine analytically the components $g_{l}^{(2)}(0)$ of harmonic $l$. The approach is similar to \,\cite{PhysRevLett.130.023601}. In contrast, we consider the atomic array with arbitrary spatial configurations. Notably, in the following calculations, we distinguish the physical quantities obtained from two braided and separate giant atoms using superscripts $B$ and $S$, respectively.

Here we give the detailed calculations of the above correlation function $g^{(2)}_{l}$. Taking the case of two braided giant atoms as an example, the motion of an excitation in this atomic array is governed by
\begin{align}
H_{{\rm eff}}^{(B)}=(\omega_{0}-i\text{\ensuremath{\Gamma}}_{1D}^{(B)})(c_{1}^{\dagger}c_{1}+c_{2}^{\dagger}c_{2})-i\Gamma_{{\rm col}}^{(B)}(c_{1}^{\dagger}c_{2}+c_{2}^{\dagger}c_{1}),\label{S49}
\end{align}
where the complex frequencies $\Gamma_{1D}^{(B)}=2\gamma_{1D}(1+e^{2i\varphi})$ and $\Gamma_{{\rm col}}^{(B)}=\gamma_{1D}(e^{3i\varphi}+3e^{i\varphi})$ are defined for simplicity. After that, the dressed Green function is given by $\boldsymbol{G}^{(B)}(\omega)=\tilde{G}_{1}^{(B)}(\omega)\mathds{1}_{2\times2}+\tilde{G}_{2}^{(B)}(\omega)\sigma_{x}$ with
\begin{align}
\tilde{G}_{n}^{(B)}(\omega)=\frac{\mathcal{G}_{n}^{(B)}(\omega)}{[\omega+i\epsilon-\omega_{0}+i(\text{\ensuremath{\Gamma}}_{1D}^{(B)}+\Gamma_{{\rm col}}^{(B)})][\omega+i\epsilon-\omega_{0}+i(\text{\ensuremath{\Gamma}}_{1D}^{(B)}-\Gamma_{{\rm col}}^{(B)})]},\label{S50}
\end{align}
where $\mathcal{G}_{1}^{(B)}(\omega)=\omega-\omega_{0}+i\text{\ensuremath{\Gamma}}_{1D}^{(B)}$ and $\mathcal{G}_{2}^{(B)}(\omega)=-i\Gamma_{{\rm col}}^{(B)}$. The inverse of the dressed vertex $\mathcal{M}$ also has the closed matrix form $\mathcal{M}^{-1}(\varepsilon)=\mathcal{M}^{-1}_{11}\mathds{1}_{2\times2}+\mathcal{M}^{-1}_{12}\sigma_{x}$ with the elements
\begin{align}
\mathcal{M}^{-1}_{11}&=\frac{1}{8}[\frac{2}{-i\Gamma^{(B)}_{1D}-\varepsilon+\omega_{0}}+\frac{1}{\Gamma^{(B)}_{{\rm col}}+\Gamma^{(B)}_{1D}-i\varepsilon+i\omega_{0}}+\frac{1}{-i(\Gamma^{(B)}_{1D}-\Gamma^{(B)}_{{\rm col}})-\varepsilon+\omega_{0}}]\nonumber\\
\mathcal{M}^{-1}_{12}&=\frac{i\Gamma^{(B)2}_{{\rm col}}}{4(\Gamma^{(B)}_{1D}-i\varepsilon+i\omega_{0})[-\Gamma^{(B)2}_{{\rm col}}+(\Gamma^{(B)}_{1D}-i\varepsilon+i\omega_{0})^{2}]}.\label{S51}
\end{align}
Besides, the external line factors can also be determined as
\begin{align}
s_{n}^{+(B)}(\omega)=\frac{\mathcal{S}_{n}^{(B)}(\omega)}{[\omega+i\epsilon-\omega_{0}+i(\Gamma_{1D}^{(B)}+\Gamma_{{\rm col}}^{(B)})][\omega+i\epsilon-\omega_{0}+i(\Gamma_{1D}^{(B)}-\Gamma_{{\rm col}}^{(B)})]},\label{S52}
\end{align}
where notations $\mathcal{S}_{1}^{(B)}(\omega)=(1+e^{2i\varphi})(\omega-\omega_{0}+i\Gamma_{1D}^{(B)}-i\Gamma_{{\rm col}}^{(B)}e^{i\varphi})$ and $\mathcal{S}_{2}^{(B)}(\omega)=(1+e^{2i\varphi})[\text{(}\omega-\omega_{0}+i\Gamma_{1D}^{(B)})e^{i\varphi}-i\Gamma_{{\rm col}}^{(B)}]$ are introduced.

Under these useful and fundamental ingredients, we are readily able to write down all the considered two-photon scattering amplitudes $S_{n}$ according to Eqs.(\ref{S36}) and (\ref{S41}), which leads to
\begin{align}
S_{0}(\tau=0)=&2r^{(B)2}(\varepsilon)-2i\gamma_{1D}^{2}\mathcal{M}_{11}(\varepsilon)[s_{1}^{+(B)2}(\varepsilon)\Sigma_{1}(0,2\varepsilon)+s_{2}^{+(B)2}(\varepsilon)\Sigma_{2}(0,2\varepsilon)]\nonumber\\
&-2i\gamma_{1D}^{2}\mathcal{M}_{12}(\varepsilon)[s_{1}^{+(B)2}(\varepsilon)\Sigma_{2}(0,2\varepsilon)+s_{2}^{+(B)2}(\varepsilon)\Sigma_{1}(0,2\varepsilon)],\label{S53}\\
S_{1}(\tau=0)=&4r^{(B)}(\varepsilon)r^{(B)}_{1}(\varepsilon)-2i\gamma_{1D}^{2}\sum_{ijk}A_{k}\{\mathcal{M}_{ij}(\varepsilon)[\text{\ensuremath{\mathcal{A}_{ki,ki}}}(0,2\varepsilon+\Omega;\Omega)+\text{\ensuremath{\mathcal{B}_{ik,ik}}}(0,2\varepsilon+\Omega;2\varepsilon)]s_{j}^{+(B)2}(\varepsilon)\nonumber\\
&+2\mathcal{M}_{ij}(\varepsilon+\frac{\Omega}{2})\Sigma_{i}(0,2\varepsilon+\Omega)s_{k}^{+(B)}(\varepsilon)s_{j}^{+(B)}(\varepsilon)G_{kj}(\varepsilon+\Omega)\}\nonumber\\
&-4i\gamma_{1D}^{2}\sum_{ijkl}\mathcal{M}_{ik}(\varepsilon+\frac{\Omega}{2})\chi_{kl}^{(+)}(\varepsilon)\mathcal{M}_{lj}(\varepsilon)\Sigma_{i}(0,2\varepsilon+\Omega)s_{j}^{+(B)2}(\varepsilon),\label{S54}\\
S_{-1}(\tau=0)=&4r^{(B)}(\varepsilon)r^{(B)}_{-1}(\varepsilon)-2i\gamma_{1D}^{2}\sum_{ijk}A_{k}^{*}\{\mathcal{M}_{ij}(\varepsilon)[\text{\ensuremath{\mathcal{A}_{ki,ki}}}(0,2\varepsilon-\Omega;-\Omega)+\text{\ensuremath{\mathcal{B}_{ik,ik}}}(0,2\varepsilon-\Omega;2\varepsilon)]s_{j}^{+(B)2}(\varepsilon)\nonumber\\
&+2\mathcal{M}_{ij}(\varepsilon-\frac{\Omega}{2})\Sigma_{i}(0,2\varepsilon-\Omega)s_{k}^{+(B)}(\varepsilon)s_{j}^{+(B)}(\varepsilon)G_{kj}(\varepsilon-\Omega)\}\nonumber\\
&-4i\gamma_{1D}^{2}\sum_{ijkl}\mathcal{M}_{ik}(\varepsilon-\frac{\Omega}{2})\chi_{kl}^{(-)}(\varepsilon)\mathcal{M}_{lj}(\varepsilon)\Sigma_{i}(0,2\varepsilon-\Omega)s_{j}^{+(B)2}(\varepsilon),\label{S55}
\end{align}
where $\Sigma_{n}(a,b)$, $\text{\ensuremath{\mathcal{A}_{mn,kl}}}(a,b;c)$ and $\text{\ensuremath{\mathcal{B}_{mn,kl}}}(a,b;c)$ are defined by
\begin{align}
\Sigma_{n}(a,b)=&\ensuremath{\int}s_{n}^{+(B)}(\omega-a)s_{n}^{+(B)}(\ensuremath{b-\omega})\ensuremath{\frac{d\omega}{2\pi}},\label{S56}\\
\text{\ensuremath{\mathcal{A}_{mn,kl}}}(a,b;c)=&\delta_{kl}\mathcal{I}_{mn,1}(a,b;c)+(1-\delta_{kl})\mathcal{I}_{mn,2}(a,b;c),\label{S57}\\
\text{\ensuremath{\mathcal{B}_{mn,kl}}}(a,b;c)=&\delta_{kl}\mathcal{K}_{mn,1}(a,b;c)+(1-\delta_{kl})\mathcal{K}_{mn,2}(a,b;c).\label{S58}
\end{align}
\begin{figure}
  \centering
  \includegraphics[width=16cm]{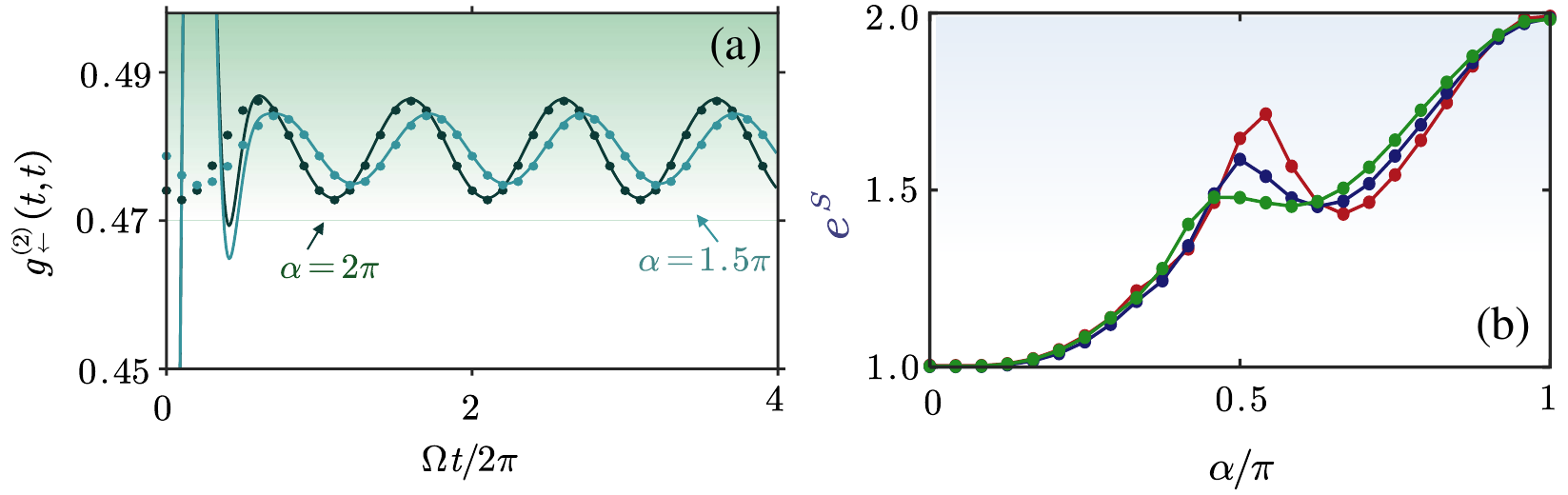}
  \caption{(a) Time dependence of the photon-photon correlation as a function of scaled time $2\pi t/\Omega$ for different relative modulation phases. The dotted and solid lines are analytical and numerical solutions, respectively. The parameters we used in panel (a) are $\varphi=0, \varphi_{l/r}=0$. (b) The dependence of $e^{S}$ on the relative modulation phase $\alpha$ by considering different atomic spacing and coupling phase that are distinguished by coloring: $\varphi=0, \varphi_{l/r}=0$ (red); $\varphi=0.75\pi, \varphi_{l/r}=0$ (blue); $\varphi=\pi/2, \varphi_{l}=0,\varphi_{r}=\pi$ (green).}\label{figS5}
\end{figure}
Note that the Eq.(\ref{S54}) is exactly the Eq.(7) in the main text. And we have introduced the residue integrals $\mathcal{I}_{mn,k}(a,b;c)$ and $\mathcal{K}_{mn,k}(a,b;c)$ in Eqs.(\ref{S57}) and (\ref{S58}), respectively, which are given by
\begin{align}
\mathcal{I}_{mn,k}(a,b;c)&=\int s_{m}^{+(B)}(\omega-a)s_{n}^{+(B)}(b-\omega)\tilde{G}^{(B)}_{k}(\omega-c)\ensuremath{\frac{d\omega}{2\pi}},\label{S59}\\
\mathcal{K}_{mn,k}(a,b;c)&=\int s_{m}^{+(B)}(\omega-a)s_{n}^{+(B)}(b-\omega)\tilde{G}^{(B)}_{k}(c-\omega)\ensuremath{\frac{d\omega}{2\pi}}.\label{S60}
\end{align}
Inserting Eqs.(\ref{S53}-\ref{S55}) into Eq.(\ref{S48}) results in the final expression for the equal-time second-order correlation function. A similar calculation of $g_{\leftarrow}^{(2)}(t,t)$ for the case of two separate giant atoms could be performed by following the same procedures. In this scenario, the definitions of both complex frequencies $\Gamma_{1D}^{(S)}=2\gamma_{1D}(1+e^{i\varphi})$ and $\Gamma_{{\rm col}}^{(S)}=\gamma_{1D}(e^{3i\varphi}+e^{i\varphi}+2e^{2i\varphi})$ are necessary.

A simple case, that is, the two giant atoms are located at the same point, i.e., $\varphi=0$ or $(2\pi)$, together with the modulations $A_{1}=A,A_{2}=Ae^{i\alpha}$, would greatly reduce the above cumbersome expressions for related scattering amplitudes, and lead to
\begin{align}
g_{0}^{(2)}(0)&=\frac{1+(\Delta/2)^{2}}{1+\Delta^{2}},\label{S61}\\
g_{1}^{(2)}(0)&=\frac{4e^{i\text{\ensuremath{\frac{\alpha}{2}}}}\cos(\text{\ensuremath{\frac{\alpha}{2}}})A g_{0}^{(2)}(0)\Delta}{\Gamma_{1D}}\times\frac{10+4\Delta^{2}-7i\Omega/\Gamma_{1D}-(\Omega/\Gamma_{1D})^{2}}{[(2\Delta)^{2}-(2i+\frac{\Omega}{\Gamma_{1D}})^{2}][\Delta^{2}-(2i+\frac{\Omega}{\Gamma_{1D}})^{2}]},\label{S62}
\end{align}
where the scaled detuning between pump frequency and atomic resonance frequency $\Delta=(\varepsilon-\omega_{0})/\Gamma_{1D}$ is defined. Note that the superscript of $\Gamma_{1D}$ is removed. This is because the interference behavior is the same regardless of whether the giant atoms are in a separate or braided setup.

The rigorous calculation of the total time-dependent zero-delay correlation function $g_{\leftarrow}^{(2)}(t,t)$ via the density matrix approach demonstrates the effectiveness of the analytical result given by Eq.(\ref{S48}). As shown in Fig.\ref{S5}(a), we numerically (solid lines) plot the time dependence of the photon-photon correlation function, for a pair of modulated braided giant atoms, as a function of the relative modulation phases $\alpha=1.5\pi, 2\pi$, which agrees well with that according to Eq.(\ref{S48}) (dotted lines) at long enough time. Notably, the analytical correlation function provided in Eq.(\ref{S48}) has been used to simulate the correlation dynamics, as shown in Fig. 4(b) of the main text.

 Moreover, we also study the dependence of the quantum entanglement, for a photon pair emitting into the sidebands $n_{1}$ and $n_{2}$,  on various parameters as shown in Fig.\ref{S5}(b). The calculations are performed by considering the addition of two detectors that are placed to the left of our qubit system.  The considered system parameters include the relative modulation phases $\alpha$, the propagation phase $\varphi$, and the coupling phase $\varphi_{l/r}$.  To quantify the frequency-encoded two-photon entanglement, we calculate the entanglement entropy\,\cite{smRevModPhys.82.277} defined by $S=-\sum_{\lambda}|\lambda|^{2}\ln |\lambda|^{2}$, where $\lambda$ are the singular values of the two-photon wave function $\Psi_{n_{1},n_{2}}$ determined numerically from the bichromatic photon-photon correlations. As we can see from  Fig.\ref{S5}(b), a non-zero value of  $\alpha$ is necessary to generate quantum entanglement. These curves of exponential of entanglement entropy $e^{S}$ exhibit non-monotonicity over $\alpha$ accompanied by local peaks and reach maximum entanglement at $\alpha=\pi$.

%%%%%%%%%%%%%%%%%%%%%%%%%%%%%%%%%%%%%%%%%%%%%%
\section{NON-MARKOVIAN DYNAMICS OF THE FLOQUET OPEN QUANTUM SYSTEM}
\setcounter{equation}{62}
\renewcommand\theequation{S\arabic{equation}}
\makeatletter
\renewcommand{\thefigure}{S\@arabic\c@figure}
\makeatother
In this section, we extend the previous discussion to the non-Markovian regime by considering the finite time delay for propagation between the modulated qubits. Once the non-Markovian effects are taken into account, multiphoton processes become exceptionally complex, rendering many approximations valid within the Markovian framework no longer applicable. Here, we address this challenge by employing matrix product states (MPS) simulations. Since this method conceals the spatial information of the bosonic field, it becomes challenging to accurately describe the dynamics of specific photon components. Therefore, we adopt a waveguide discretization scheme to capture the primary non-Markovian dynamical features of the propagating fields.

\subsection{Matrix product states simulations}
In this subsection, we present an exact non-Markovian description for the dynamics of qubits + field system based on the quantum stochastic Schr$\rm{\ddot{o}}$dinger equation (QSSE)\,\cite{smPhysRevLett.116.093601, smPhysRevResearch.3.023030}.  The considered quantum network consists of a pair of dynamically modulated atoms separated by a finite distance $d$, defined as $d\equiv x_{1}-x_{2}$. The overall dynamics of the quantum network is governed by the Schr$\rm{\ddot{o}}$dinger equation $i\frac{\partial}{\partial t}\ket{\Psi (t)}=H_{\rm{tot}}\ket{\Psi(t)}$. Here, $H_{\rm{tot}}$ represents the total Hamiltonian, expressed as $H_{\rm{tot}}=H_{\rm{sys}}+H_{\rm{bath}}+H_{\rm{int}}$, where
\begin{align}
H_{\rm{sys}}&=\sum_{n=1}^{N}[\omega_{n}(t)\sigma_{n}^{\dagger}\sigma_{n}-\frac{\Omega_{n}}{2}(\sigma_{n}e^{i\omega_{d}t}+{\rm H.c.})],\nonumber\\
H_{\rm{bath}}&=\sum_{\lambda=L,R}\int d\omega\,\omega b_{\lambda}^{\dagger}(\omega)b_{\lambda}(\omega),\nonumber\\
H_{{\rm int}}&=i\sum_{\lambda=L,R}\sum_{n=1}^{N}\int d\omega\sqrt{\frac{\gamma_{\lambda}}{2\pi}}b_{\lambda}^{\dagger}(\omega)\sigma_{n}e^{-i\omega x_{n}/v_{\lambda}}+{\rm H.c.}.\label{S63}
\end{align}
The qubits are driven by a coherent driving field with Rabi frequency $\Omega_{n}$ at the driving frequency $\omega_{d}$, while their energies are modulated around the resonant frequency as $\omega_{n}(t)=\omega_{0}+\Delta_{n}(t)$. For simplicity, the decay rates into different output channels are assumed to be identical, i.e., $\gamma_{R}=\gamma_{L}=\gamma$. Under the rotating transformation $U_{I}(t)=\exp(-iH_{\rm{bath}}t-i\sum_{n}\omega_{0}\sigma_{n}^{\dagger}\sigma_{n})$, the Schr$\rm{\ddot{o}}$dinger equation can be written as $i\frac{\partial}{\partial t}\ket{\Psi_{I} (t)}=[H_{{\rm sys},I}(t)+H_{{\rm int},I}(t)]\ket{\Psi_{I}(t)}$, where the wave function has the form of $\ket{\Psi_{I}(t)}\equiv U^{\dagger}_{I}(t)\ket{\Psi(t)}$, and the system Hamiltonian and the interaction Hamiltonian in this picture read
\begin{align}
H_{{\rm sys},I}(t)&=\sum_{n=1}^{N}[\Delta_{n}(t)\sigma_{n}^{\dagger}\sigma_{n}-\frac{\Omega_{n}}{2}(\sigma_{n}+{\rm H.c.})],\nonumber\\
H_{{\rm int},I}(t)&=i\sum_{\lambda=L,R}\sum_{n=1}^{N}\sqrt{\gamma}b_{\lambda}^{\dagger}(t-x_{n}/v_{\lambda})\sigma_{n}e^{-i\omega_{d} x_{n}/v_{\lambda}}+{\rm H.c.},\label{S64}
\end{align}
where the quantum noise operators $b_{\lambda}(t)=\frac{1}{\sqrt{2\pi}}\int d\omega b_{\lambda}(\omega)e^{-i(\omega-\omega_{d})t}$ have been defined, satisfying the commutation relations $[b_{\lambda}(t),b^{\dagger}_{\lambda'}(t')]=\delta_{\lambda,\lambda'}\delta(t-t')$. Finally, redefining
\begin{align}
b_{R}(t)\rightarrow b_{R}(t+x_{1}/v_{R})e^{-i\omega_{d} x_{1}/v_{R}};\,\,\,\,\,\, b_{L}(t)\rightarrow b_{L}(t+x_{2}/v_{L})e^{-i\omega_{d} x_{2}/v_{L}},\label{S65}
\end{align}
we obtain
\begin{align}
H_{{\rm int},I}(t)&=i\sqrt{\gamma}\sum_{n}[b_{R}^{\dagger}(t-x_{n}/v_{R})\sigma_{n}e^{-i\omega_{d} x_{n}/v_{R}}+b_{L}^{\dagger}(t-x_{n}/v_{L})\sigma_{n}e^{-i\omega_{d} x_{n}/v_{L}}-{\rm H.c.}]\nonumber\\
&=i\sqrt{\gamma}\sum_{n}[b_{R}^{\dagger}(t-x_{n}/v_{R}+x_{1}/v_{R})\sigma_{n}e^{-i\omega_{d} (x_{n}-x_{1})/v_{R}}+b_{L}^{\dagger}(t-x_{n}/v_{L}+x_{2}/v_{L})\sigma_{n}e^{-i\omega_{d} (x_{n}-x_{2})/v_{L}}-{\rm H.c.}]\nonumber\\
&=i\sqrt{\gamma}[(b_{R}^{\dagger}(t)+b_{L}^{\dagger}(t+\tau)e^{i\varphi})\sigma_{1}-{\rm H.c.}]+i\sqrt{\gamma}[(b_{R}^{\dagger}(t-\tau)e^{i\varphi}+b_{L}^{\dagger}(t))\sigma_{1}-{\rm H.c.}],\label{S66}
\end{align}
where $\varphi\equiv -\omega_{d}\tau$ denotes the photon propagation phase and $\tau=d/v_{R}=-d/v_{L}$ is the time delay associated with photon exchange. It is clear that the terms $b_{L}^{\dagger}(t+\tau)$ and $b_{R}^{\dagger}(t-\tau)$ make the problem non-Markovian. To apply the tensor network algorithm, we have to discretize the annihilation operators $b_{\lambda}(t)$ in the time domain as the time-bin noise operators $\Delta B_{\lambda}(t_{k})\equiv\int_{t_{k}}^{t_{k+1}}\,dt b_{\lambda}(t)$. These operators satisfying the bosonic commutation relations $[\Delta B_{\lambda}(t_{k}),\Delta B^{\dagger}_{\lambda'}(t_{k'})]=\Delta t \delta_{\lambda,\lambda'}\delta_{kk'}$ where $\Delta t=t_{k+1}-t_{k}$. For convenience, we define $\Delta t$ as a fractional unit of the discretized time $t_{k}$, such that $\tau$ can be expressed as $\tau=l\Delta t$. By retaining the first-order approximation in the time step $\Delta t$, the time evolution of the system can be expressed as $\ket{\Psi_{I}(t_{k+1})}=U_{k}\ket{\Psi_{I}(t_{k})}=\exp[-iH_{{\rm sys},I}(t)\Delta t +V_{k,1}(t_{k})+V_{k,2}(t_{k})]\ket{\Psi_{I}(t_{k})}$ with
\begin{align}
V_{k,1}(t_{k})&=\sqrt{\gamma}[\Delta B_{R}^{\dagger}(t_{k})+\Delta B_{L}^{\dagger}(t_{k-l})e^{i\varphi}]\sigma_{1}-{\rm H.c.}\nonumber\\
V_{k,2}(t_{k})&=\sqrt{\gamma}[\Delta B_{R}^{\dagger}(t_{k-l})e^{i\varphi}+\Delta B_{L}^{\dagger}(t_{k})]\sigma_{2}-{\rm H.c.}.\label{S67}
\end{align}

In the time bin representation, we can express arbitrary quantum states $\ket{\Psi_{I}(t_{k})}$ into the following MPS form
\begin{align}
\ket{\Psi_{I}(t_{k})}=\sum\limits_{i_{s},i_{1},...,i_{N}}A^{i_{s}}_{a_{1}}A^{i_{1}}_{a_{1},a_{2}}...A^{i_{N-1}}_{a_{N-1},a_{N}}A^{i_{N}}_{a_{N}}\ket{i_{s},i_{1},...,i_{N}}.\label{S68}
\end{align}
where the first term corresponds to the system bin, which in our case contains two modulated qubits, while the remaining $N$ terms represent the time bins containing both the right and left propagating modes. The diagrammatic representation for such an MPS is shown in Fig.\ref{S6}(a) in w hich each tensor has one physical index (depicted as open-ended lines) and two bond indices  (depicted as connecting lines). Notably, we consider an initial state with empty excitations, such that the state $\ket{\Psi_{I}(0)}$ is completely uncorrelated, and the dimension of the bond indices is initially $1$. Moreover, any $N_{s}$-body operator $O$ can be expressed similarly in a local representation as
\begin{align}
O=\sum\limits_{i_{1},...,i_{N_{s}},i'_{1},...,i'_{N_{s}}}O^{i_{1},i'_{1}}_{a_{0},a_{1}}O^{i_{2},i'_{2}}_{a_{1},a_{2}}...O^{i_{N_{s}},i'_{N_{s}}}_{a_{N_{s}-1},a_{N_{s}}}\ket{i'_{1},...,i'_{N_{s}}}\bra{i_{1},...,i'_{N_{s}}}.\label{S69}
\end{align}
\begin{figure}
  \centering
  \includegraphics[width=16cm]{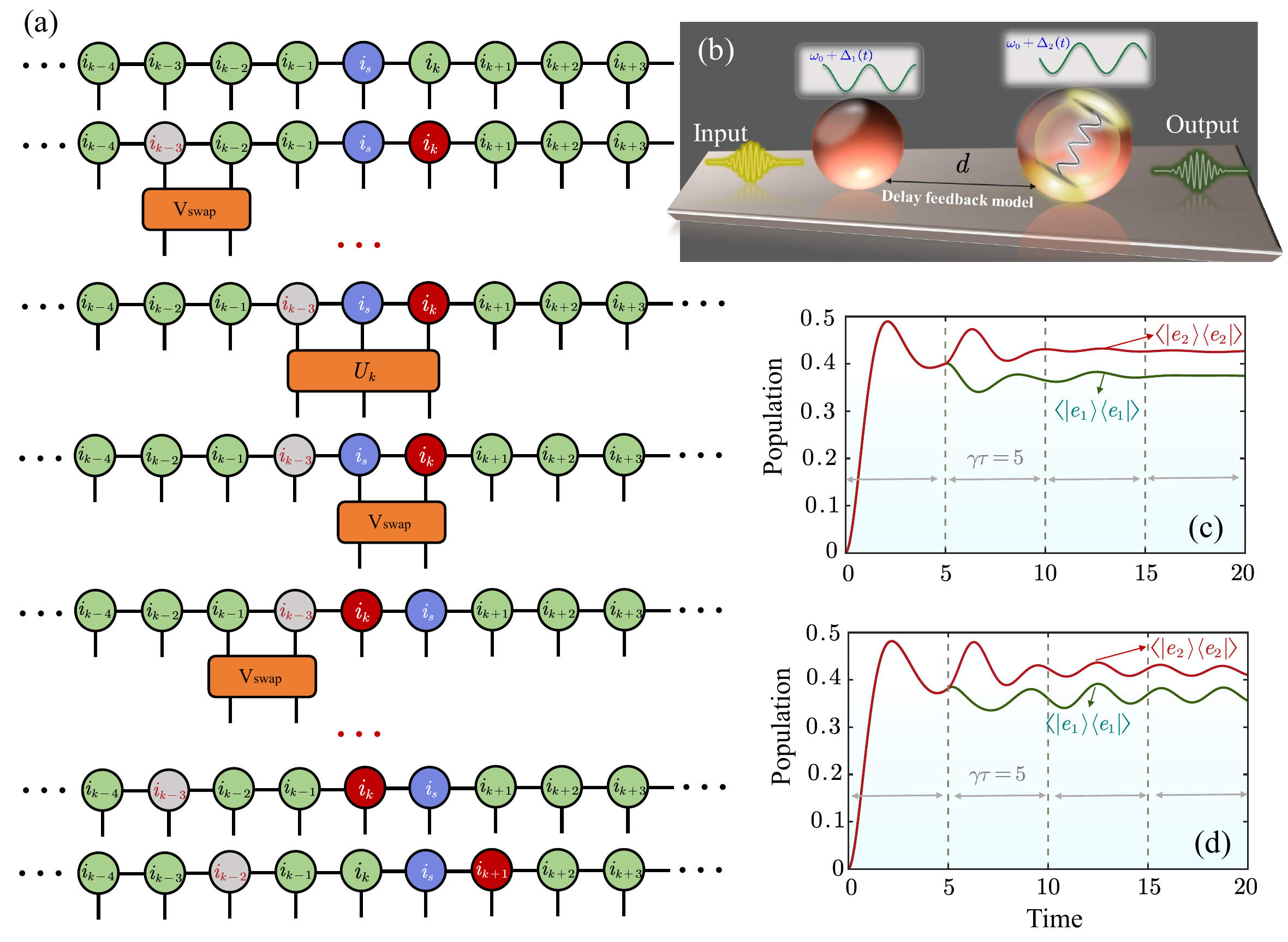}
  \caption{(a) Diagrammatic representation of the update process for the $k$th step of the MPS, in which the system bin, time bin and feedback bin are represented by blue, green, and gray balls, respectively. The tensor representations of the matrix product operators, including swap gates and the time evolution operators, are depicted by orange boxes.  The considered quantum optics model (b) is composed by a pair of dynamically modulated qubits with time-delay feedback. (c) and (d) show the time evolution of the atomic population without and with modulation, respectively. The atoms are initially in their ground state and driven by a coherent field with Rabi frequencies $\Omega_{1}=1.5$ and $ \Omega_{2}=1.5e^{-i\varphi}$, while the waveguide is in the vacuum state. The parameters used here are $\Delta t=0.1,\gamma\tau=5, \varphi=0.5\pi, A=0.5\gamma, \Omega =\gamma$. }\label{figS6}
\end{figure}

We are now in a position to briefly illustrate the update of the MPS for $l=3$ illustrated in Fig.\ref{S6}(a). In order to map the state $\ket{\Psi_{I}(t_{k})}$ to $\ket{\Psi_{I}(t_{k+1})}$, the contraction operation between local tensors, including the system bin (blue ball), the feedback bin (gray ball), the current time bin (green ball), and the tensor representation of the matrix product operator (MPO) corresponding to the unitary operation $U_{k}$, should be applied. This contraction operation is a long-range operation. Since the delay bin is separated from the system bin by $l$ time bins, a series of swap gates must be sequentially applied to move the delay bin to the left of the system bin. Once the time evolution operation is completed, the system bin becomes entangled with time bins $i_{k}$ and $i_{k-l}$, with bond indices exceeding $1$. After that, we swap the system bin with time bin $i_{k+1}$ to facilitate the next step of time evolution. Meanwhile, the feedback bin is similarly moved back to its original position through a series of local swap operations. Thus, we have described a complete procedure for updating the MPS, where iterating this process repeatedly enables the entire time evolution.

In Fig.\ref{S6}(c), we present the dynamics of the atomic populations $\langle\ket {e_{1}}\bra {e_{1}}\rangle$ and $\langle\ket {e_{2}}\bra{e_{2}}\rangle$ without modulation, obtained using the MPS simulation described above. The two atoms separated by a distance $d$ undergo independent Rabi oscillations before the instant $d/v_{g}$, as they do not experience the influence of delayed photons. Once $t>\tau$, the interference between the driving laser field and the time-delayed electromagnetic field causes the destructive and constructive effects in the atomic populations for qubits $1$ and $2$, respectively. When the dynamical modulation is applied, the steady-state population dynamics exhibit pronounced periodic behavior, as shown in Fig.\ref{S6}(d).

\subsection{Waveguide spatial discretization approach}
It is important to note that the time-bin noise operators $\Delta B_{\lambda}(t_{k})$ obscure the positional information of photons propagating through the waveguide. As a result, this treatment does not capture the detailed dynamics of photons during their interaction with the waveguide-coupled atoms. Here, we alleviate it by modeling the continuum of bosonic modes in the $1$D waveguide as   an effective coupled-cavity array\,\cite{smPhysRevLett.122.073601}. In this setup, we consider $N_{c}$ identical cavities, each with a resonant frequency $\omega_{c}$ and nearest-neighbour coupling rate $J$, with $N_{c}$ sufficiently large to avoid the non-Markovian behavior arising from the edge effect. The Hamiltonian in Eq.(\ref{S20}) can be approximated by the following discretized form
\begin{align}
H_{{\rm dis}}(t)=\sum_{n=1}^{2}\omega_{n}(t)\sigma_{n}^{\dagger}\sigma_{n}+\omega_{c}\sum\limits_{j=1}^{N_{c}}a_{j}^{\dagger}a_{j}-J\sum\limits_{j=1}^{N_{c}}(a_{j}^{\dagger}a_{j+1}+a_{j+1}^{\dagger}a_{j})+g\sum\limits_{n=1}^{2}
(\sigma_{n}^{\dagger}a_{j_{n}}+a_{j_{n}}^{\dagger}\sigma_{n}),\label{S70}
\end{align}
where $a_{j}$ represents the bosonic annihilation operator associated with the $j$th coupled cavity, while the index $j_{n}$ denotes the positional location of the optical cavity coupled to the $n$th qubit. This 1D structured reservoir provides a finite band of width $4J$ with dispersion law $\omega_{k}=-2J\cos(k)$.  The dynamics described by Eq.(\ref{S70}) can be approximately mapped to the continuous waveguide model, provided that the coupling and modulation amplitudes are sufficiently weak, the modulation frequency is low, and the atomic resonance energy aligns with the central band energy $\omega_{c}$.

\begin{figure}
  \centering
  \includegraphics[width=16cm]{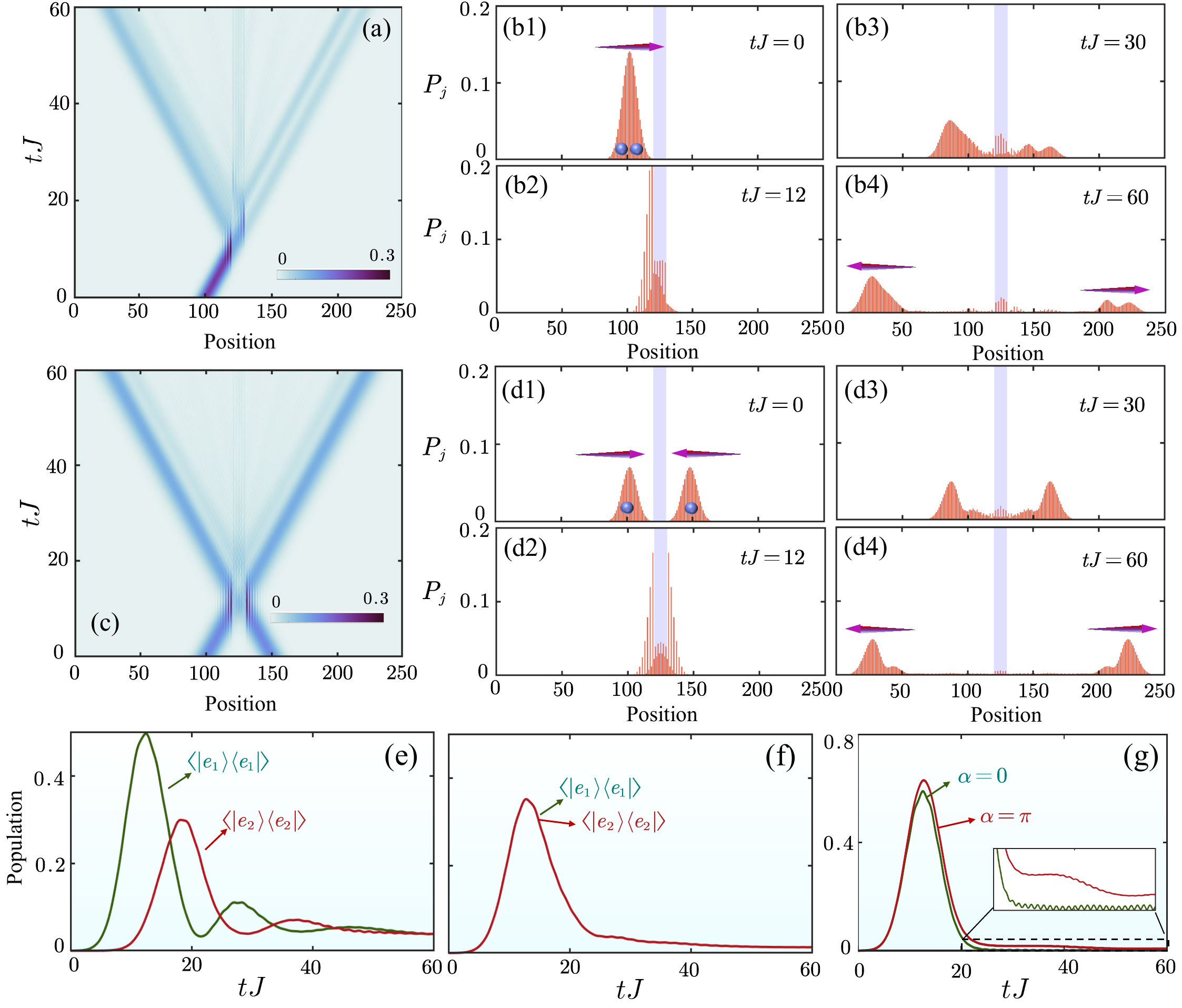}
  \caption{(a) Density plot of the lattice occupation $P_{j}(t)$ as a function of position $j$ and scaled time $tJ$, considering a two-photon wavepacket incident from the left, centered at $n_{0}-25$ with width $\sigma=8$.  Here, $n_{0}$  indexes the cavity at the center of the one-dimensional coupled cavity array. Panels (b1)-(b4) are the snapshots of the expectation values $P_{j}(t)$ at instants $tJ=0,12,30,60$. Consider two independent single-photon Gaussian wave packets initially localized at $n_{0}-25$ and $n_{0}+25$, propagating to the right and to the left, respectively. The corresponding density plot of $P_{j}(t)$ is shown in panel (c) followed by four snapshots as shown in panels (d1)-(d4). Note that the two dynamically modulated atoms are coupled to the $(n_{0}-5)$th and $(n_{0}+5)$th  cavities with modulation $\omega_{1}=\omega_{c}+A\cos(\Omega t),\omega_{2}=\Omega_{c}+A\cos(\omega t+\alpha)$, respectively. The atomic dynamics corresponding to the two-photon scattering simulations shown in panels (a) and (c) are presented respectively in panels (e) and (f). Panel (g) shows the dependence of time evolution of atomic population $P_{e}(t)=\langle\ket {e_{1}}\bra {e_{1}}\rangle(t)+\langle\ket {e_{2}}\bra {e_{2}}\rangle(t)$ on the relative modulation phase $\alpha$, with both qubits placed in the $n_{0}$th cavity. Note that the region between the two atoms in the coupled cavity array is shaded in light purple. The parameters used here are $\sigma=8, g=0.6J, A=0.4 J,\Omega =5g^{2}/J, N_{c}=249$. }\label{figS7}
\end{figure}

Such a discrete model conserves the total excitation number $\langle\sum_{n}\sigma^{\dagger}_{n}\sigma_{n}+\sum_{j}a_{j}^{\dagger}a_{j}\rangle$, and enables the efficient simulation of  multiphoton scattering, where the photonic dynamics can be easily monitored. In the following, we consider a pair of dynamically modulated qubits excited by two photon wave packets, and thus choose the initial wave function as\,\cite{smPhysRevLett.104.023602,smLongo_2009}
\begin{align}
\ket{\Psi(t=0)}=\mathcal{N}\sum_{j}\sum_{j'}[\psi_{1}^{(k_{0},\sigma)}(j,n_{c})\psi_{2}^{(q_{0},\sigma')}(j',m_{c})a_{j}^{\dagger}a_{j'}^{\dagger}+1\leftrightarrow 2]\ket{G},\label{S71}
\end{align}
where $\mathcal{N}$ is the normalization constant, and $\psi_{s}^{(k_{0},\sigma)}(j,n_{c})$  represents the single-photon wave function described by a Gaussian envelope:
\begin{align}
\psi_{s}^{(k_{0},\sigma)}(j,n_{c}) = \frac{1}{(\pi\sigma^{2})^{1/4}}e^{\frac{(j-n_{c})^{2}}{2\sigma^{2}}}e^{ik_{0}n},\label{S72}
\end{align}
with $k_{0},n_{c}$, and $\sigma$ denoting the carrier wave's wavenumber, the wavepacket's center (in unit of longitudinal size of each fictitious cavity), and its width, respectively. In two-excitation subspace, a general state of the system at time $t>0$ takes the form of
\begin{align}
\ket{\Psi(t) } =\sum_{j}\sum_{j'}\Phi_{j,j'}(t)a_{j}^{\dagger}a_{j'}^{\dagger}\ket{G}+\sum_{j}\sum_{n}C_{nj}(t)a_{j}^{\dagger}\sigma_{n}^{\dagger}\ket{G}+\sum_{n}\sum_{n'}W_{nn'}(t)\sigma_{n}^{\dagger}\sigma_{n'}^{\dagger}\ket{G},\label{S73}
\end{align}
where $\Phi_{j,j'}(t)$ denote the probability amplitudes for the two photons being located in the $j$th and $j'$th optical cavities; and $C_{nj}(t)$  are the probability amplitudes of finding one photon in the $j$th cavity while the $n$th atom is excited, and $W_{nn'}(t)$ describe the probability amplitudes for both the $n$th and $n'$th atoms being excited at time $t$.

The time evolution of the quantum mechanical state $\ket{\Psi(t)}$ obeys the time-dependent Schr$\rm{\ddot{o}}$dinger equation $i\frac{\partial}{\partial t}\ket{\Psi(t)}=H_{{\rm dis}}(t)\ket{\Psi(t)}$, from which the physical quantities, including the atomic population $\langle\ket{e_{n}}\bra{e_{n}}\rangle(t)=\bra{\Psi(t)}\sigma_{n}^\dagger\sigma_{n}\ket{\Psi(t)}$ and the occupation $P_{j}(t)=\bra{\Psi(t)}a_{j}^\dagger a_{j}\ket{\Psi(t)}$ for a given lattice site, can be obtained. We simulate the dynamics of two-photon wavepacket scattered by a pair of qubits with modulation $\omega_{1,2}=\omega_{c}+A\cos(\Omega t)$ where the qubits are separated by $10$ cavities for $N_{c}=249$ . We examine two types of two-photon scattering processes: (i) a two-photon wave packet incident from the left, and (ii) two independent single-photon waveguides incident from both sides of the coupled cavity. Figures \ref{S7}(a) and (c) present the detailed dynamics, with four representative instants selected to illustrate the transient distribution of the lattice. The incident two-photon wave packet propagates with group velocity $\partial_{k} \omega_{k}$
through the $1$D lattice, interacting with the dynamically modulated atoms, leading to reflection and transmission, along with changes in the shape of the scattered photon wave packets. We also investigate the effect of relative modulation phase on the system dynamics in Fig.\ref{S7}(g), where the qubits are placed in the same cavity. It shows that when the atoms are modulated in phase ($\alpha=0$), they rapidly de-excite after being excited, whereas the de-excitation process of atoms modulated in anti-phase ($\alpha=\pi$) occurs more gradually.

%\bibliography{referSM}

\begin{thebibliography}{64}%
\makeatletter
\providecommand \@ifxundefined [1]{%
 \@ifx{#1\undefined}
}%
\providecommand \@ifnum [1]{%
 \ifnum #1\expandafter \@firstoftwo
 \else \expandafter \@secondoftwo
 \fi
}%
\providecommand \@ifx [1]{%
 \ifx #1\expandafter \@firstoftwo
 \else \expandafter \@secondoftwo
 \fi
}%
\providecommand \natexlab [1]{#1}%
\providecommand \enquote  [1]{``#1''}%
\providecommand \bibnamefont  [1]{#1}%
\providecommand \bibfnamefont [1]{#1}%
\providecommand \citenamefont [1]{#1}%
\providecommand \href@noop [0]{\@secondoftwo}%
\providecommand \href [0]{\begingroup \@sanitize@url \@href}%
\providecommand \@href[1]{\@@startlink{#1}\@@href}%
\providecommand \@@href[1]{\endgroup#1\@@endlink}%
\providecommand \@sanitize@url [0]{\catcode `\\12\catcode `\$12\catcode
  `\&12\catcode `\#12\catcode `\^12\catcode `\_12\catcode `\%12\relax}%
\providecommand \@@startlink[1]{}%
\providecommand \@@endlink[0]{}%
\providecommand \url  [0]{\begingroup\@sanitize@url \@url }%
\providecommand \@url [1]{\endgroup\@href {#1}{\urlprefix }}%
\providecommand \urlprefix  [0]{URL }%
\providecommand \Eprint [0]{\href }%
\providecommand \doibase [0]{http://dx.doi.org/}%
\providecommand \selectlanguage [0]{\@gobble}%
\providecommand \bibinfo  [0]{\@secondoftwo}%
\providecommand \bibfield  [0]{\@secondoftwo}%
\providecommand \translation [1]{[#1]}%
\providecommand \BibitemOpen [0]{}%
\providecommand \bibitemStop [0]{}%
\providecommand \bibitemNoStop [0]{.\EOS\space}%
\providecommand \EOS [0]{\spacefactor3000\relax}%
\providecommand \BibitemShut  [1]{\csname bibitem#1\endcsname}%
\let\auto@bib@innerbib\@empty
%</preamble>
\bibitem [{\citenamefont {Fortier}\ and\ \citenamefont
  {Baumann}(2019)}]{Fortier2019}%
  \BibitemOpen
  \bibfield  {author} {\bibinfo {author} {\bibfnamefont {T.}~\bibnamefont
  {Fortier}}\ and\ \bibinfo {author} {\bibfnamefont {E.}~\bibnamefont
  {Baumann}},\ }\href {\doibase 10.1038/s42005-019-0249-y} {\bibfield
  {journal} {\bibinfo  {journal} {Commun. Phys.}\ }\textbf {\bibinfo {volume}
  {2}},\ \bibinfo {pages} {153} (\bibinfo {year} {2019})}\BibitemShut {NoStop}%
\bibitem [{\citenamefont {Chang}\ \emph {et~al.}(2022)\citenamefont {Chang},
  \citenamefont {Liu},\ and\ \citenamefont {Bowers}}]{Chang2022}%
  \BibitemOpen
  \bibfield  {author} {\bibinfo {author} {\bibfnamefont {L.}~\bibnamefont
  {Chang}}, \bibinfo {author} {\bibfnamefont {S.}~\bibnamefont {Liu}}, \ and\
  \bibinfo {author} {\bibfnamefont {J.~E.}\ \bibnamefont {Bowers}},\ }\href
  {\doibase 10.1038/s41566-021-00945-1} {\bibfield  {journal} {\bibinfo
  {journal} {Nat. Photon.}\ }\textbf {\bibinfo {volume} {16}},\ \bibinfo
  {pages} {95} (\bibinfo {year} {2022})}\BibitemShut {NoStop}%
\bibitem [{\citenamefont {Wang}\ \emph {et~al.}(2023)\citenamefont {Wang},
  \citenamefont {Huang}, \citenamefont {Qiu},\ and\ \citenamefont
  {Xiong}}]{WANG2023114137}%
  \BibitemOpen
  \bibfield  {author} {\bibinfo {author} {\bibfnamefont {X.}~\bibnamefont
  {Wang}}, \bibinfo {author} {\bibfnamefont {K.-W.}\ \bibnamefont {Huang}},
  \bibinfo {author} {\bibfnamefont {Q.-Y.}\ \bibnamefont {Qiu}}, \ and\
  \bibinfo {author} {\bibfnamefont {H.}~\bibnamefont {Xiong}},\ }\href
  {\doibase https://doi.org/10.1016/j.chaos.2023.114137} {\bibfield  {journal}
  {\bibinfo  {journal} {Chaos Solitons. Fractals.}\ }\textbf {\bibinfo {volume}
  {176}},\ \bibinfo {pages} {114137} (\bibinfo {year} {2023})}\BibitemShut
  {NoStop}%
\bibitem [{\citenamefont {Raussendorf}\ and\ \citenamefont
  {Briegel}(2001)}]{PhysRevLett.86.5188}%
  \BibitemOpen
  \bibfield  {author} {\bibinfo {author} {\bibfnamefont {R.}~\bibnamefont
  {Raussendorf}}\ and\ \bibinfo {author} {\bibfnamefont {H.~J.}\ \bibnamefont
  {Briegel}},\ }\href {\doibase 10.1103/PhysRevLett.86.5188} {\bibfield
  {journal} {\bibinfo  {journal} {Phys. Rev. Lett.}\ }\textbf {\bibinfo
  {volume} {86}},\ \bibinfo {pages} {5188} (\bibinfo {year}
  {2001})}\BibitemShut {NoStop}%
\bibitem [{\citenamefont {Menicucci}\ \emph {et~al.}(2008)\citenamefont
  {Menicucci}, \citenamefont {Flammia},\ and\ \citenamefont
  {Pfister}}]{PhysRevLett.101.130501}%
  \BibitemOpen
  \bibfield  {author} {\bibinfo {author} {\bibfnamefont {N.~C.}\ \bibnamefont
  {Menicucci}}, \bibinfo {author} {\bibfnamefont {S.~T.}\ \bibnamefont
  {Flammia}}, \ and\ \bibinfo {author} {\bibfnamefont {O.}~\bibnamefont
  {Pfister}},\ }\href {\doibase 10.1103/PhysRevLett.101.130501} {\bibfield
  {journal} {\bibinfo  {journal} {Phys. Rev. Lett.}\ }\textbf {\bibinfo
  {volume} {101}},\ \bibinfo {pages} {130501} (\bibinfo {year}
  {2008})}\BibitemShut {NoStop}%
\bibitem [{\citenamefont {Pysher}\ \emph {et~al.}(2011)\citenamefont {Pysher},
  \citenamefont {Miwa}, \citenamefont {Shahrokhshahi}, \citenamefont
  {Bloomer},\ and\ \citenamefont {Pfister}}]{PhysRevLett.107.030505}%
  \BibitemOpen
  \bibfield  {author} {\bibinfo {author} {\bibfnamefont {M.}~\bibnamefont
  {Pysher}}, \bibinfo {author} {\bibfnamefont {Y.}~\bibnamefont {Miwa}},
  \bibinfo {author} {\bibfnamefont {R.}~\bibnamefont {Shahrokhshahi}}, \bibinfo
  {author} {\bibfnamefont {R.}~\bibnamefont {Bloomer}}, \ and\ \bibinfo
  {author} {\bibfnamefont {O.}~\bibnamefont {Pfister}},\ }\href {\doibase
  10.1103/PhysRevLett.107.030505} {\bibfield  {journal} {\bibinfo  {journal}
  {Phys. Rev. Lett.}\ }\textbf {\bibinfo {volume} {107}},\ \bibinfo {pages}
  {030505} (\bibinfo {year} {2011})}\BibitemShut {NoStop}%
\bibitem [{\citenamefont {Du}\ \emph {et~al.}(2023{\natexlab{a}})\citenamefont
  {Du}, \citenamefont {Wang}, \citenamefont {Liu}, \citenamefont {Yang},\ and\
  \citenamefont {Zhang}}]{Du2023}%
  \BibitemOpen
  \bibfield  {author} {\bibinfo {author} {\bibfnamefont {P.}~\bibnamefont
  {Du}}, \bibinfo {author} {\bibfnamefont {Y.}~\bibnamefont {Wang}}, \bibinfo
  {author} {\bibfnamefont {K.}~\bibnamefont {Liu}}, \bibinfo {author}
  {\bibfnamefont {R.}~\bibnamefont {Yang}}, \ and\ \bibinfo {author}
  {\bibfnamefont {J.}~\bibnamefont {Zhang}},\ }\href {\doibase
  10.1364/OE.479420} {\bibfield  {journal} {\bibinfo  {journal} {Opt. Express}\
  }\textbf {\bibinfo {volume} {31}},\ \bibinfo {pages} {7535} (\bibinfo {year}
  {2023}{\natexlab{a}})}\BibitemShut {NoStop}%
\bibitem [{\citenamefont {Kues}\ \emph {et~al.}(2017)\citenamefont {Kues},
  \citenamefont {Reimer}, \citenamefont {Roztocki}, \citenamefont {Cortés},\
  and\ \citenamefont {Morandotti}}]{Kues2017}%
  \BibitemOpen
  \bibfield  {author} {\bibinfo {author} {\bibfnamefont {M.}~\bibnamefont
  {Kues}}, \bibinfo {author} {\bibfnamefont {C.}~\bibnamefont {Reimer}},
  \bibinfo {author} {\bibfnamefont {P.}~\bibnamefont {Roztocki}}, \bibinfo
  {author} {\bibfnamefont {L.~R.}\ \bibnamefont {Cortés}}, \ and\ \bibinfo
  {author} {\bibfnamefont {R.}~\bibnamefont {Morandotti}},\ }\href {\doibase
  10.1038/nature22986} {\bibfield  {journal} {\bibinfo  {journal} {Nature}\
  }\textbf {\bibinfo {volume} {546}},\ \bibinfo {pages} {622} (\bibinfo {year}
  {2017})}\BibitemShut {NoStop}%
\bibitem [{\citenamefont {Lukens}\ and\ \citenamefont
  {Lougovski}(2017)}]{Lukens17}%
  \BibitemOpen
  \bibfield  {author} {\bibinfo {author} {\bibfnamefont {J.~M.}\ \bibnamefont
  {Lukens}}\ and\ \bibinfo {author} {\bibfnamefont {P.}~\bibnamefont
  {Lougovski}},\ }\href {\doibase 10.1364/OPTICA.4.000008} {\bibfield
  {journal} {\bibinfo  {journal} {Optica}\ }\textbf {\bibinfo {volume} {4}},\
  \bibinfo {pages} {8} (\bibinfo {year} {2017})}\BibitemShut {NoStop}%
\bibitem [{\citenamefont {Qiu}\ \emph {et~al.}(2024)\citenamefont {Qiu},
  \citenamefont {Lu}, \citenamefont {He}, \citenamefont {Wu},\ and\
  \citenamefont {L\"u}}]{PhysRevB.110.L220301}%
  \BibitemOpen
  \bibfield  {author} {\bibinfo {author} {\bibfnamefont {Q.-Y.}\ \bibnamefont
  {Qiu}}, \bibinfo {author} {\bibfnamefont {Z.-G.}\ \bibnamefont {Lu}},
  \bibinfo {author} {\bibfnamefont {Q.}~\bibnamefont {He}}, \bibinfo {author}
  {\bibfnamefont {Y.}~\bibnamefont {Wu}}, \ and\ \bibinfo {author}
  {\bibfnamefont {X.-Y.}\ \bibnamefont {L\"u}},\ }\href {\doibase
  10.1103/PhysRevB.110.L220301} {\bibfield  {journal} {\bibinfo  {journal}
  {Phys. Rev. B}\ }\textbf {\bibinfo {volume} {110}},\ \bibinfo {pages}
  {L220301} (\bibinfo {year} {2024})}\BibitemShut {NoStop}%
\bibitem [{\citenamefont {Sweeney}\ \emph {et~al.}(2014)\citenamefont
  {Sweeney}, \citenamefont {Carter}, \citenamefont {Bracker}, \citenamefont
  {Kim}, \citenamefont {Kim}, \citenamefont {Yang},\ and\ \citenamefont
  {Gammon}}]{Sweeney2014}%
  \BibitemOpen
  \bibfield  {author} {\bibinfo {author} {\bibfnamefont {T.~M.}\ \bibnamefont
  {Sweeney}}, \bibinfo {author} {\bibfnamefont {S.~G.}\ \bibnamefont {Carter}},
  \bibinfo {author} {\bibfnamefont {A.~S.}\ \bibnamefont {Bracker}}, \bibinfo
  {author} {\bibfnamefont {M.}~\bibnamefont {Kim}}, \bibinfo {author}
  {\bibfnamefont {C.~S.}\ \bibnamefont {Kim}}, \bibinfo {author} {\bibfnamefont
  {L.}~\bibnamefont {Yang}}, \ and\ \bibinfo {author} {\bibfnamefont
  {D.}~\bibnamefont {Gammon}},\ }\href {\doibase 10.1038/nphoton.2014.84}
  {\bibfield  {journal} {\bibinfo  {journal} {Nature}\ }\textbf {\bibinfo
  {volume} {8}},\ \bibinfo {pages} {442–447} (\bibinfo {year}
  {2014})}\BibitemShut {NoStop}%
\bibitem [{\citenamefont {Fotso}\ \emph {et~al.}(2016)\citenamefont {Fotso},
  \citenamefont {Feiguin}, \citenamefont {Awschalom},\ and\ \citenamefont
  {Dobrovitski}}]{PhysRevLett.116.033603}%
  \BibitemOpen
  \bibfield  {author} {\bibinfo {author} {\bibfnamefont {H.~F.}\ \bibnamefont
  {Fotso}}, \bibinfo {author} {\bibfnamefont {A.~E.}\ \bibnamefont {Feiguin}},
  \bibinfo {author} {\bibfnamefont {D.~D.}\ \bibnamefont {Awschalom}}, \ and\
  \bibinfo {author} {\bibfnamefont {V.~V.}\ \bibnamefont {Dobrovitski}},\
  }\href {\doibase 10.1103/PhysRevLett.116.033603} {\bibfield  {journal}
  {\bibinfo  {journal} {Phys. Rev. Lett.}\ }\textbf {\bibinfo {volume} {116}},\
  \bibinfo {pages} {033603} (\bibinfo {year} {2016})}\BibitemShut {NoStop}%
\bibitem [{\citenamefont {Specht}\ \emph {et~al.}(2009)\citenamefont {Specht},
  \citenamefont {Bochmann}, \citenamefont {Mücke}, \citenamefont {Weber},
  \citenamefont {Figueroa}, \citenamefont {Moehring},\ and\ \citenamefont
  {Rempe}}]{Specht2009}%
  \BibitemOpen
  \bibfield  {author} {\bibinfo {author} {\bibfnamefont {H.~P.}\ \bibnamefont
  {Specht}}, \bibinfo {author} {\bibfnamefont {J.}~\bibnamefont {Bochmann}},
  \bibinfo {author} {\bibfnamefont {M.}~\bibnamefont {Mücke}}, \bibinfo
  {author} {\bibfnamefont {B.}~\bibnamefont {Weber}}, \bibinfo {author}
  {\bibfnamefont {E.}~\bibnamefont {Figueroa}}, \bibinfo {author}
  {\bibfnamefont {D.~L.}\ \bibnamefont {Moehring}}, \ and\ \bibinfo {author}
  {\bibfnamefont {G.}~\bibnamefont {Rempe}},\ }\href {\doibase
  10.1038/nphoton.2009.115} {\bibfield  {journal} {\bibinfo  {journal} {Nat.
  Photon.}\ }\textbf {\bibinfo {volume} {3}},\ \bibinfo {pages} {469} (\bibinfo
  {year} {2009})}\BibitemShut {NoStop}%
\bibitem [{\citenamefont {Lavoie}\ \emph {et~al.}(2013)\citenamefont {Lavoie},
  \citenamefont {Donohue}, \citenamefont {Wright}, \citenamefont {Fedrizzi},\
  and\ \citenamefont {Resch}}]{Lavoie2013}%
  \BibitemOpen
  \bibfield  {author} {\bibinfo {author} {\bibfnamefont {J.}~\bibnamefont
  {Lavoie}}, \bibinfo {author} {\bibfnamefont {J.~M.}\ \bibnamefont {Donohue}},
  \bibinfo {author} {\bibfnamefont {L.~G.}\ \bibnamefont {Wright}}, \bibinfo
  {author} {\bibfnamefont {A.}~\bibnamefont {Fedrizzi}}, \ and\ \bibinfo
  {author} {\bibfnamefont {K.~J.}\ \bibnamefont {Resch}},\ }\href {\doibase
  10.1038/nphoton.2013.47} {\bibfield  {journal} {\bibinfo  {journal} {Nat.
  Photon.}\ }\textbf {\bibinfo {volume} {7}},\ \bibinfo {pages} {363} (\bibinfo
  {year} {2013})}\BibitemShut {NoStop}%
\bibitem [{\citenamefont {Matsuda}(2016)}]{sciadv.1501223}%
  \BibitemOpen
  \bibfield  {author} {\bibinfo {author} {\bibfnamefont {N.}~\bibnamefont
  {Matsuda}},\ }\href {\doibase 10.1126/sciadv.1501223} {\bibfield  {journal}
  {\bibinfo  {journal} {Sci. Adv.}\ }\textbf {\bibinfo {volume} {2}},\ \bibinfo
  {pages} {e1501223} (\bibinfo {year} {2016})}\BibitemShut {NoStop}%
\bibitem [{\citenamefont {Lukin}\ \emph {et~al.}(2020)\citenamefont {Lukin},
  \citenamefont {White}, \citenamefont {Trivedi}, \citenamefont {Wrachtrup},
  \citenamefont {Figueroa}, \citenamefont {Kaiser},\ and\ \citenamefont
  {Vučković}}]{Lukin2020}%
  \BibitemOpen
  \bibfield  {author} {\bibinfo {author} {\bibfnamefont {D.~M.}\ \bibnamefont
  {Lukin}}, \bibinfo {author} {\bibfnamefont {A.~D.}\ \bibnamefont {White}},
  \bibinfo {author} {\bibfnamefont {R.}~\bibnamefont {Trivedi}}, \bibinfo
  {author} {\bibfnamefont {J.}~\bibnamefont {Wrachtrup}}, \bibinfo {author}
  {\bibfnamefont {E.}~\bibnamefont {Figueroa}}, \bibinfo {author}
  {\bibfnamefont {F.}~\bibnamefont {Kaiser}}, \ and\ \bibinfo {author}
  {\bibfnamefont {J.}~\bibnamefont {Vučković}},\ }\href {\doibase
  10.1038/s41534-020-00310-0} {\bibfield  {journal} {\bibinfo  {journal} {npj
  Quantum Inf.}\ }\textbf {\bibinfo {volume} {6}},\ \bibinfo {pages} {80}
  (\bibinfo {year} {2020})}\BibitemShut {NoStop}%
\bibitem [{\citenamefont {Gerke}\ \emph {et~al.}(2015)\citenamefont {Gerke},
  \citenamefont {Sperling}, \citenamefont {Vogel}, \citenamefont {Cai},
  \citenamefont {Roslund}, \citenamefont {Treps},\ and\ \citenamefont
  {Fabre}}]{PhysRevLett.114.050501}%
  \BibitemOpen
  \bibfield  {author} {\bibinfo {author} {\bibfnamefont {S.}~\bibnamefont
  {Gerke}}, \bibinfo {author} {\bibfnamefont {J.}~\bibnamefont {Sperling}},
  \bibinfo {author} {\bibfnamefont {W.}~\bibnamefont {Vogel}}, \bibinfo
  {author} {\bibfnamefont {Y.}~\bibnamefont {Cai}}, \bibinfo {author}
  {\bibfnamefont {J.}~\bibnamefont {Roslund}}, \bibinfo {author} {\bibfnamefont
  {N.}~\bibnamefont {Treps}}, \ and\ \bibinfo {author} {\bibfnamefont
  {C.}~\bibnamefont {Fabre}},\ }\href {\doibase 10.1103/PhysRevLett.114.050501}
  {\bibfield  {journal} {\bibinfo  {journal} {Phys. Rev. Lett.}\ }\textbf
  {\bibinfo {volume} {114}},\ \bibinfo {pages} {050501} (\bibinfo {year}
  {2015})}\BibitemShut {NoStop}%
\bibitem [{\citenamefont {Jolin}\ \emph {et~al.}(2023)\citenamefont {Jolin},
  \citenamefont {Andersson}, \citenamefont {Hern\'andez}, \citenamefont
  {Strandberg}, \citenamefont {Quijandr\'{\i}a}, \citenamefont {Aumentado},
  \citenamefont {Borgani}, \citenamefont {Thol\'en},\ and\ \citenamefont
  {Haviland}}]{PhysRevLett.130.120601}%
  \BibitemOpen
  \bibfield  {author} {\bibinfo {author} {\bibfnamefont {S.~W.}\ \bibnamefont
  {Jolin}}, \bibinfo {author} {\bibfnamefont {G.}~\bibnamefont {Andersson}},
  \bibinfo {author} {\bibfnamefont {J.~C.~R.}\ \bibnamefont {Hern\'andez}},
  \bibinfo {author} {\bibfnamefont {I.}~\bibnamefont {Strandberg}}, \bibinfo
  {author} {\bibfnamefont {F.}~\bibnamefont {Quijandr\'{\i}a}}, \bibinfo
  {author} {\bibfnamefont {J.}~\bibnamefont {Aumentado}}, \bibinfo {author}
  {\bibfnamefont {R.}~\bibnamefont {Borgani}}, \bibinfo {author} {\bibfnamefont
  {M.~O.}\ \bibnamefont {Thol\'en}}, \ and\ \bibinfo {author} {\bibfnamefont
  {D.~B.}\ \bibnamefont {Haviland}},\ }\href {\doibase
  10.1103/PhysRevLett.130.120601} {\bibfield  {journal} {\bibinfo  {journal}
  {Phys. Rev. Lett.}\ }\textbf {\bibinfo {volume} {130}},\ \bibinfo {pages}
  {120601} (\bibinfo {year} {2023})}\BibitemShut {NoStop}%
\bibitem [{\citenamefont {Rivera~Hernández}\ \emph {et~al.}(2024)\citenamefont
  {Rivera~Hernández}, \citenamefont {Lingua}, \citenamefont {Jolin},\ and\
  \citenamefont {Haviland}}]{APL0203426}%
  \BibitemOpen
  \bibfield  {author} {\bibinfo {author} {\bibfnamefont {J.~C.}\ \bibnamefont
  {Rivera~Hernández}}, \bibinfo {author} {\bibfnamefont {F.}~\bibnamefont
  {Lingua}}, \bibinfo {author} {\bibfnamefont {S.~W.}\ \bibnamefont {Jolin}}, \
  and\ \bibinfo {author} {\bibfnamefont {D.~B.}\ \bibnamefont {Haviland}},\
  }\href {\doibase 10.1063/5.0203426} {\bibfield  {journal} {\bibinfo
  {journal} {APL Quantum}\ }\textbf {\bibinfo {volume} {1}},\ \bibinfo {pages}
  {036101} (\bibinfo {year} {2024})}\BibitemShut {NoStop}%
\bibitem [{\citenamefont {del Valle}\ \emph {et~al.}(2012)\citenamefont {del
  Valle}, \citenamefont {Gonzalez-Tudela}, \citenamefont {Laussy},
  \citenamefont {Tejedor},\ and\ \citenamefont
  {Hartmann}}]{PhysRevLett.109.183601}%
  \BibitemOpen
  \bibfield  {author} {\bibinfo {author} {\bibfnamefont {E.}~\bibnamefont {del
  Valle}}, \bibinfo {author} {\bibfnamefont {A.}~\bibnamefont
  {Gonzalez-Tudela}}, \bibinfo {author} {\bibfnamefont {F.~P.}\ \bibnamefont
  {Laussy}}, \bibinfo {author} {\bibfnamefont {C.}~\bibnamefont {Tejedor}}, \
  and\ \bibinfo {author} {\bibfnamefont {M.~J.}\ \bibnamefont {Hartmann}},\
  }\href {\doibase 10.1103/PhysRevLett.109.183601} {\bibfield  {journal}
  {\bibinfo  {journal} {Phys. Rev. Lett.}\ }\textbf {\bibinfo {volume} {109}},\
  \bibinfo {pages} {183601} (\bibinfo {year} {2012})}\BibitemShut {NoStop}%
\bibitem [{\citenamefont {Shatokhin}\ and\ \citenamefont
  {Kilin}(2016)}]{PhysRevA.94.033835}%
  \BibitemOpen
  \bibfield  {author} {\bibinfo {author} {\bibfnamefont {V.~N.}\ \bibnamefont
  {Shatokhin}}\ and\ \bibinfo {author} {\bibfnamefont {S.~Y.}\ \bibnamefont
  {Kilin}},\ }\href {\doibase 10.1103/PhysRevA.94.033835} {\bibfield  {journal}
  {\bibinfo  {journal} {Phys. Rev. A}\ }\textbf {\bibinfo {volume} {94}},\
  \bibinfo {pages} {033835} (\bibinfo {year} {2016})}\BibitemShut {NoStop}%
\bibitem [{\citenamefont {Schmidt}\ \emph {et~al.}(2021)\citenamefont
  {Schmidt}, \citenamefont {Esteban}, \citenamefont {Giedke}, \citenamefont
  {Aizpurua},\ and\ \citenamefont {González-Tudela}}]{Schmidt_2021}%
  \BibitemOpen
  \bibfield  {author} {\bibinfo {author} {\bibfnamefont {M.~K.}\ \bibnamefont
  {Schmidt}}, \bibinfo {author} {\bibfnamefont {R.}~\bibnamefont {Esteban}},
  \bibinfo {author} {\bibfnamefont {G.}~\bibnamefont {Giedke}}, \bibinfo
  {author} {\bibfnamefont {J.}~\bibnamefont {Aizpurua}}, \ and\ \bibinfo
  {author} {\bibfnamefont {A.}~\bibnamefont {González-Tudela}},\ }\href
  {\doibase 10.1088/2058-9565/abe569} {\bibfield  {journal} {\bibinfo
  {journal} {Quantum Sci. Technol.}\ }\textbf {\bibinfo {volume} {6}},\
  \bibinfo {pages} {034005} (\bibinfo {year} {2021})}\BibitemShut {NoStop}%
\bibitem [{\citenamefont {Ilin}\ \emph {et~al.}(2023)\citenamefont {Ilin},
  \citenamefont {Poshakinskiy}, \citenamefont {Poddubny},\ and\ \citenamefont
  {Iorsh}}]{PhysRevLett.130.023601}%
  \BibitemOpen
  \bibfield  {author} {\bibinfo {author} {\bibfnamefont {D.}~\bibnamefont
  {Ilin}}, \bibinfo {author} {\bibfnamefont {A.~V.}\ \bibnamefont
  {Poshakinskiy}}, \bibinfo {author} {\bibfnamefont {A.~N.}\ \bibnamefont
  {Poddubny}}, \ and\ \bibinfo {author} {\bibfnamefont {I.}~\bibnamefont
  {Iorsh}},\ }\href {\doibase 10.1103/PhysRevLett.130.023601} {\bibfield
  {journal} {\bibinfo  {journal} {Phys. Rev. Lett.}\ }\textbf {\bibinfo
  {volume} {130}},\ \bibinfo {pages} {023601} (\bibinfo {year}
  {2023})}\BibitemShut {NoStop}%
\bibitem [{SM()}]{SM}%
  \BibitemOpen
  \href@noop {} {}\bibinfo {note} {See Supplemental Material for additional
  details about the Floquet optimization algorithm, derivation of the general
  master equation, the Green function approach for determining time-dependent
  photon-photon correlations, as well as the detailed illustration of MPS
  simulations and waveguide spatial discretization approach.}\BibitemShut
  {Stop}%
\bibitem [{\citenamefont {Zhang}\ and\ \citenamefont
  {Gong}(2019)}]{PhysRevB.100.235452}%
  \BibitemOpen
  \bibfield  {author} {\bibinfo {author} {\bibfnamefont {S.}~\bibnamefont
  {Zhang}}\ and\ \bibinfo {author} {\bibfnamefont {J.}~\bibnamefont {Gong}},\
  }\href {\doibase 10.1103/PhysRevB.100.235452} {\bibfield  {journal} {\bibinfo
   {journal} {Phys. Rev. B}\ }\textbf {\bibinfo {volume} {100}},\ \bibinfo
  {pages} {235452} (\bibinfo {year} {2019})}\BibitemShut {NoStop}%
\bibitem [{\citenamefont {Cao}\ \emph {et~al.}(2019)\citenamefont {Cao},
  \citenamefont {Zhang}, \citenamefont {Li}, \citenamefont {Zhou},
  \citenamefont {Zhang},\ and\ \citenamefont {Chaovalitwongse}}]{IEEE1256}%
  \BibitemOpen
  \bibfield  {author} {\bibinfo {author} {\bibfnamefont {Y.}~\bibnamefont
  {Cao}}, \bibinfo {author} {\bibfnamefont {H.}~\bibnamefont {Zhang}}, \bibinfo
  {author} {\bibfnamefont {W.}~\bibnamefont {Li}}, \bibinfo {author}
  {\bibfnamefont {M.}~\bibnamefont {Zhou}}, \bibinfo {author} {\bibfnamefont
  {Y.}~\bibnamefont {Zhang}}, \ and\ \bibinfo {author} {\bibfnamefont {W.~A.}\
  \bibnamefont {Chaovalitwongse}},\ }\href {\doibase 10.1109/TEVC.2018.2885075}
  {\bibfield  {journal} {\bibinfo  {journal} {IEEE Trans. on Evol. Comput.}\
  }\textbf {\bibinfo {volume} {23}},\ \bibinfo {pages} {718} (\bibinfo {year}
  {2019})}\BibitemShut {NoStop}%
\bibitem [{\citenamefont {Hu}\ \emph {et~al.}(2017)\citenamefont {Hu},
  \citenamefont {Li}, \citenamefont {Liu}, \citenamefont {Li},\ and\
  \citenamefont {Xu}}]{PhysRevLett.119.173201}%
  \BibitemOpen
  \bibfield  {author} {\bibinfo {author} {\bibfnamefont {H.}~\bibnamefont
  {Hu}}, \bibinfo {author} {\bibfnamefont {N.}~\bibnamefont {Li}}, \bibinfo
  {author} {\bibfnamefont {P.}~\bibnamefont {Liu}}, \bibinfo {author}
  {\bibfnamefont {R.}~\bibnamefont {Li}}, \ and\ \bibinfo {author}
  {\bibfnamefont {Z.}~\bibnamefont {Xu}},\ }\href {\doibase
  10.1103/PhysRevLett.119.173201} {\bibfield  {journal} {\bibinfo  {journal}
  {Phys. Rev. Lett.}\ }\textbf {\bibinfo {volume} {119}},\ \bibinfo {pages}
  {173201} (\bibinfo {year} {2017})}\BibitemShut {NoStop}%
\bibitem [{\citenamefont {Yin}\ \emph {et~al.}(2021)\citenamefont {Yin},
  \citenamefont {Liu},\ and\ \citenamefont {Zeng}}]{Appl124}%
  \BibitemOpen
  \bibfield  {author} {\bibinfo {author} {\bibfnamefont {H.}~\bibnamefont
  {Yin}}, \bibinfo {author} {\bibfnamefont {D.}~\bibnamefont {Liu}}, \ and\
  \bibinfo {author} {\bibfnamefont {F.}~\bibnamefont {Zeng}},\ }\href {\doibase
  10.1063/5.0063124} {\bibfield  {journal} {\bibinfo  {journal} {Appl. Phys.
  Lett.}\ }\textbf {\bibinfo {volume} {119}},\ \bibinfo {pages} {151105}
  (\bibinfo {year} {2021})}\BibitemShut {NoStop}%
\bibitem [{\citenamefont {Hoyt}\ \emph {et~al.}(2000)\citenamefont {Hoyt},
  \citenamefont {Sheik-Bahae}, \citenamefont {Epstein}, \citenamefont
  {Edwards},\ and\ \citenamefont {Anderson}}]{PhysRevLett.85.3600}%
  \BibitemOpen
  \bibfield  {author} {\bibinfo {author} {\bibfnamefont {C.~W.}\ \bibnamefont
  {Hoyt}}, \bibinfo {author} {\bibfnamefont {M.}~\bibnamefont {Sheik-Bahae}},
  \bibinfo {author} {\bibfnamefont {R.~I.}\ \bibnamefont {Epstein}}, \bibinfo
  {author} {\bibfnamefont {B.~C.}\ \bibnamefont {Edwards}}, \ and\ \bibinfo
  {author} {\bibfnamefont {J.~E.}\ \bibnamefont {Anderson}},\ }\href {\doibase
  10.1103/PhysRevLett.85.3600} {\bibfield  {journal} {\bibinfo  {journal}
  {Phys. Rev. Lett.}\ }\textbf {\bibinfo {volume} {85}},\ \bibinfo {pages}
  {3600} (\bibinfo {year} {2000})}\BibitemShut {NoStop}%
\bibitem [{\citenamefont {Delić}\ \emph {et~al.}(2020)\citenamefont {Delić},
  \citenamefont {Reisenbauer}, \citenamefont {Dare}, \citenamefont {Grass},
  \citenamefont {Vuletić}, \citenamefont {Kiesel},\ and\ \citenamefont
  {Aspelmeyer}}]{science3993}%
  \BibitemOpen
  \bibfield  {author} {\bibinfo {author} {\bibfnamefont {U.}~\bibnamefont
  {Delić}}, \bibinfo {author} {\bibfnamefont {M.}~\bibnamefont {Reisenbauer}},
  \bibinfo {author} {\bibfnamefont {K.}~\bibnamefont {Dare}}, \bibinfo {author}
  {\bibfnamefont {D.}~\bibnamefont {Grass}}, \bibinfo {author} {\bibfnamefont
  {V.}~\bibnamefont {Vuletić}}, \bibinfo {author} {\bibfnamefont
  {N.}~\bibnamefont {Kiesel}}, \ and\ \bibinfo {author} {\bibfnamefont
  {M.}~\bibnamefont {Aspelmeyer}},\ }\href {\doibase 10.1126/science.aba3993}
  {\bibfield  {journal} {\bibinfo  {journal} {Science}\ }\textbf {\bibinfo
  {volume} {367}},\ \bibinfo {pages} {892} (\bibinfo {year}
  {2020})}\BibitemShut {NoStop}%
\bibitem [{\citenamefont {Trivedi}\ \emph {et~al.}(2018)\citenamefont
  {Trivedi}, \citenamefont {Fischer}, \citenamefont {Xu}, \citenamefont {Fan},\
  and\ \citenamefont {Vuckovic}}]{PhysRevB.98.144112}%
  \BibitemOpen
  \bibfield  {author} {\bibinfo {author} {\bibfnamefont {R.}~\bibnamefont
  {Trivedi}}, \bibinfo {author} {\bibfnamefont {K.}~\bibnamefont {Fischer}},
  \bibinfo {author} {\bibfnamefont {S.}~\bibnamefont {Xu}}, \bibinfo {author}
  {\bibfnamefont {S.}~\bibnamefont {Fan}}, \ and\ \bibinfo {author}
  {\bibfnamefont {J.}~\bibnamefont {Vuckovic}},\ }\href {\doibase
  10.1103/PhysRevB.98.144112} {\bibfield  {journal} {\bibinfo  {journal} {Phys.
  Rev. B}\ }\textbf {\bibinfo {volume} {98}},\ \bibinfo {pages} {144112}
  (\bibinfo {year} {2018})}\BibitemShut {NoStop}%
\bibitem [{\citenamefont {Trivedi}\ \emph {et~al.}(2020)\citenamefont
  {Trivedi}, \citenamefont {White}, \citenamefont {Fan},\ and\ \citenamefont
  {Vu\ifmmode \check{c}\else \v{c}\fi{}kovi\ifmmode~\acute{c}\else
  \'{c}\fi{}}}]{PhysRevA.102.033707}%
  \BibitemOpen
  \bibfield  {author} {\bibinfo {author} {\bibfnamefont {R.}~\bibnamefont
  {Trivedi}}, \bibinfo {author} {\bibfnamefont {A.}~\bibnamefont {White}},
  \bibinfo {author} {\bibfnamefont {S.}~\bibnamefont {Fan}}, \ and\ \bibinfo
  {author} {\bibfnamefont {J.}~\bibnamefont {Vu\ifmmode \check{c}\else
  \v{c}\fi{}kovi\ifmmode~\acute{c}\else \'{c}\fi{}}},\ }\href {\doibase
  10.1103/PhysRevA.102.033707} {\bibfield  {journal} {\bibinfo  {journal}
  {Phys. Rev. A}\ }\textbf {\bibinfo {volume} {102}},\ \bibinfo {pages}
  {033707} (\bibinfo {year} {2020})}\BibitemShut {NoStop}%
\bibitem [{\citenamefont {Cui}\ \emph {et~al.}(2020)\citenamefont {Cui},
  \citenamefont {Seshadreesan}, \citenamefont {Guha},\ and\ \citenamefont
  {Fan}}]{PhysRevLett.124.190502}%
  \BibitemOpen
  \bibfield  {author} {\bibinfo {author} {\bibfnamefont {C.}~\bibnamefont
  {Cui}}, \bibinfo {author} {\bibfnamefont {K.~P.}\ \bibnamefont
  {Seshadreesan}}, \bibinfo {author} {\bibfnamefont {S.}~\bibnamefont {Guha}},
  \ and\ \bibinfo {author} {\bibfnamefont {L.}~\bibnamefont {Fan}},\ }\href
  {\doibase 10.1103/PhysRevLett.124.190502} {\bibfield  {journal} {\bibinfo
  {journal} {Phys. Rev. Lett.}\ }\textbf {\bibinfo {volume} {124}},\ \bibinfo
  {pages} {190502} (\bibinfo {year} {2020})}\BibitemShut {NoStop}%
\bibitem [{\citenamefont {Zhang}\ \emph {et~al.}(2023)\citenamefont {Zhang},
  \citenamefont {Cui}, \citenamefont {Yan}, \citenamefont {Guo}, \citenamefont
  {Wang},\ and\ \citenamefont {Fan}}]{Zhang2023}%
  \BibitemOpen
  \bibfield  {author} {\bibinfo {author} {\bibfnamefont {L.}~\bibnamefont
  {Zhang}}, \bibinfo {author} {\bibfnamefont {C.}~\bibnamefont {Cui}}, \bibinfo
  {author} {\bibfnamefont {J.}~\bibnamefont {Yan}}, \bibinfo {author}
  {\bibfnamefont {Y.}~\bibnamefont {Guo}}, \bibinfo {author} {\bibfnamefont
  {J.}~\bibnamefont {Wang}}, \ and\ \bibinfo {author} {\bibfnamefont
  {L.}~\bibnamefont {Fan}},\ }\href {\doibase 10.1038/s41534-023-00725-5}
  {\bibfield  {journal} {\bibinfo  {journal} {npj Quantum Inf.}\ }\textbf
  {\bibinfo {volume} {9}},\ \bibinfo {pages} {57} (\bibinfo {year}
  {2023})}\BibitemShut {NoStop}%
\bibitem [{\citenamefont {Pupeza}\ \emph {et~al.}(2021)\citenamefont {Pupeza},
  \citenamefont {Zhang}, \citenamefont {Högner},\ and\ \citenamefont
  {Ye}}]{Pupeza2021}%
  \BibitemOpen
  \bibfield  {author} {\bibinfo {author} {\bibfnamefont {I.}~\bibnamefont
  {Pupeza}}, \bibinfo {author} {\bibfnamefont {C.}~\bibnamefont {Zhang}},
  \bibinfo {author} {\bibfnamefont {M.}~\bibnamefont {Högner}}, \ and\
  \bibinfo {author} {\bibfnamefont {J.}~\bibnamefont {Ye}},\ }\href {\doibase
  10.1038/s41566-020-00741-3} {\bibfield  {journal} {\bibinfo  {journal} {Nat.
  Photon.}\ }\textbf {\bibinfo {volume} {15}},\ \bibinfo {pages} {175}
  (\bibinfo {year} {2021})}\BibitemShut {NoStop}%
\bibitem [{\citenamefont {Picqué}\ and\ \citenamefont
  {Hänsch}(2019)}]{Picque2019}%
  \BibitemOpen
  \bibfield  {author} {\bibinfo {author} {\bibfnamefont {N.}~\bibnamefont
  {Picqué}}\ and\ \bibinfo {author} {\bibfnamefont {T.~W.}\ \bibnamefont
  {Hänsch}},\ }\href {\doibase 10.1038/s41566-018-0347-5} {\bibfield
  {journal} {\bibinfo  {journal} {Nat. Photon.}\ }\textbf {\bibinfo {volume}
  {13}},\ \bibinfo {pages} {146} (\bibinfo {year} {2019})}\BibitemShut
  {NoStop}%
\bibitem [{\citenamefont {Zhang}\ \emph {et~al.}(2021)\citenamefont {Zhang},
  \citenamefont {i~Carceller}, \citenamefont {Kjaergaard},\ and\ \citenamefont
  {S\o{}rensen}}]{PhysRevLett.127.233601}%
  \BibitemOpen
  \bibfield  {author} {\bibinfo {author} {\bibfnamefont {Y.-X.}\ \bibnamefont
  {Zhang}}, \bibinfo {author} {\bibfnamefont {C.~R.}\ \bibnamefont
  {i~Carceller}}, \bibinfo {author} {\bibfnamefont {M.}~\bibnamefont
  {Kjaergaard}}, \ and\ \bibinfo {author} {\bibfnamefont {A.~S.}\ \bibnamefont
  {S\o{}rensen}},\ }\href {\doibase 10.1103/PhysRevLett.127.233601} {\bibfield
  {journal} {\bibinfo  {journal} {Phys. Rev. Lett.}\ }\textbf {\bibinfo
  {volume} {127}},\ \bibinfo {pages} {233601} (\bibinfo {year}
  {2021})}\BibitemShut {NoStop}%
\bibitem [{\citenamefont {Joshi}\ \emph {et~al.}(2023)\citenamefont {Joshi},
  \citenamefont {Yang},\ and\ \citenamefont
  {Mirhosseini}}]{PhysRevX.13.021039}%
  \BibitemOpen
  \bibfield  {author} {\bibinfo {author} {\bibfnamefont {C.}~\bibnamefont
  {Joshi}}, \bibinfo {author} {\bibfnamefont {F.}~\bibnamefont {Yang}}, \ and\
  \bibinfo {author} {\bibfnamefont {M.}~\bibnamefont {Mirhosseini}},\ }\href
  {\doibase 10.1103/PhysRevX.13.021039} {\bibfield  {journal} {\bibinfo
  {journal} {Phys. Rev. X}\ }\textbf {\bibinfo {volume} {13}},\ \bibinfo
  {pages} {021039} (\bibinfo {year} {2023})}\BibitemShut {NoStop}%
\bibitem [{\citenamefont {Du}\ \emph {et~al.}(2023{\natexlab{b}})\citenamefont
  {Du}, \citenamefont {Zhang},\ and\ \citenamefont {Li}}]{DuL2023}%
  \BibitemOpen
  \bibfield  {author} {\bibinfo {author} {\bibfnamefont {L.}~\bibnamefont
  {Du}}, \bibinfo {author} {\bibfnamefont {Y.}~\bibnamefont {Zhang}}, \ and\
  \bibinfo {author} {\bibfnamefont {Y.}~\bibnamefont {Li}},\ }\href {\doibase
  10.1007/s11467-022-1215-9} {\bibfield  {journal} {\bibinfo  {journal} {Front.
  of Phys.}\ }\textbf {\bibinfo {volume} {18}},\ \bibinfo {pages} {12301}
  (\bibinfo {year} {2023}{\natexlab{b}})}\BibitemShut {NoStop}%
\bibitem [{\citenamefont {Roccati}\ and\ \citenamefont
  {Cilluffo}(2024)}]{PhysRevLett.133.063603}%
  \BibitemOpen
  \bibfield  {author} {\bibinfo {author} {\bibfnamefont {F.}~\bibnamefont
  {Roccati}}\ and\ \bibinfo {author} {\bibfnamefont {D.}~\bibnamefont
  {Cilluffo}},\ }\href {\doibase 10.1103/PhysRevLett.133.063603} {\bibfield
  {journal} {\bibinfo  {journal} {Phys. Rev. Lett.}\ }\textbf {\bibinfo
  {volume} {133}},\ \bibinfo {pages} {063603} (\bibinfo {year}
  {2024})}\BibitemShut {NoStop}%
\bibitem [{\citenamefont {Kockum}\ \emph {et~al.}(2018)\citenamefont {Kockum},
  \citenamefont {Johansson},\ and\ \citenamefont
  {Nori}}]{PhysRevLett.120.140404}%
  \BibitemOpen
  \bibfield  {author} {\bibinfo {author} {\bibfnamefont {A.~F.}\ \bibnamefont
  {Kockum}}, \bibinfo {author} {\bibfnamefont {G.}~\bibnamefont {Johansson}}, \
  and\ \bibinfo {author} {\bibfnamefont {F.}~\bibnamefont {Nori}},\ }\href
  {\doibase 10.1103/PhysRevLett.120.140404} {\bibfield  {journal} {\bibinfo
  {journal} {Phys. Rev. Lett.}\ }\textbf {\bibinfo {volume} {120}},\ \bibinfo
  {pages} {140404} (\bibinfo {year} {2018})}\BibitemShut {NoStop}%
\bibitem [{\citenamefont {Qiu}\ \emph {et~al.}(2023)\citenamefont {Qiu},
  \citenamefont {Wu},\ and\ \citenamefont {Lü}}]{Qiu2023}%
  \BibitemOpen
  \bibfield  {author} {\bibinfo {author} {\bibfnamefont {Q.-Y.}\ \bibnamefont
  {Qiu}}, \bibinfo {author} {\bibfnamefont {Y.}~\bibnamefont {Wu}}, \ and\
  \bibinfo {author} {\bibfnamefont {X.-Y.}\ \bibnamefont {Lü}},\ }\href
  {\doibase 10.1007/s11433-022-1990-x} {\bibfield  {journal} {\bibinfo
  {journal} {Sci. China Phys. Mech. Astron.}\ }\textbf {\bibinfo {volume}
  {66}},\ \bibinfo {pages} {224212} (\bibinfo {year} {2023})}\BibitemShut
  {NoStop}%
\bibitem [{\citenamefont {Qiu}\ and\ \citenamefont
  {L\"u}(2024)}]{PhysRevResearch.6.033243}%
  \BibitemOpen
  \bibfield  {author} {\bibinfo {author} {\bibfnamefont {Q.-Y.}\ \bibnamefont
  {Qiu}}\ and\ \bibinfo {author} {\bibfnamefont {X.-Y.}\ \bibnamefont {L\"u}},\
  }\href {\doibase 10.1103/PhysRevResearch.6.033243} {\bibfield  {journal}
  {\bibinfo  {journal} {Phys. Rev. Res.}\ }\textbf {\bibinfo {volume} {6}},\
  \bibinfo {pages} {033243} (\bibinfo {year} {2024})}\BibitemShut {NoStop}%
\bibitem [{\citenamefont {Horodecki}\ \emph {et~al.}(2009)\citenamefont
  {Horodecki}, \citenamefont {Horodecki}, \citenamefont {Horodecki},\ and\
  \citenamefont {Horodecki}}]{RevModPhys.81.865}%
  \BibitemOpen
  \bibfield  {author} {\bibinfo {author} {\bibfnamefont {R.}~\bibnamefont
  {Horodecki}}, \bibinfo {author} {\bibfnamefont {P.}~\bibnamefont
  {Horodecki}}, \bibinfo {author} {\bibfnamefont {M.}~\bibnamefont
  {Horodecki}}, \ and\ \bibinfo {author} {\bibfnamefont {K.}~\bibnamefont
  {Horodecki}},\ }\href {\doibase 10.1103/RevModPhys.81.865} {\bibfield
  {journal} {\bibinfo  {journal} {Rev. Mod. Phys.}\ }\textbf {\bibinfo {volume}
  {81}},\ \bibinfo {pages} {865} (\bibinfo {year} {2009})}\BibitemShut
  {NoStop}%
\bibitem [{\citenamefont {Nishioka}(2018)}]{RevModPhys.90.035007}%
  \BibitemOpen
  \bibfield  {author} {\bibinfo {author} {\bibfnamefont {T.}~\bibnamefont
  {Nishioka}},\ }\href {\doibase 10.1103/RevModPhys.90.035007} {\bibfield
  {journal} {\bibinfo  {journal} {Rev. Mod. Phys.}\ }\textbf {\bibinfo {volume}
  {90}},\ \bibinfo {pages} {035007} (\bibinfo {year} {2018})}\BibitemShut
  {NoStop}%
\bibitem [{\citenamefont {Poshakinskiy}\ and\ \citenamefont
  {Poddubny}(2021)}]{PhysRevLett.127.173601}%
  \BibitemOpen
  \bibfield  {author} {\bibinfo {author} {\bibfnamefont {A.~V.}\ \bibnamefont
  {Poshakinskiy}}\ and\ \bibinfo {author} {\bibfnamefont {A.~N.}\ \bibnamefont
  {Poddubny}},\ }\href {\doibase 10.1103/PhysRevLett.127.173601} {\bibfield
  {journal} {\bibinfo  {journal} {Phys. Rev. Lett.}\ }\textbf {\bibinfo
  {volume} {127}},\ \bibinfo {pages} {173601} (\bibinfo {year}
  {2021})}\BibitemShut {NoStop}%
\bibitem [{\citenamefont {Poshakinskiy}\ \emph {et~al.}(2021)\citenamefont
  {Poshakinskiy}, \citenamefont {Zhong}, \citenamefont {Ke}, \citenamefont
  {Ke}, \citenamefont {Olekhno}, \citenamefont {Olekhno}, \citenamefont {Lee},
  \citenamefont {Kivshar},\ and\ \citenamefont {Poddubny}}]{Poshakinskiy2021}%
  \BibitemOpen
  \bibfield  {author} {\bibinfo {author} {\bibfnamefont {A.~V.}\ \bibnamefont
  {Poshakinskiy}}, \bibinfo {author} {\bibfnamefont {J.}~\bibnamefont {Zhong}},
  \bibinfo {author} {\bibfnamefont {Y.}~\bibnamefont {Ke}}, \bibinfo {author}
  {\bibfnamefont {Y.}~\bibnamefont {Ke}}, \bibinfo {author} {\bibfnamefont
  {N.~A.}\ \bibnamefont {Olekhno}}, \bibinfo {author} {\bibfnamefont {N.~A.}\
  \bibnamefont {Olekhno}}, \bibinfo {author} {\bibfnamefont {C.}~\bibnamefont
  {Lee}}, \bibinfo {author} {\bibfnamefont {Y.~S.}\ \bibnamefont {Kivshar}}, \
  and\ \bibinfo {author} {\bibfnamefont {A.~N.}\ \bibnamefont {Poddubny}},\
  }\href {\doibase 10.1038/s41534-021-00372-8} {\bibfield  {journal} {\bibinfo
  {journal} {npj Quantum Inf.}\ }\textbf {\bibinfo {volume} {7}},\ \bibinfo
  {pages} {34} (\bibinfo {year} {2021})}\BibitemShut {NoStop}%
\bibitem [{\citenamefont {Lu}\ \emph {et~al.}(2023)\citenamefont {Lu},
  \citenamefont {Shang}, \citenamefont {Wu},\ and\ \citenamefont
  {L\"u}}]{PhysRevA.108.053703}%
  \BibitemOpen
  \bibfield  {author} {\bibinfo {author} {\bibfnamefont {Z.-G.}\ \bibnamefont
  {Lu}}, \bibinfo {author} {\bibfnamefont {C.}~\bibnamefont {Shang}}, \bibinfo
  {author} {\bibfnamefont {Y.}~\bibnamefont {Wu}}, \ and\ \bibinfo {author}
  {\bibfnamefont {X.-Y.}\ \bibnamefont {L\"u}},\ }\href {\doibase
  10.1103/PhysRevA.108.053703} {\bibfield  {journal} {\bibinfo  {journal}
  {Phys. Rev. A}\ }\textbf {\bibinfo {volume} {108}},\ \bibinfo {pages}
  {053703} (\bibinfo {year} {2023})}\BibitemShut {NoStop}%
\bibitem [{\citenamefont {Poshakinskiy}\ and\ \citenamefont
  {Poddubny}(2016)}]{PhysRevA.93.033856}%
  \BibitemOpen
  \bibfield  {author} {\bibinfo {author} {\bibfnamefont {A.~V.}\ \bibnamefont
  {Poshakinskiy}}\ and\ \bibinfo {author} {\bibfnamefont {A.~N.}\ \bibnamefont
  {Poddubny}},\ }\href {\doibase 10.1103/PhysRevA.93.033856} {\bibfield
  {journal} {\bibinfo  {journal} {Phys. Rev. A}\ }\textbf {\bibinfo {volume}
  {93}},\ \bibinfo {pages} {033856} (\bibinfo {year} {2016})}\BibitemShut
  {NoStop}%
\bibitem [{\citenamefont {Ke}\ \emph {et~al.}(2019)\citenamefont {Ke},
  \citenamefont {Poshakinskiy}, \citenamefont {Lee}, \citenamefont {Kivshar},\
  and\ \citenamefont {Poddubny}}]{PhysRevLett.123.253601}%
  \BibitemOpen
  \bibfield  {author} {\bibinfo {author} {\bibfnamefont {Y.}~\bibnamefont
  {Ke}}, \bibinfo {author} {\bibfnamefont {A.~V.}\ \bibnamefont
  {Poshakinskiy}}, \bibinfo {author} {\bibfnamefont {C.}~\bibnamefont {Lee}},
  \bibinfo {author} {\bibfnamefont {Y.~S.}\ \bibnamefont {Kivshar}}, \ and\
  \bibinfo {author} {\bibfnamefont {A.~N.}\ \bibnamefont {Poddubny}},\ }\href
  {\doibase 10.1103/PhysRevLett.123.253601} {\bibfield  {journal} {\bibinfo
  {journal} {Phys. Rev. Lett.}\ }\textbf {\bibinfo {volume} {123}},\ \bibinfo
  {pages} {253601} (\bibinfo {year} {2019})}\BibitemShut {NoStop}%
\bibitem [{\citenamefont {Vyatkin}\ \emph {et~al.}(2023)\citenamefont
  {Vyatkin}, \citenamefont {Poshakinskiy},\ and\ \citenamefont
  {Poddubny}}]{PhysRevA.108.023715}%
  \BibitemOpen
  \bibfield  {author} {\bibinfo {author} {\bibfnamefont {E.~S.}\ \bibnamefont
  {Vyatkin}}, \bibinfo {author} {\bibfnamefont {A.~V.}\ \bibnamefont
  {Poshakinskiy}}, \ and\ \bibinfo {author} {\bibfnamefont {A.~N.}\
  \bibnamefont {Poddubny}},\ }\href {\doibase 10.1103/PhysRevA.108.023715}
  {\bibfield  {journal} {\bibinfo  {journal} {Phys. Rev. A}\ }\textbf {\bibinfo
  {volume} {108}},\ \bibinfo {pages} {023715} (\bibinfo {year}
  {2023})}\BibitemShut {NoStop}%
\bibitem [{\citenamefont {Sheremet}\ \emph {et~al.}(2023)\citenamefont
  {Sheremet}, \citenamefont {Petrov}, \citenamefont {Iorsh}, \citenamefont
  {Poshakinskiy},\ and\ \citenamefont {Poddubny}}]{RevModPhys.95.015002}%
  \BibitemOpen
  \bibfield  {author} {\bibinfo {author} {\bibfnamefont {A.~S.}\ \bibnamefont
  {Sheremet}}, \bibinfo {author} {\bibfnamefont {M.~I.}\ \bibnamefont
  {Petrov}}, \bibinfo {author} {\bibfnamefont {I.~V.}\ \bibnamefont {Iorsh}},
  \bibinfo {author} {\bibfnamefont {A.~V.}\ \bibnamefont {Poshakinskiy}}, \
  and\ \bibinfo {author} {\bibfnamefont {A.~N.}\ \bibnamefont {Poddubny}},\
  }\href {\doibase 10.1103/RevModPhys.95.015002} {\bibfield  {journal}
  {\bibinfo  {journal} {Rev. Mod. Phys.}\ }\textbf {\bibinfo {volume} {95}},\
  \bibinfo {pages} {015002} (\bibinfo {year} {2023})}\BibitemShut {NoStop}%
\bibitem [{\citenamefont {Sinha}\ \emph {et~al.}(2020)\citenamefont {Sinha},
  \citenamefont {Meystre}, \citenamefont {Goldschmidt}, \citenamefont {Fatemi},
  \citenamefont {Rolston},\ and\ \citenamefont
  {Solano}}]{PhysRevLett.124.043603}%
  \BibitemOpen
  \bibfield  {author} {\bibinfo {author} {\bibfnamefont {K.}~\bibnamefont
  {Sinha}}, \bibinfo {author} {\bibfnamefont {P.}~\bibnamefont {Meystre}},
  \bibinfo {author} {\bibfnamefont {E.~A.}\ \bibnamefont {Goldschmidt}},
  \bibinfo {author} {\bibfnamefont {F.~K.}\ \bibnamefont {Fatemi}}, \bibinfo
  {author} {\bibfnamefont {S.~L.}\ \bibnamefont {Rolston}}, \ and\ \bibinfo
  {author} {\bibfnamefont {P.}~\bibnamefont {Solano}},\ }\href {\doibase
  10.1103/PhysRevLett.124.043603} {\bibfield  {journal} {\bibinfo  {journal}
  {Phys. Rev. Lett.}\ }\textbf {\bibinfo {volume} {124}},\ \bibinfo {pages}
  {043603} (\bibinfo {year} {2020})}\BibitemShut {NoStop}%
\bibitem [{\citenamefont {Zhang}(2023)}]{PhysRevLett.131.193603}%
  \BibitemOpen
  \bibfield  {author} {\bibinfo {author} {\bibfnamefont {Y.-X.}\ \bibnamefont
  {Zhang}},\ }\href {\doibase 10.1103/PhysRevLett.131.193603} {\bibfield
  {journal} {\bibinfo  {journal} {Phys. Rev. Lett.}\ }\textbf {\bibinfo
  {volume} {131}},\ \bibinfo {pages} {193603} (\bibinfo {year}
  {2023})}\BibitemShut {NoStop}%
\bibitem [{\citenamefont {Lanuza}\ and\ \citenamefont
  {Schneble}(2024)}]{PhysRevResearch.6.033196}%
  \BibitemOpen
  \bibfield  {author} {\bibinfo {author} {\bibfnamefont {A.}~\bibnamefont
  {Lanuza}}\ and\ \bibinfo {author} {\bibfnamefont {D.}~\bibnamefont
  {Schneble}},\ }\href {\doibase 10.1103/PhysRevResearch.6.033196} {\bibfield
  {journal} {\bibinfo  {journal} {Phys. Rev. Res.}\ }\textbf {\bibinfo {volume}
  {6}},\ \bibinfo {pages} {033196} (\bibinfo {year} {2024})}\BibitemShut
  {NoStop}%
\bibitem [{\citenamefont {Pichler}\ and\ \citenamefont
  {Zoller}(2016)}]{PhysRevLett.116.093601}%
  \BibitemOpen
  \bibfield  {author} {\bibinfo {author} {\bibfnamefont {H.}~\bibnamefont
  {Pichler}}\ and\ \bibinfo {author} {\bibfnamefont {P.}~\bibnamefont
  {Zoller}},\ }\href {\doibase 10.1103/PhysRevLett.116.093601} {\bibfield
  {journal} {\bibinfo  {journal} {Phys. Rev. Lett.}\ }\textbf {\bibinfo
  {volume} {116}},\ \bibinfo {pages} {093601} (\bibinfo {year}
  {2016})}\BibitemShut {NoStop}%
\bibitem [{\citenamefont {Arranz~Regidor}\ \emph {et~al.}(2021)\citenamefont
  {Arranz~Regidor}, \citenamefont {Crowder}, \citenamefont {Carmichael},\ and\
  \citenamefont {Hughes}}]{PhysRevResearch.3.023030}%
  \BibitemOpen
  \bibfield  {author} {\bibinfo {author} {\bibfnamefont {S.}~\bibnamefont
  {Arranz~Regidor}}, \bibinfo {author} {\bibfnamefont {G.}~\bibnamefont
  {Crowder}}, \bibinfo {author} {\bibfnamefont {H.}~\bibnamefont {Carmichael}},
  \ and\ \bibinfo {author} {\bibfnamefont {S.}~\bibnamefont {Hughes}},\ }\href
  {\doibase 10.1103/PhysRevResearch.3.023030} {\bibfield  {journal} {\bibinfo
  {journal} {Phys. Rev. Res.}\ }\textbf {\bibinfo {volume} {3}},\ \bibinfo
  {pages} {023030} (\bibinfo {year} {2021})}\BibitemShut {NoStop}%
\bibitem [{\citenamefont {Calaj\'o}\ \emph {et~al.}(2019)\citenamefont
  {Calaj\'o}, \citenamefont {Fang}, \citenamefont {Baranger},\ and\
  \citenamefont {Ciccarello}}]{PhysRevLett.122.073601}%
  \BibitemOpen
  \bibfield  {author} {\bibinfo {author} {\bibfnamefont {G.}~\bibnamefont
  {Calaj\'o}}, \bibinfo {author} {\bibfnamefont {Y.-L.~L.}\ \bibnamefont
  {Fang}}, \bibinfo {author} {\bibfnamefont {H.~U.}\ \bibnamefont {Baranger}},
  \ and\ \bibinfo {author} {\bibfnamefont {F.}~\bibnamefont {Ciccarello}},\
  }\href {\doibase 10.1103/PhysRevLett.122.073601} {\bibfield  {journal}
  {\bibinfo  {journal} {Phys. Rev. Lett.}\ }\textbf {\bibinfo {volume} {122}},\
  \bibinfo {pages} {073601} (\bibinfo {year} {2019})}\BibitemShut {NoStop}%
\bibitem [{\citenamefont {Longo}\ \emph {et~al.}(2010)\citenamefont {Longo},
  \citenamefont {Schmitteckert},\ and\ \citenamefont
  {Busch}}]{PhysRevLett.104.023602}%
  \BibitemOpen
  \bibfield  {author} {\bibinfo {author} {\bibfnamefont {P.}~\bibnamefont
  {Longo}}, \bibinfo {author} {\bibfnamefont {P.}~\bibnamefont
  {Schmitteckert}}, \ and\ \bibinfo {author} {\bibfnamefont {K.}~\bibnamefont
  {Busch}},\ }\href {\doibase 10.1103/PhysRevLett.104.023602} {\bibfield
  {journal} {\bibinfo  {journal} {Phys. Rev. Lett.}\ }\textbf {\bibinfo
  {volume} {104}},\ \bibinfo {pages} {023602} (\bibinfo {year}
  {2010})}\BibitemShut {NoStop}%
\bibitem [{\citenamefont {Longo}\ \emph {et~al.}(2009)\citenamefont {Longo},
  \citenamefont {Schmitteckert},\ and\ \citenamefont {Busch}}]{Longo_2009}%
  \BibitemOpen
  \bibfield  {author} {\bibinfo {author} {\bibfnamefont {P.}~\bibnamefont
  {Longo}}, \bibinfo {author} {\bibfnamefont {P.}~\bibnamefont
  {Schmitteckert}}, \ and\ \bibinfo {author} {\bibfnamefont {K.}~\bibnamefont
  {Busch}},\ }\href {\doibase 10.1088/1464-4258/11/11/114009} {\bibfield
  {journal} {\bibinfo  {journal} {J. Opt. A: Pure Appl. Opt.}\ }\textbf
  {\bibinfo {volume} {11}},\ \bibinfo {pages} {114009} (\bibinfo {year}
  {2009})}\BibitemShut {NoStop}%
\bibitem [{\citenamefont {Andersson}\ \emph {et~al.}(2019)\citenamefont
  {Andersson}, \citenamefont {Suri}, \citenamefont {Guo}, \citenamefont
  {Aref},\ and\ \citenamefont {Delsing}}]{Andersson2019}%
  \BibitemOpen
  \bibfield  {author} {\bibinfo {author} {\bibfnamefont {G.}~\bibnamefont
  {Andersson}}, \bibinfo {author} {\bibfnamefont {B.}~\bibnamefont {Suri}},
  \bibinfo {author} {\bibfnamefont {L.}~\bibnamefont {Guo}}, \bibinfo {author}
  {\bibfnamefont {T.}~\bibnamefont {Aref}}, \ and\ \bibinfo {author}
  {\bibfnamefont {P.}~\bibnamefont {Delsing}},\ }\href {\doibase
  10.1038/s41567-019-0605-6} {\bibfield  {journal} {\bibinfo  {journal} {Nat.
  Phys.}\ }\textbf {\bibinfo {volume} {15}},\ \bibinfo {pages} {1123} (\bibinfo
  {year} {2019})}\BibitemShut {NoStop}%
\bibitem [{\citenamefont {Kannan}\ \emph {et~al.}(2020)\citenamefont {Kannan},
  \citenamefont {Ruckriegel}, \citenamefont {Campbell}, \citenamefont
  {Frisk~Kockum}, \citenamefont {Braum\"uller}, \citenamefont {Kim},
  \citenamefont {Kjaergaard}, \citenamefont {Krantz}, \citenamefont {Melville},
  \citenamefont {Niedzielski}, \citenamefont {Veps\"al\"ainen}, \citenamefont
  {Winik}, \citenamefont {Yoder}, \citenamefont {Nori}, \citenamefont
  {Orlando}, \citenamefont {Gustavsson},\ and\ \citenamefont
  {Oliver}}]{Kannan2020}%
  \BibitemOpen
  \bibfield  {author} {\bibinfo {author} {\bibfnamefont {B.}~\bibnamefont
  {Kannan}}, \bibinfo {author} {\bibfnamefont {M.~J.}\ \bibnamefont
  {Ruckriegel}}, \bibinfo {author} {\bibfnamefont {D.~L.}\ \bibnamefont
  {Campbell}}, \bibinfo {author} {\bibfnamefont {A.}~\bibnamefont
  {Frisk~Kockum}}, \bibinfo {author} {\bibfnamefont {J.}~\bibnamefont
  {Braum\"uller}}, \bibinfo {author} {\bibfnamefont {D.~K.}\ \bibnamefont
  {Kim}}, \bibinfo {author} {\bibfnamefont {M.}~\bibnamefont {Kjaergaard}},
  \bibinfo {author} {\bibfnamefont {P.}~\bibnamefont {Krantz}}, \bibinfo
  {author} {\bibfnamefont {A.}~\bibnamefont {Melville}}, \bibinfo {author}
  {\bibfnamefont {B.~M.}\ \bibnamefont {Niedzielski}}, \bibinfo {author}
  {\bibfnamefont {A.}~\bibnamefont {Veps\"al\"ainen}}, \bibinfo {author}
  {\bibfnamefont {R.}~\bibnamefont {Winik}}, \bibinfo {author} {\bibfnamefont
  {J.~L.}\ \bibnamefont {Yoder}}, \bibinfo {author} {\bibfnamefont
  {F.}~\bibnamefont {Nori}}, \bibinfo {author} {\bibfnamefont {T.~P.}\
  \bibnamefont {Orlando}}, \bibinfo {author} {\bibfnamefont {S.}~\bibnamefont
  {Gustavsson}}, \ and\ \bibinfo {author} {\bibfnamefont {W.~D.}\ \bibnamefont
  {Oliver}},\ }\href {\doibase 10.1038/s41586-020-2529-9} {\bibfield  {journal}
  {\bibinfo  {journal} {Nature}\ }\textbf {\bibinfo {volume} {583}},\ \bibinfo
  {pages} {775–779} (\bibinfo {year} {2020})}\BibitemShut {NoStop}%
\bibitem [{\citenamefont {Redchenko}\ \emph {et~al.}(2023)\citenamefont
  {Redchenko}, \citenamefont {Poshakinskiy}, \citenamefont {Sett},
  \citenamefont {Žemlička}, \citenamefont {Poddubny},\ and\ \citenamefont
  {Fink}}]{Redchenko2023}%
  \BibitemOpen
  \bibfield  {author} {\bibinfo {author} {\bibfnamefont {E.~S.}\ \bibnamefont
  {Redchenko}}, \bibinfo {author} {\bibfnamefont {A.~V.}\ \bibnamefont
  {Poshakinskiy}}, \bibinfo {author} {\bibfnamefont {R.}~\bibnamefont {Sett}},
  \bibinfo {author} {\bibfnamefont {M.}~\bibnamefont {Žemlička}}, \bibinfo
  {author} {\bibfnamefont {A.~N.}\ \bibnamefont {Poddubny}}, \ and\ \bibinfo
  {author} {\bibfnamefont {J.~M.}\ \bibnamefont {Fink}},\ }\href {\doibase
  10.1038/s41467-023-38761-6} {\bibfield  {journal} {\bibinfo  {journal} {Nat.
  Commun.}\ }\textbf {\bibinfo {volume} {14}},\ \bibinfo {pages} {2998}
  (\bibinfo {year} {2023})}\BibitemShut {NoStop}%
\bibitem [{\citenamefont {Li}\ \emph {et~al.}(2013)\citenamefont {Li},
  \citenamefont {Silveri}, \citenamefont {Kumar}, \citenamefont {Pirkkalainen},
  \citenamefont {Vepsäläinen}, \citenamefont {Chien}, \citenamefont
  {Tuorila}, \citenamefont {Sillanpää}, \citenamefont {Hakonen},
  \citenamefont {Thuneberg},\ and\ \citenamefont {Paraoanu}}]{LiJian2020}%
  \BibitemOpen
  \bibfield  {author} {\bibinfo {author} {\bibfnamefont {J.}~\bibnamefont
  {Li}}, \bibinfo {author} {\bibfnamefont {M.}~\bibnamefont {Silveri}},
  \bibinfo {author} {\bibfnamefont {K.}~\bibnamefont {Kumar}}, \bibinfo
  {author} {\bibfnamefont {J.-M.}\ \bibnamefont {Pirkkalainen}}, \bibinfo
  {author} {\bibfnamefont {A.}~\bibnamefont {Vepsäläinen}}, \bibinfo {author}
  {\bibfnamefont {W.}~\bibnamefont {Chien}}, \bibinfo {author} {\bibfnamefont
  {J.}~\bibnamefont {Tuorila}}, \bibinfo {author} {\bibfnamefont
  {M.}~\bibnamefont {Sillanpää}}, \bibinfo {author} {\bibfnamefont
  {P.}~\bibnamefont {Hakonen}}, \bibinfo {author} {\bibfnamefont
  {E.}~\bibnamefont {Thuneberg}}, \ and\ \bibinfo {author} {\bibfnamefont
  {G.}~\bibnamefont {Paraoanu}},\ }\href {\doibase 10.1038/ncomms2383}
  {\bibfield  {journal} {\bibinfo  {journal} {Nature}\ }\textbf {\bibinfo
  {volume} {4}},\ \bibinfo {pages} {1402} (\bibinfo {year} {2013})}\BibitemShut
  {NoStop}%
\end{thebibliography}

\begin{thebibliography}{18}%
\makeatletter
\providecommand \@ifxundefined [1]{%
 \@ifx{#1\undefined}
}%
\providecommand \@ifnum [1]{%
 \ifnum #1\expandafter \@firstoftwo
 \else \expandafter \@secondoftwo
 \fi
}%
\providecommand \@ifx [1]{%
 \ifx #1\expandafter \@firstoftwo
 \else \expandafter \@secondoftwo
 \fi
}%
\providecommand \natexlab [1]{#1}%
\providecommand \enquote  [1]{``#1''}%
\providecommand \bibnamefont  [1]{#1}%
\providecommand \bibfnamefont [1]{#1}%
\providecommand \citenamefont [1]{#1}%
\providecommand \href@noop [0]{\@secondoftwo}%
\providecommand \href [0]{\begingroup \@sanitize@url \@href}%
\providecommand \@href[1]{\@@startlink{#1}\@@href}%
\providecommand \@@href[1]{\endgroup#1\@@endlink}%
\providecommand \@sanitize@url [0]{\catcode `\\12\catcode `\$12\catcode
  `\&12\catcode `\#12\catcode `\^12\catcode `\_12\catcode `\%12\relax}%
\providecommand \@@startlink[1]{}%
\providecommand \@@endlink[0]{}%
\providecommand \url  [0]{\begingroup\@sanitize@url \@url }%
\providecommand \@url [1]{\endgroup\@href {#1}{\urlprefix }}%
\providecommand \urlprefix  [0]{URL }%
\providecommand \Eprint [0]{\href }%
\providecommand \doibase [0]{http://dx.doi.org/}%
\providecommand \selectlanguage [0]{\@gobble}%
\providecommand \bibinfo  [0]{\@secondoftwo}%
\providecommand \bibfield  [0]{\@secondoftwo}%
\providecommand \translation [1]{[#1]}%
\providecommand \BibitemOpen [0]{}%
\providecommand \bibitemStop [0]{}%
\providecommand \bibitemNoStop [0]{.\EOS\space}%
\providecommand \EOS [0]{\spacefactor3000\relax}%
\providecommand \BibitemShut  [1]{\csname bibitem#1\endcsname}%
\let\auto@bib@innerbib\@empty
%</preamble>
\bibitem [{\citenamefont {Trivedi}\ \emph {et~al.}(2018)\citenamefont
  {Trivedi}, \citenamefont {Fischer}, \citenamefont {Xu}, \citenamefont {Fan},\
  and\ \citenamefont {Vuckovic}}]{smPhysRevB.98.144112}%
  \BibitemOpen
  \bibfield  {author} {\bibinfo {author} {\bibfnamefont {R.}~\bibnamefont
  {Trivedi}}, \bibinfo {author} {\bibfnamefont {K.}~\bibnamefont {Fischer}},
  \bibinfo {author} {\bibfnamefont {S.}~\bibnamefont {Xu}}, \bibinfo {author}
  {\bibfnamefont {S.}~\bibnamefont {Fan}}, \ and\ \bibinfo {author}
  {\bibfnamefont {J.}~\bibnamefont {Vuckovic}},\ }\enquote {\bibinfo {title}
  {Few-photon scattering and emission from low-dimensional quantum systems},}\
  \href {\doibase 10.1103/PhysRevB.98.144112} {\bibfield  {journal} {\bibinfo
  {journal} {Phys. Rev. B}\ }\textbf {\bibinfo {volume} {98}},\ \bibinfo
  {pages} {144112} (\bibinfo {year} {2018})}\BibitemShut {NoStop}%
\bibitem [{\citenamefont {Trivedi}\ \emph {et~al.}(2020)\citenamefont
  {Trivedi}, \citenamefont {White}, \citenamefont {Fan},\ and\ \citenamefont
  {Vu\ifmmode \check{c}\else \v{c}\fi{}kovi\ifmmode~\acute{c}\else
  \'{c}\fi{}}}]{smPhysRevA.102.033707}%
  \BibitemOpen
  \bibfield  {author} {\bibinfo {author} {\bibfnamefont {R.}~\bibnamefont
  {Trivedi}}, \bibinfo {author} {\bibfnamefont {A.}~\bibnamefont {White}},
  \bibinfo {author} {\bibfnamefont {S.}~\bibnamefont {Fan}}, \ and\ \bibinfo
  {author} {\bibfnamefont {J.}~\bibnamefont {Vu\ifmmode \check{c}\else
  \v{c}\fi{}kovi\ifmmode~\acute{c}\else \'{c}\fi{}}},\ }\enquote {\bibinfo
  {title} {Analytic and geometric properties of scattering from periodically
  modulated quantum-optical systems},}\ \href {\doibase
  10.1103/PhysRevA.102.033707} {\bibfield  {journal} {\bibinfo  {journal}
  {Phys. Rev. A}\ }\textbf {\bibinfo {volume} {102}},\ \bibinfo {pages}
  {033707} (\bibinfo {year} {2020})}\BibitemShut {NoStop}%
\bibitem [{\citenamefont {Pichler}\ \emph {et~al.}(2015)\citenamefont
  {Pichler}, \citenamefont {Ramos}, \citenamefont {Daley},\ and\ \citenamefont
  {Zoller}}]{smPhysRevA.91.042116}%
  \BibitemOpen
  \bibfield  {author} {\bibinfo {author} {\bibfnamefont {H.}~\bibnamefont
  {Pichler}}, \bibinfo {author} {\bibfnamefont {T.}~\bibnamefont {Ramos}},
  \bibinfo {author} {\bibfnamefont {A.~J.}\ \bibnamefont {Daley}}, \ and\
  \bibinfo {author} {\bibfnamefont {P.}~\bibnamefont {Zoller}},\ }\enquote
  {\bibinfo {title} {Quantum optics of chiral spin networks},}\ \href {\doibase
  10.1103/PhysRevA.91.042116} {\bibfield  {journal} {\bibinfo  {journal} {Phys.
  Rev. A}\ }\textbf {\bibinfo {volume} {91}},\ \bibinfo {pages} {042116}
  (\bibinfo {year} {2015})}\BibitemShut {NoStop}%
\bibitem [{\citenamefont {Mok}\ \emph {et~al.}(2020)\citenamefont {Mok},
  \citenamefont {Aghamalyan}, \citenamefont {You}, \citenamefont {Haug},
  \citenamefont {Zhang}, \citenamefont {Png},\ and\ \citenamefont
  {Kwek}}]{smPhysRevResearch.2.013369}%
  \BibitemOpen
  \bibfield  {author} {\bibinfo {author} {\bibfnamefont {W.-K.}\ \bibnamefont
  {Mok}}, \bibinfo {author} {\bibfnamefont {D.}~\bibnamefont {Aghamalyan}},
  \bibinfo {author} {\bibfnamefont {J.-B.}\ \bibnamefont {You}}, \bibinfo
  {author} {\bibfnamefont {T.}~\bibnamefont {Haug}}, \bibinfo {author}
  {\bibfnamefont {W.}~\bibnamefont {Zhang}}, \bibinfo {author} {\bibfnamefont
  {C.~E.}\ \bibnamefont {Png}}, \ and\ \bibinfo {author} {\bibfnamefont
  {L.-C.}\ \bibnamefont {Kwek}},\ }\enquote {\bibinfo {title} {Long-distance
  dissipation-assisted transport of entangled states via a chiral waveguide},}\
  \href {\doibase 10.1103/PhysRevResearch.2.013369} {\bibfield  {journal}
  {\bibinfo  {journal} {Phys. Rev. Res.}\ }\textbf {\bibinfo {volume} {2}},\
  \bibinfo {pages} {013369} (\bibinfo {year} {2020})}\BibitemShut {NoStop}%
\bibitem [{\citenamefont {Adam}\ and\ \citenamefont
  {Seke}(1981)}]{smPhysRevA.23.3118}%
  \BibitemOpen
  \bibfield  {author} {\bibinfo {author} {\bibfnamefont {G.}~\bibnamefont
  {Adam}}\ and\ \bibinfo {author} {\bibfnamefont {J.}~\bibnamefont {Seke}},\
  }\enquote {\bibinfo {title} {Closed equations of motion for a system of $n$
  two-level atoms in the case of spontaneous emission},}\ \href {\doibase
  10.1103/PhysRevA.23.3118} {\bibfield  {journal} {\bibinfo  {journal} {Phys.
  Rev. A}\ }\textbf {\bibinfo {volume} {23}},\ \bibinfo {pages} {3118}
  (\bibinfo {year} {1981})}\BibitemShut {NoStop}%
\bibitem [{\citenamefont {Milonni}\ and\ \citenamefont
  {Knight}(1974)}]{smPhysRevA.10.1096}%
  \BibitemOpen
  \bibfield  {author} {\bibinfo {author} {\bibfnamefont {P.~W.}\ \bibnamefont
  {Milonni}}\ and\ \bibinfo {author} {\bibfnamefont {P.~L.}\ \bibnamefont
  {Knight}},\ }\enquote {\bibinfo {title} {Retardation in the resonant
  interaction of two identical atoms},}\ \href {\doibase
  10.1103/PhysRevA.10.1096} {\bibfield  {journal} {\bibinfo  {journal} {Phys.
  Rev. A}\ }\textbf {\bibinfo {volume} {10}},\ \bibinfo {pages} {1096}
  (\bibinfo {year} {1974})}\BibitemShut {NoStop}%
\bibitem [{\citenamefont {Ilin}\ \emph {et~al.}(2023)\citenamefont {Ilin},
  \citenamefont {Poshakinskiy}, \citenamefont {Poddubny},\ and\ \citenamefont
  {Iorsh}}]{smPhysRevLett.130.023601}%
  \BibitemOpen
  \bibfield  {author} {\bibinfo {author} {\bibfnamefont {D.}~\bibnamefont
  {Ilin}}, \bibinfo {author} {\bibfnamefont {A.~V.}\ \bibnamefont
  {Poshakinskiy}}, \bibinfo {author} {\bibfnamefont {A.~N.}\ \bibnamefont
  {Poddubny}}, \ and\ \bibinfo {author} {\bibfnamefont {I.}~\bibnamefont
  {Iorsh}},\ }\enquote {\bibinfo {title} {Frequency combs with parity-protected
  cross-correlations and entanglement from dynamically modulated qubit
  arrays},}\ \href {\doibase 10.1103/PhysRevLett.130.023601} {\bibfield
  {journal} {\bibinfo  {journal} {Phys. Rev. Lett.}\ }\textbf {\bibinfo
  {volume} {130}},\ \bibinfo {pages} {023601} (\bibinfo {year}
  {2023})}\BibitemShut {NoStop}%
\bibitem [{\citenamefont {Vyatkin}\ \emph {et~al.}(2023)\citenamefont
  {Vyatkin}, \citenamefont {Poshakinskiy},\ and\ \citenamefont
  {Poddubny}}]{smPhysRevA.108.023715}%
  \BibitemOpen
  \bibfield  {author} {\bibinfo {author} {\bibfnamefont {E.~S.}\ \bibnamefont
  {Vyatkin}}, \bibinfo {author} {\bibfnamefont {A.~V.}\ \bibnamefont
  {Poshakinskiy}}, \ and\ \bibinfo {author} {\bibfnamefont {A.~N.}\
  \bibnamefont {Poddubny}},\ }\enquote {\bibinfo {title} {Resonant parametric
  photon generation in waveguide-coupled quantum emitter arrays},}\ \href
  {\doibase 10.1103/PhysRevA.108.023715} {\bibfield  {journal} {\bibinfo
  {journal} {Phys. Rev. A}\ }\textbf {\bibinfo {volume} {108}},\ \bibinfo
  {pages} {023715} (\bibinfo {year} {2023})}\BibitemShut {NoStop}%
\bibitem [{\citenamefont {Zhong}\ \emph {et~al.}(2020)\citenamefont {Zhong},
  \citenamefont {Olekhno}, \citenamefont {Ke}, \citenamefont {Poshakinskiy},
  \citenamefont {Lee}, \citenamefont {Kivshar},\ and\ \citenamefont
  {Poddubny}}]{smPhysRevLett.124.093604}%
  \BibitemOpen
  \bibfield  {author} {\bibinfo {author} {\bibfnamefont {J.}~\bibnamefont
  {Zhong}}, \bibinfo {author} {\bibfnamefont {N.~A.}\ \bibnamefont {Olekhno}},
  \bibinfo {author} {\bibfnamefont {Y.}~\bibnamefont {Ke}}, \bibinfo {author}
  {\bibfnamefont {A.~V.}\ \bibnamefont {Poshakinskiy}}, \bibinfo {author}
  {\bibfnamefont {C.}~\bibnamefont {Lee}}, \bibinfo {author} {\bibfnamefont
  {Y.~S.}\ \bibnamefont {Kivshar}}, \ and\ \bibinfo {author} {\bibfnamefont
  {A.~N.}\ \bibnamefont {Poddubny}},\ }\enquote {\bibinfo {title}
  {Photon-mediated localization in two-level qubit arrays},}\ \href {\doibase
  10.1103/PhysRevLett.124.093604} {\bibfield  {journal} {\bibinfo  {journal}
  {Phys. Rev. Lett.}\ }\textbf {\bibinfo {volume} {124}},\ \bibinfo {pages}
  {093604} (\bibinfo {year} {2020})}\BibitemShut {NoStop}%
\bibitem [{\citenamefont {Poshakinskiy}\ \emph {et~al.}(2021)\citenamefont
  {Poshakinskiy}, \citenamefont {Zhong},\ and\ \citenamefont
  {Poddubny}}]{smPhysRevLett.126.203602}%
  \BibitemOpen
  \bibfield  {author} {\bibinfo {author} {\bibfnamefont {A.~V.}\ \bibnamefont
  {Poshakinskiy}}, \bibinfo {author} {\bibfnamefont {J.}~\bibnamefont {Zhong}},
  \ and\ \bibinfo {author} {\bibfnamefont {A.~N.}\ \bibnamefont {Poddubny}},\
  }\enquote {\bibinfo {title} {Quantum chaos driven by long-range
  waveguide-mediated interactions},}\ \href {\doibase
  10.1103/PhysRevLett.126.203602} {\bibfield  {journal} {\bibinfo  {journal}
  {Phys. Rev. Lett.}\ }\textbf {\bibinfo {volume} {126}},\ \bibinfo {pages}
  {203602} (\bibinfo {year} {2021})}\BibitemShut {NoStop}%
\bibitem [{\citenamefont {Poshakinskiy}\ and\ \citenamefont
  {Poddubny}(2016)}]{smPhysRevA.93.033856}%
  \BibitemOpen
  \bibfield  {author} {\bibinfo {author} {\bibfnamefont {A.~V.}\ \bibnamefont
  {Poshakinskiy}}\ and\ \bibinfo {author} {\bibfnamefont {A.~N.}\ \bibnamefont
  {Poddubny}},\ }\enquote {\bibinfo {title} {Biexciton-mediated superradiant
  photon blockade},}\ \href {\doibase 10.1103/PhysRevA.93.033856} {\bibfield
  {journal} {\bibinfo  {journal} {Phys. Rev. A}\ }\textbf {\bibinfo {volume}
  {93}},\ \bibinfo {pages} {033856} (\bibinfo {year} {2016})}\BibitemShut
  {NoStop}%
\bibitem [{\citenamefont {Ke}\ \emph {et~al.}(2019)\citenamefont {Ke},
  \citenamefont {Poshakinskiy}, \citenamefont {Lee}, \citenamefont {Kivshar},\
  and\ \citenamefont {Poddubny}}]{smPhysRevLett.123.253601}%
  \BibitemOpen
  \bibfield  {author} {\bibinfo {author} {\bibfnamefont {Y.}~\bibnamefont
  {Ke}}, \bibinfo {author} {\bibfnamefont {A.~V.}\ \bibnamefont
  {Poshakinskiy}}, \bibinfo {author} {\bibfnamefont {C.}~\bibnamefont {Lee}},
  \bibinfo {author} {\bibfnamefont {Y.~S.}\ \bibnamefont {Kivshar}}, \ and\
  \bibinfo {author} {\bibfnamefont {A.~N.}\ \bibnamefont {Poddubny}},\
  }\enquote {\bibinfo {title} {Inelastic scattering of photon pairs in qubit
  arrays with subradiant states},}\ \href {\doibase
  10.1103/PhysRevLett.123.253601} {\bibfield  {journal} {\bibinfo  {journal}
  {Phys. Rev. Lett.}\ }\textbf {\bibinfo {volume} {123}},\ \bibinfo {pages}
  {253601} (\bibinfo {year} {2019})}\BibitemShut {NoStop}%
\bibitem [{\citenamefont {Eisert}\ \emph {et~al.}(2010)\citenamefont {Eisert},
  \citenamefont {Cramer},\ and\ \citenamefont {Plenio}}]{smRevModPhys.82.277}%
  \BibitemOpen
  \bibfield  {author} {\bibinfo {author} {\bibfnamefont {J.}~\bibnamefont
  {Eisert}}, \bibinfo {author} {\bibfnamefont {M.}~\bibnamefont {Cramer}}, \
  and\ \bibinfo {author} {\bibfnamefont {M.~B.}\ \bibnamefont {Plenio}},\
  }\enquote {\bibinfo {title} {Colloquium: Area laws for the entanglement
  entropy},}\ \href {\doibase 10.1103/RevModPhys.82.277} {\bibfield  {journal}
  {\bibinfo  {journal} {Rev. Mod. Phys.}\ }\textbf {\bibinfo {volume} {82}},\
  \bibinfo {pages} {277} (\bibinfo {year} {2010})}\BibitemShut {NoStop}%
\bibitem [{\citenamefont {Pichler}\ and\ \citenamefont
  {Zoller}(2016)}]{smPhysRevLett.116.093601}%
  \BibitemOpen
  \bibfield  {author} {\bibinfo {author} {\bibfnamefont {H.}~\bibnamefont
  {Pichler}}\ and\ \bibinfo {author} {\bibfnamefont {P.}~\bibnamefont
  {Zoller}},\ }\enquote {\bibinfo {title} {Photonic circuits with time delays
  and quantum feedback},}\ \href {\doibase 10.1103/PhysRevLett.116.093601}
  {\bibfield  {journal} {\bibinfo  {journal} {Phys. Rev. Lett.}\ }\textbf
  {\bibinfo {volume} {116}},\ \bibinfo {pages} {093601} (\bibinfo {year}
  {2016})}\BibitemShut {NoStop}%
\bibitem [{\citenamefont {Arranz~Regidor}\ \emph {et~al.}(2021)\citenamefont
  {Arranz~Regidor}, \citenamefont {Crowder}, \citenamefont {Carmichael},\ and\
  \citenamefont {Hughes}}]{smPhysRevResearch.3.023030}%
  \BibitemOpen
  \bibfield  {author} {\bibinfo {author} {\bibfnamefont {S.}~\bibnamefont
  {Arranz~Regidor}}, \bibinfo {author} {\bibfnamefont {G.}~\bibnamefont
  {Crowder}}, \bibinfo {author} {\bibfnamefont {H.}~\bibnamefont {Carmichael}},
  \ and\ \bibinfo {author} {\bibfnamefont {S.}~\bibnamefont {Hughes}},\
  }\enquote {\bibinfo {title} {Modeling quantum light-matter interactions in
  waveguide qed with retardation, nonlinear interactions, and a time-delayed
  feedback: Matrix product states versus a space-discretized waveguide
  model},}\ \href {\doibase 10.1103/PhysRevResearch.3.023030} {\bibfield
  {journal} {\bibinfo  {journal} {Phys. Rev. Res.}\ }\textbf {\bibinfo {volume}
  {3}},\ \bibinfo {pages} {023030} (\bibinfo {year} {2021})}\BibitemShut
  {NoStop}%
\bibitem [{\citenamefont {Calaj\'o}\ \emph {et~al.}(2019)\citenamefont
  {Calaj\'o}, \citenamefont {Fang}, \citenamefont {Baranger},\ and\
  \citenamefont {Ciccarello}}]{smPhysRevLett.122.073601}%
  \BibitemOpen
  \bibfield  {author} {\bibinfo {author} {\bibfnamefont {G.}~\bibnamefont
  {Calaj\'o}}, \bibinfo {author} {\bibfnamefont {Y.-L.~L.}\ \bibnamefont
  {Fang}}, \bibinfo {author} {\bibfnamefont {H.~U.}\ \bibnamefont {Baranger}},
  \ and\ \bibinfo {author} {\bibfnamefont {F.}~\bibnamefont {Ciccarello}},\
  }\enquote {\bibinfo {title} {Exciting a bound state in the continuum through
  multiphoton scattering plus delayed quantum feedback},}\ \href {\doibase
  10.1103/PhysRevLett.122.073601} {\bibfield  {journal} {\bibinfo  {journal}
  {Phys. Rev. Lett.}\ }\textbf {\bibinfo {volume} {122}},\ \bibinfo {pages}
  {073601} (\bibinfo {year} {2019})}\BibitemShut {NoStop}%
\bibitem [{\citenamefont {Longo}\ \emph {et~al.}(2010)\citenamefont {Longo},
  \citenamefont {Schmitteckert},\ and\ \citenamefont
  {Busch}}]{smPhysRevLett.104.023602}%
  \BibitemOpen
  \bibfield  {author} {\bibinfo {author} {\bibfnamefont {P.}~\bibnamefont
  {Longo}}, \bibinfo {author} {\bibfnamefont {P.}~\bibnamefont
  {Schmitteckert}}, \ and\ \bibinfo {author} {\bibfnamefont {K.}~\bibnamefont
  {Busch}},\ }\enquote {\bibinfo {title} {Few-photon transport in
  low-dimensional systems: Interaction-induced radiation trapping},}\ \href
  {\doibase 10.1103/PhysRevLett.104.023602} {\bibfield  {journal} {\bibinfo
  {journal} {Phys. Rev. Lett.}\ }\textbf {\bibinfo {volume} {104}},\ \bibinfo
  {pages} {023602} (\bibinfo {year} {2010})}\BibitemShut {NoStop}%
\bibitem [{\citenamefont {Longo}\ \emph {et~al.}(2009)\citenamefont {Longo},
  \citenamefont {Schmitteckert},\ and\ \citenamefont {Busch}}]{smLongo_2009}%
  \BibitemOpen
  \bibfield  {author} {\bibinfo {author} {\bibfnamefont {P.}~\bibnamefont
  {Longo}}, \bibinfo {author} {\bibfnamefont {P.}~\bibnamefont
  {Schmitteckert}}, \ and\ \bibinfo {author} {\bibfnamefont {K.}~\bibnamefont
  {Busch}},\ }\enquote {\bibinfo {title} {Dynamics of photon transport through
  quantum impurities in dispersion-engineered one-dimensional systems},}\ \href
  {\doibase 10.1088/1464-4258/11/11/114009} {\bibfield  {journal} {\bibinfo
  {journal} {J. Opt. A: Pure Appl. Opt.}\ }\textbf {\bibinfo {volume} {11}},\
  \bibinfo {pages} {114009} (\bibinfo {year} {2009})}\BibitemShut {NoStop}%
\end{thebibliography}
\end{document}